\documentclass[sigconf, nonacm]{acmart}

\newcommand\vldbdoi{}
\newcommand\vldbpages{}
\newcommand\vldbvolume{16}
\newcommand\vldbissue{10}
\newcommand\vldbyear{2023}
\newcommand\vldbauthors{\authors}
\newcommand\vldbtitle{\shorttitle} 
\newcommand\vldbavailabilityurl{URL_TO_YOUR_ARTIFACTS}
\newcommand\vldbpagestyle{plain} 

\usepackage{flushend}
\usepackage[a-1b]{pdfx}   
\usepackage{balance}

\usepackage{color}
\usepackage[labelfont=bf]{caption}
\usepackage{subcaption}
\usepackage{hyperref}
\usepackage{makecell}
\usepackage{fixltx2e}
\usepackage{rotating,multirow}
\usepackage{etoolbox,xspace}
\usepackage{algorithmicx}
\usepackage{algorithm}
\usepackage{fontawesome}
\usepackage{algpseudocode}
\algtext*{EndWhile}
\algtext*{EndIf}
\algtext*{EndFunction}
\algtext*{EndFor}
\algrenewcommand\algorithmicrequire{\textbf{Input:}}
\algrenewcommand\algorithmicensure{\textbf{Output:}}

\usepackage{amsthm}

\usepackage{tabularx}
\usepackage{blindtext}
\usepackage{ragged2e}  
\usepackage{booktabs}  
\usepackage{textcomp}
\usepackage{url}
\usepackage{comment}
\usepackage{enumitem}
\usepackage{nicematrix}
\usepackage[simplified]{pgf-umlcd}
\usepackage{tikz}
\usepackage{diagbox}

\newcolumntype{Y}{>{\RaggedRight\arraybackslash}X}

\newtheorem{definition}{Definition}

\newcommand{\eop}{\hspace*{\fill}\mbox{$\Box$}}     
\newenvironment{proof}{\paragraph{Proof:}}{\hfill$\square$}
\newcounter{example}
\renewcommand{\theexample}{\arabic{example}}
\newenvironment{example}{
        \vspace{1ex}
        \refstepcounter{example}
        {\noindent\bf Example \theexample:}}
	{\eop\vspace{1ex}}

\newcommand{\squishlist}{
 \begin{list}{$\bullet$}
  { \setlength{\itemsep}{0pt}
     \setlength{\parsep}{1pt}
     \setlength{\topsep}{1pt}
     \setlength{\partopsep}{0pt}
     \setlength{\leftmargin}{1em}
     \setlength{\labelwidth}{1em}
     \setlength{\labelsep}{0.5em} } }

\newcommand{\squishend}{
  \end{list}
}

\definecolor{americanrose}{rgb}{1.0, 0.01, 0.24}
\definecolor{airforceblue}{rgb}{0.36, 0.54, 0.66}
\definecolor{ao(english)}{rgb}{0.0, 0.5, 0.0}
\definecolor{ao}{rgb}{0.0, 0.0, 1.0}

\newcommand{\nima}[1]{\textcolor{red}{Nima: #1}}

\newcommand{\abol}[1]{\textcolor{red}{Abol: #1}}
\newcommand{\fn}[1]{\textcolor{red}{FN: #1}}

\newcommand\fd[1]{\textcolor{red}{(Further Discussion) #1}}
\newcommand\review[1]{\vspace{1ex}\noindent{\em #1}}
\newcommand\response[1]{\textcolor{blue}{#1}}

\newcommand{\rev}[1]{\textcolor{black}{#1}}
\newcommand{\eat}[1]{}

\newcommand{\techrep}[1]{#1}
\newcommand{\submit}[1]{}

\usepackage{xspace}
\newcommand{\stitle}[1]{\vspace{2mm}\noindent{\bf #1:}}

\newcommand{\gee}{\mathcal{G}}
\newcommand{\dee}{\mathcal{D}}

\newcommand{\nofly}{{\sc NoFlyCompas}\xspace}
\newcommand{\fmatch}{{\sc FacultyMatch}\xspace}

\renewcommand{\marginpar}[2][]{}

\usepackage{bbm, dsfont}

\newcommand{\spheading}
[2][5em]{
  \rotatebox{90}{\parbox{#1}{\raggedright #2}}}
\newcommand{\spheadingp}
[2][6.5em]{
  \rotatebox{90}{\parbox{#1}{\raggedright #2}}}

\begin{document}

\submit{


\begin{center}
    {\LARGE \bf Paper ID 1374 Author Response: \\
Through the Fairness Lens: Experimental Analysis and Evaluation of Entity Matching
}
\end{center}

\vspace{3mm}\noindent
We would like to thank the reviewers and the meta-reviewer for their careful evaluation of our paper and their insightful comments, which helped us significantly improve the paper. Below we outline the revisions made in response to each review comment. In the revised version of our manuscript, we have highlighted all changes in \textcolor{blue}{blue}.
The summary of changes to the manuscript are:
\textcolor{blue}{
\begin{itemize}[leftmargin=*]
    \item {\bf New experiment results} to evaluate the sensitivity of the models to the choice of threshold value (\S~\ref{sec:exp:sensitivity}).
    \item \textbf{Clarified and formalized} the study (\S~\ref{sec:3}). Also added a new motivating example for additional clarification (Example~\ref{eg:loyal}).
    \item \textbf{Improved presentation of the experimental analysis} (\S~\ref{sec:exp}).
    \item {Better articulation of the \textbf{differences between fairness in entity matching vs. fairness in general ML}} (\S~\ref{sec:measuresdiscussion}).
    \item \textbf{Thorough concluding remarks} (\S~\ref{sec:discussion}) and added a set of actionable rules of thumb for responsible EM (Table~\ref{tab:Rules}).
    \item \textbf{Improved presentation and writing}
        \begin{itemize}
            \item We fixed all the presentation issues raised by the reviewers.
            \item We made several careful passes over the paper to improve its writing and presentation.
        \end{itemize}
\end{itemize}
}
In the rest of our author response document, we provide clarifications on some of the issues raised by reviewers in more detail and explain the set of actions we took to resolve them:

\section*{Reviewer \#1}\label{rev1}

\review{
W1: Experimental section is hard to follow and lacks some additional information.
}

\response{
Thank you for your comment. Addressing the detailed comments by the reviewer (and other reviewers), we improved the readability of the experiments section (\S~\ref{sec:exp}).
}

\review{
W2: There is no experiment on the matching threshold.
}

\response{
In light of the reviewer's comment, we conducted thorough experiments to evaluate the sensitivity of the proposed approaches on the choice of matching threshold, considering all neural and non-neural matchers.
At a high level, our experiments confirm that while the overall accuracy of the matchers is more smooth and less sensitive to the choice of threshold, fairness of the matchers (TPRP and PPVP) is more sensitive; this is particularly true for neural matchers. 
The new experiment is added in \S~\ref{sec:exp:sensitivity}.
We also updated \S~\ref{sec:discussion} and the lessons learned accordingly. 
}

\eat{
\response{\fd{Add this as an insight.}
Based on the new experiments we ran on the effect of choice of similarity threshold on accuracy and the fairness of the matchers, we observed that the neural matchers are more sensitive to the choice of the threshold while non-neural matchers are significantly less sensitive. This suggests the classic problem of multi-objective optimization with multiple potential solutions such as letting the user from the non-dominated Pareto frontier, a weighted sum over the accuracy and fairness, or constraining one objective and optimizing the other and repeating this for the other one. There are two messages: 1) There exist thresholds where some of the neural matchers could be heavily biased 2) We seem to find some thresholds where neural matchers show reasonably high accuracy while being relatively bias-free. This underlines the importance of doing an evaluation on both accuracy and fairness before setting thresholds for a given task in production. Given a model and a dataset, at some thresholds, the model may show high bias whereas, in the neighboring thresholds, it may show low bias, which underscores the amount of sensitivity resulting from the choice of threshold. In other words, these models are not robust in terms of fairness and fairness is not a smooth function. If you care about fairness you have to evaluate it on your dataset or a sample of your dataset in order to balance accuracy and fairness in a way that it's acceptable to you.}
\nima{This was a summary of a discussion in one of our meetings, so I thought to put it somewhere in case we want to add it to the paper and/or author response.} \abol{not sure if we want to delete/keep/summarize this. But I left it here for now. Is there any part from this that you would like to add to 5.3.4 (or other parts of the paper?)}
\abol{Specifically, let's not make a big deal out of the multi-objective optimization. It is a side-track and in my view might become dangerous.}
}

\review{
D2: Why do you need the abstraction for $S_A$ and $S_B$? It does not add much to understanding formalism.
Two sets of entities $A$ and $B$ should be clear enough.
}
\response{Resolved in \S~\ref{sec:background}, as suggested by the reviewer.}

\review{
D3: I am not sure whether choosing the manufacturer as a sensitive attribute makes sense. Fairness experiments should really focus on real protected attributes and not made-up ones.
}

\response{
In our experiments, we used two types of datasets.
The two semi-synthetic datasets were designed to focus on realistic protected attributes (race and country).
We additionally used six existing EM benchmarking datasets for which ground truth labels were available to study fairness of matchers on diverse (textual, dirty, etc.) datasets.
However, since these benchmark datasets were not designed for auditing fairness in EM, they did not have any natural sensitive attributes unfortunately.
We compromised by treating categorical attributes like manufacturer and genre as sensitive attributes to illustrate the behavior of the matchers.
There is clearly a need for benchmark datasets for evaluating fairness in entity matching, and our two semi-synthetic social datasets are a first step in this direction.
\eat{
existing EM benchmarking datasets were not designed for auditing fairness in EM. 
As highlighted in our first insight, there is a clear need for benchmark datasets for evaluating fairness in entity matching.
That being said, 
the selected textual datasets from WDC repository are benchmark datasets for EM tasks and the best ones we could find trying to cover a variety of types of datasets in our experiments. 
We tried our best to choose the best sensitive attributes with some social ties existing in the data. 
For example, in {\sc iTunesAmazon} dataset, we selected {\small \tt gener} as the senstive attribute due to its correlation with race (e.g., rap is more popular among black singers).
We leave the work on generating benchmark textual datasets for fairness to future works.
}
}

\review{
D4: ``Next, we handpick some of the generated features based on which we declare matching conditions.'' This needs more clarification, potentially link to a repository where it is well-documented which attributes were treated how.
}

\response{
In light of the reviewer's suggestion, we added the link to our repository in which we explain which features were chosen and what settings were considered for the rule-based matching. Furthermore, we explained how the rules were chosen and added references to the existing work on rule selection in rule-based EM methods.}

\review{
D5: Tables 5 and 6: it would be better to highlight the outperforming numbers.
}
\response{Resolved.}

\review{
D6: Similarity of 0.5 seems not to be justified and quite generous. Considering the motivation that talks about similar names in specific groups, this threshold needs to be systematically evaluated. The example with ``browne'' also shows that the matching threshold is way too low.
}

\response{
The threshold of 0.5 has a probabilistic interpretation of having a higher likelihood of being a match rather than being non-match. 
In other words, we follow a randomized rounding logic, where non-integer values in range $(0,1)$ are rounded to the closer integer.
The choice of the threshold in our experiments is an a-priori task with respect to the probabilistic interpretation of the threshold.
We added this clarification to the paper. 
Besides, a-posteriori, we empirically observed a reasonably high accuracy of around 0.5 among all the matchers in our new experiments. 
Furthermore, it is often the case that using a higher threshold one may be able to capture mismatches suggested by the reviewer, however, that will result in a low recall and a low overall F1 that would basically violate the reason behind using the matcher.
}

\review{
D7: Figure 2 and 3: There should be a description on how to interpret these type of graph. I even asked some other colleagues and we can only make guesses.
}

\response{We thank the reviewer for their comment. Each colored point in the plots is indicative of a matcher that is unfair with respect to a particular group (x-axis) and a fairness measure (y-axis) for a specific dataset and a given fairness threshold.
Per the reviewer's suggestion, we added the explanation and visual cues on how to interpret the plots to Figure~\ref{fig:DBLP-ACM-single}.}

\review{
D8: 5.3.1 and 5.3.2 are a bit confusing. They seem to be based on the experiments moved to 42 But what about summary of 5.2?
}

\response{
As the reviewer correctly pointed out, \S~\ref{sec:evalcorrecteness} was summarizing some experiments that had been moved to [42] (now \cite{techrep}). However, the main results supporting \S~\ref{sec:evalmeasures} and \S~\ref{sec:social_exp} were in the submitted manuscript.  
In light of the reviewer's comment, we made an extra-careful pass over these three sections and made sure to make it clear, independent from the technical report.
}

\review{
D9: Why is there an entire discussion on intersectional subgroups (the lattice etc.) when the experiments only considered groups?
}

\response{
Per the reviewer's suggestion, we removed the discussion on intersectional fairness and will leave it to future works.
}

\review{
D10: I do not understand the lesson in (vi). Are you suggesting other ensemble learning techniques are needed or that the existing ones 23 and 46 are good enough? It seems that 46 was not included in this benchmark- why not?
}

\response{
The mentioned studies [23] (now~\cite{jurek2017novel}) and [46](now~\cite{yi2017method})
develop ensemble learning techniques with the objective of improving the overall EM accuracy, but do not seek to resolve EM unfairness. 
Our suggestion is to develop ensemble techniques to improve fairness, for which new techniques are needed.}
\eat{
Our suggestion is based on our observation in the experiments, where some of the matchers performed better for some groups while the others were more accurate for the other groups.
Consequently, we suggested the idea of using an ensemble of (neural and non-neural) matchers, while using the more accurate one for each group with the hope that it eventually may result in less disparities.
The mentioned studies 
develop ensemble learning techniques with the objective of improving the overall EM accuracy, but does not seek to resolve EM unfairness. As a result, their design is not similar to the idea we suggested and hence are not relevant.
}
\response{
We did not include [46](now~\cite{yi2017method}) in the benchmark since 
our study only focused on the state-of-the-art EM systems, while
in [46](now~\cite{yi2017method}), an ensemble of models is created based on different similarity metric schemes. 
}

\review{
R1: I think the paper could be more accessible and crisp if the authors focus on fewer core messages in their evaluation to leave clear boundaries for future research. For example, the authors could leave out the discussion on intersectional groups for later studies and instead focus on a deeper discussion of their evaluation results. Discussions on accuracy vs. Fairness and different matching thresholds could be carried out in more depth and more explicitly. Right now the reader has to jump back and forth between the tech report and the paper at hand to get the full picture on a particular experimental setting.
}

\response{
Thank you again for the insightful comments and suggestions. 
As mentioned earlier, following your suggestions, we moved the discussions on the intersectional groups for the future and focused the discussions more on accuracy v.s. fairness. 
We made sure to make the experiment section (\S~\ref{sec:exp}) independent from the technical report as much as possible.
}

\review{There is also shortening potential with regard to some ``lessons learned'': We need more benchmarks.
}

\response{
Thank you for your suggestion. We shortened the need-for-benchmark section, as suggested by the reviewer. We also made a careful pass over the paper to shorten different sections when possible. It also helped us fit the revised manuscript into the page limit.
}

\section*{Reviewer \#2}\label{rev2}

\review{
W1: \textbf{The lack of clarity and formalization of fairness and bias:}}

\review{D1-1: The authors provided different types and measures of fairness. The paper lacks a comparative framework to help decide based on the fairness type of a set of measures. This enables researchers to identify the proper measures to compare a set of EM approaches. It was hard to follow the discussion in Section 3.}

\response{Thank you for your comment. We would like to recall that in \S~\ref{sec:measuresdiscussion}, we have clarified insights for selecting the fairness measures. The choice of fairness measure is usually context-specific and depends on the costs of TPs, FPs, etc. That being said, due to the inherent class imbalance in EM tasks and the fact that {true matches} are rare events, we recommend using {\bf TPRP} and {\bf PPVP} measures, as those show more potential in revealing the unfairness associated with the correct match predictions among the true matches among different groups. In our evaluation framework, we only focused on the notion of {\em (sub)group fairness} and will leave {\em individual} and {\em causal} fairness to future works. Per the reviewer's request, we ensured to add more clarification to \S~\ref{sec:3} on how to select the fairness type and measures.}

\review{D1-2: Moreover, the paper did not clearly define fairness or bias. This definition enables researchers to determine the scope of studying fairness.}

\response{In this study (similar to many of the existing works on algorithmic fairness and responsible data science) bias and unfairness are used interchangeably to refer to when a matcher is not fair. 
The formal definitions of fairness measures and their descriptions are provided in Table~\ref{tbl:measures}, and explained in \S~\ref{sec:single_pairwise_lens}. Measuring unfairness is explained in \S~\ref{sec:measuring_unfairness}. The scope of the studying fairness is the measures that provided in Table~\ref{tbl:measures}. Per reviewer's suggestion, we made a careful pass over \S~\ref{sec:3} to ensure that these points are clear.}

\review{For example, the authors mentioned, “One of the reasons Ditto was unfair for VLDBJ is that, similar to the following example, it is common to publish extended versions of previously published papers in this venue.” Is any false positive due to bias or unfairness? Is there a case where the model will predict VLDB instead of VLDBJ?
The framework should help researchers determine the scope of fairness and the suitable measures to be used.
}

\response{
At a high level, as formalized in \S~\ref{sec:measuring_unfairness}, unfairness refers to the 
(performance) {\em difference} (based on any correctness metric) between different demographic groups. 
For example, the FP rate does not indicate unfairness by itself but having {\em different} FP rates for different groups is considered unfairness. 
We want to remind that the model's task is to determine whether two entities are a match/non-match and they do not predict a class label associated with a (sensitive or non-sensitive) attribute.
}

\review{
D2/W2: \textbf{The lack of thorough concluding remarks:}

The evaluation section (\S~\ref{sec:exp}) provided several experiments for the different EM approaches. The authors provided brief explanations of the results. There are no thorough concluding remarks that recommend a particular approach for a specific situation or a dataset with certain characteristics.
Moreover, the authors provided no insights to overcome bias in these approaches.
The authors, instead, gave general and intuitive statements, such as:\\
 -- `` No matcher is unfair across all datasets and no matcher is unfair across all measures.'' $\Rightarrow$ It would be more interesting if the authors discussed which matcher to be used in which situation or dataset.
 }
 
\response{Thank you for the comment. Following this comment, we updated the concluding remarks to recommend particular approaches based on a dataset specifications and provided some rules of thumb (Table~\ref{tab:Rules}) for responsible EM.
We removed the statement highlighted by the reviewer.
}

\review{
-- ``we observed that neural matchers are more accurate than non-neural matchers on textual and dirty data.'' and ``non-neural matchers are more accurate than neural matchers on structured data.'' $\Rightarrow$ How does the format of the data lead to these differences? Does the fairness depend more on the used features and the coverage of the training dataset?
}

\response{
Thank you for the comment.
On the performance, it has also been reported in~\cite{mudgal2018deep} that non-neural models performed better for structured datasets because of considering the data set structure.
On fairness, at a high level, 
not fully considering the dataset structures and heavily relying on semantic similarities (which can be different between different groups) and (biased) pretrained models hurts the fairness of neural matchers.
On the other hand, heavily relying on problematic proxies hurts the fairness of non-neural matchers.
In light of the reviewer's comment, in addition to \S~\ref{sec:evalmatchers} and \S~\ref{sec:discussion}, we added proper clarifications to different parts of the paper, including ``summary of results'' in \S~\ref{sec:comprehensive_results}.
We also reflected these in our ``rules of thumb'' (Table~\ref{tab:Rules}).
}

\review{
D3/W3: \textbf{The presentation of the paper needs much work:}

The paper would be easier to follow if the authors added more structure to assist the readers in detecting the crucial details of the study, such as algorithms to help researchers reproduce the results of this study.
The paper contains strong statements without references. It is important to revise and strengthen those statements by providing appropriate citations to support them. For example:
\begin{itemize}
    \item ``Furthermore, such techniques are computationally expensive (demanding a blocking phase to reduce the search space) and are less explainable on account of using black-box classification methods.'' $\Rightarrow$ The authors have to support both claims, namely computationally expensive and less explainable, with references.
\end{itemize}
}

\response{We thank the reviewer and apologize for the presentation issues. We added guidelines for practitioners in \S~\ref{sec:3} and the rules of thumb for responsible EM.
We made sure to add all reproducibility documentation in our GitHub repository (link provided in the paper), explaining how to use our fairness evaluation framework and reproduce our results. 
We also created the pseudocode of the steps to use the framework (due to the space limitations it has been added to the technical report).
Our repository also includes well-documented code and examples on how to use the framework to evaluate any EM system. We also made sure to revise and provide references for the claims specified by the reviewer.} 

\section*{Reviewer \#3}\label{rev3}

\review{
W1: The motivation of this study can be strengthened.
}
\response{Thank you for the suggestion. Per the reviewer's request, we updated our introduction section with a discussion and another example motivating our study.}

\review{
W2: It is unclear whether fairness in the EM problem is substantially different from fairness in ML classification (there is a class imbalance problem, but that is present in ML classification in general too).
}

\response{While we agree with the reviewer on the class imbalance problem in some of classification problems, the degree of imbalance is often constant while in the context of EM the positive to negative ratio can be as low as $n$ to $n^2$, i.e., $O(1/n)$. As a result of such a high degree of imbalance in this setting, measures such as PPVP and TPRP are more likely to reveal unfairnesses in EM, while this is not true for ML in general.
Besides, as discussed in \S~\ref{sec:3}, the pairwise nature of the EM distinguishes it from the rest of the classification problems.}

\review{
W3: Limited actionable insights. For individuals whose names are common (e.g., Joe Smith), it is unclear what algorithmically can be done to improve the situation for them.
}

\response{
Thank you for your comment.
Following your suggestion, we updated \S~\ref{sec:discussion} to make the insights more actionable. In particular, based on our findings in this paper, we added a set of "rules of thumb" (Table~\ref{tab:Rules}) for responsible EM.
For individuals whose names are common (e.g., Joe Smith) one should (i) use the matchers that consider extra information such table structures, and 
(ii) make sure that a matcher is not putting a lot of weights on a small set of attributes, especially the problematic ones such as names. 
} 

\review{
D1: The motivating no-fly-list example feels a bit contrived when such matchings will likely uniquely identify information such as drivers-license/passport ID or date-of-birth info. I can also imagine other social settings, where over-matching of EM may actually be advantageous (e.g., distribution of welfare through fuzzy name matching), to the benefit of historically disadvantaged groups.
}

\response{ 
We agree that in some scenarios overmatching can be beneficial to historically disadvantaged groups.
Our original motivating example had the purpose of highlighting potential unfairness issues in scenarios when unique identifiers  such as drivers-license/passport ID are not available.
We have added a second motivating example (Example~\ref{eg:loyal}) to the paper illustrating potential unfairness issues in a different scenario involving advertising, where overmatching benefits the privileged groups, and indirectly hurting the under-privileged groups.
}

\eat{


The example provided by the reviewer is also a case where overmatching is beneficial. In particular, that examples relate to reverse-discrimination where the welfare is distributed to help disadvantaged groups.
As another example (Added to the paper -- Example 2), 
suppose a business $A$ acquires a list of their potential loyal customers collected from other businesses. 
Business $A$ matches their customers against the list and send exclusive offers to the positive cases in order to attract them.
Unlike the previous example, in this case over-representation in the loyal-customer's list is beneficial. On the other hand, unlike the no-fly list, the privileged group(s) are likely to be over-represented or in other words minorities are under-represented.
As a result, a lower (true and false) positive ratio is expected to happen for minority groups and they end up receiving much less of the exclusive offers.
}

}

\title{Through the Fairness Lens:\\
Experimental Analysis and Evaluation of Entity Matching} 

\author{Nima Shahbazi}
\affiliation{%
  \institution{University of Illinois Chicago}
}
\email{nshahb3@uic.edu}

\author{Nikola Danevski}
\affiliation{%
  \institution{University of Rochester}
}
\email{ndanevsk@u.rochester.edu}

\author{Fatemeh Nargesian}
\affiliation{%
  \institution{University of Rochester}
}
\email{fnargesian@rochester.edu }

\author{Abolfazl Asudeh}
\affiliation{%
  \institution{University of Illinois Chicago}
}
\email{asudeh@uic.edu}

\author{Divesh Srivastava}
\affiliation{%
  \institution{AT\&T Chief Data Office}
}
\email{divesh@research.att.com}

\begin{abstract}
Entity matching (EM) is a challenging problem studied by different communities for over half a century.
Algorithmic fairness has also become a timely topic to address machine bias and its societal impacts.
Despite extensive research on these two topics, little attention has been paid to the fairness of entity matching.

Towards addressing this gap, we perform an extensive experimental evaluation of a variety of EM techniques in this paper. 
We generated two social datasets from publicly available datasets for the purpose of auditing EM through the lens of fairness.
Our findings underscore potential unfairness under two common conditions in real-world societies: (i) when some demographic groups are overrepresented, and (ii) when names are more similar in some groups compared to others.
Among our many findings, it is noteworthy to mention that while various fairness definitions are valuable for different settings, due to EM's class imbalance nature, measures such as positive predictive value parity and true positive rate parity are, in general, more capable of revealing EM unfairness.
\end{abstract}

\maketitle
\submit{\setcounter{page}{1}}

\pagestyle{\vldbpagestyle}
\begingroup\small\noindent\raggedright\textbf{PVLDB Reference Format:}\\
\vldbauthors. \vldbtitle. PVLDB, \vldbvolume(\vldbissue): \vldbpages, \vldbyear.\\
\href{https://doi.org/\vldbdoi}{doi:\vldbdoi}
\endgroup
\begingroup
\renewcommand\thefootnote{}\footnote{\noindent
This work is licensed under the Creative Commons BY-NC-ND 4.0 International License. Visit \url{https://creativecommons.org/licenses/by-nc-nd/4.0/} to view a copy of this license. For any use beyond those covered by this license, obtain permission by emailing \href{mailto:info@vldb.org}{info@vldb.org}. Copyright is held by the owner/author(s). Publication rights licensed to the VLDB Endowment. \\
\raggedright Proceedings of the VLDB Endowment, Vol. \vldbvolume, No. \vldbissue\ %
ISSN 2150-8097. \\
\href{https://doi.org/\vldbdoi}{doi:\vldbdoi} \\
}\addtocounter{footnote}{-1}\endgroup

\ifdefempty{\vldbavailabilityurl}{}{
\vspace{.3cm}
\begingroup\small\noindent\raggedright\textbf{PVLDB Artifact Availability:}\\
\rev{The source code, data, and/or other artifacts have been made available at \url{https://github.com/UIC-InDeXLab/fair\_entity\_matching}}.
\endgroup
}

\submit{\vspace{-2mm}}
\section{Introduction}

Entity matching (EM) seeks to match pairs of entity records from (the 
same or different) data sources that refer to the same real-world entity.
EM is very useful in many applications domains, including 
(a)~healthcare, where matching of patient records from different 
healthcare facilities (e.g., emergency rooms, hospitals, etc.) can be
used to determine if they refer to the same real-world person; 
(b)~airline security, where airline passenger records are matched 
against no-fly list records to identify people who should be
prevented from boarding flights or should undergo additional screening; 
(c)~e-commerce, where product records from different retailers' websites
can be matched to identify popular products and fraudulent knockoffs; 
and so on.

EM is a challenging problem that has been extensively investigated 
for over half a century by different communities, e.g., statistics, 
databases (DB), natural language processing (NLP), and machine learning 
(ML), resulting in a variety of techniques proposed in the literature 
for addressing this problem. 
These challenges arise because entities in autonomous data sources can 
be represented in a variety of ways (e.g., highly structured records 
versus textual descriptions), using different conventions (e.g., the 
many ways in which person names and postal addresses are represented), 
data quality issues (e.g., misspellings, missing values), and so on.
A consequence is that, despite significant advances in recent years 
(especially with recent neural techniques like {\sc Ditto}~\cite{li2020deep}), 
EM techniques still result in both false positives (non-matching record pairs that are declared as matches) and false negatives 
(matching record pairs that are declared as non-matches).
These errors can have serious consequences in practice,
\rev{as seen in the following examples.}

\begin{example}\label{eg:nofly}
    {\bf (No-fly list)} Consider the airline security application, which aims to identify passengers that are likely to be dangerous (e.g. terrorists) for screening and potentially preventing them from boarding the flights.
    Using a dataset of criminal records called the no-fly list, passenger names (and other information) are matched against the no-fly list for this purpose. 
    False positives in airline security can lead to significantly inconveniencing passengers.
    On the other hand, false negatives can result in known terrorists being permitted to board flights with undesirable consequences.
    Due to historical biases, the no-fly list datasets could over-represent some minority groups in comparison to society's population distribution. This, as we shall evaluate in our experiments, can result in higher false positive rates for those demographic groups. 
    Another potential issue is that some demographic groups have more similar names. Hence, passengers from those groups may have a higher chance of having the same or similar information to those of known terrorists, which in turn will cause higher false positive rates for them.
\end{example}

\rev{
In Example~\ref{eg:nofly}, getting (falsely) matched is harmful.
We next show an example where not getting matched can be harmful.
}

\begin{example}\label{eg:loyal}
\rev{\marginpar{R3.D1 R3.W1}
    {\bf (High-value customers list)}  
    Upselling to potentially high-value customers is critical for many businesses such as the fashion industry, airlines, and tourism.
    Suppose business $A$ acquires a list $L$ of names of high-value customers from other businesses.
    Business $A$ uses EM techniques to match its own customers against list $L$ and sends exclusive offers to upsell to the matching cases in order to prioritize them.
    Unlike the previous example, in this case, over-representation in the high-value customer's list $L$ is beneficial. However, unlike the no-fly list, the privileged group(s) are likely to be over-represented and minority groups under-represented in the list $L$.
    As a result, a lower (true and false) positive match rate is expected for minority groups and they end up receiving fewer exclusive offers, resulting in biased advertising~\cite{miller2019targeted,AdBiasIBM, AdBias1}. 
    }
\end{example}

When such disparities (e.g., false positives) occur in a systematic way for
some demographic (sub-)groups, 
thereby disadvantaging them over others, concerns about the \emph{fairness} of EM techniques arise.
While the fairness of ML models has been the topic of much  recent work in the literature~\cite{fairmlbook,dwork2012fairness,feldman2015certifying,zhang2021fairrover,feldman2015certifying,zafar2017fairness,hardt2016equality,zemel2013learning,kusner2017counterfactual}, not much attention has 
been paid to the fairness of EM techniques.

In this paper, we seek to address this gap in the literature and 
perform an extensive experimental evaluation and analysis of a variety 
of EM techniques on a range of datasets through the 
\emph{fairness lens}. 
Traditionally, blocking may precede matching to reduce the space of possible matching candidates from quadratic to sub-quadratic, e.g., linear. 
A rich body of research focuses on  blocking algorithms~\cite{papadakis2016comparative,papadakis2020blocking,li2020survey,christophides2020overview}. In this paper, our goal is to audit off-the-shelf entity matching systems used in practice.  As such, our evaluation and analysis are performed on end-to-end matching systems which may include their own built-in blocking algorithms.

\submit{\vspace{-1mm}}
\stitle{Summary of Contributions} In summary, we make the following technical contributions in this paper:
\submit{\vspace{-1mm}}
\begin{itemize}[leftmargin=*]
\item 
Given the pairwise nature of EM, we propose the use of 
\emph{single fairness} and \emph{pairwise fairness} to evaluate 
entity matchers.
We adopt 11 popular fairness measures from the literature for this task
and analyze their suitability for EM.

\item
We select a suite of 13 EM techniques (including 1 declarative rule-based 
technique, 7 non-neural ML techniques, and 5 neural ML techniques) and 6 benchmark datasets (including 2 structured datasets, 2 textual datasets,
and 2 dirty datasets) that have been used in prior work on entity 
matching for fairness evaluation.

Using publicly available individual-level data, we also created and used two semi-synthetic matching {\em social datasets} for fairness evaluation. These datasets are shared publicly as benchmarks for auditing the fairness of matchers.
\item
We evaluated all combinations of EM techniques, datasets, 
and fairness measures and analyzed the outcomes\techrep{ to obtain 
generalizable results}. 
We classified the results into four cases based on
whether an (EM technique, dataset, or fairness measure) yielded
(i) accurate or inaccurate matching results, and 
(ii) fair or unfair matching results.
\techrep{Interestingly, all four classes contained configurations involving ML-based classifiers.}

Some of our findings in this study are as follows:

\begin{itemize}[leftmargin=*]
    \item Our results on social data confirm matching unfairness when (i) there are higher similarities among records of a certain group, (ii) the representation of demographic groups in data is biased.
    \item Our results underscore that responsible EM requires training data \rev{\marginpar{R2.D1}that is representative of different possibilities from various (demographic) groups.}
    \item While different fairness measures are valuable for different settings, due to the class-imbalance property of EM, measures such as {\em positive predictive value parity} and {\em true positive rate parity} are more capable of revealing EM unfairness.
    \item 
    Significantly relying on proxy attributes such as name, can cause unfairness in non-neural models.
    On the other hand, relying on pre-trained language models and embeddings, or not fully considering the dataset structure can cause unfairness in neural matchers.
\end{itemize}

\rev{
We use our findings and lessons learned to put together a set of rules of thumb for responsible entity matching (Table~\ref{tab:Rules}).
}
\submit{\vspace{-1mm}}
\end{itemize}
\submit{\vspace{-4mm}}
\section{Related Work}
Fairness in entity resolution (ER) has briefly been studied in the literature. In \cite{efthymiou2021fairer}, a constraint-based formulation for fairness is proposed to mitigate bias in ER tasks by ensuring that all (sub-)groups have the same opportunity to be resolved. Furthermore, \cite{makri2022towards} proposes a (sub-)group-based training for different ethnicities in order to increase both accuracy and fairness in SVM-based ER which is consistent with our suggestion to use ensemble learning for EM. Finally, in a parallel work \cite{nilforoushan2022entity}, the authors propose an AUC-based fairness definition for EM and ER tasks and try to resolve the bias issues through a data augmentation solution.
To the best of our knowledge, we are the first to comprehensively audit off-the-shelf entity matching models for fairness and propose proper measures, datasets, and comparison angles fitting the problem settings given the inherent differences with typical machine learning tasks\techrep{ that have so far been studied}.
\submit{\vspace{-2mm}}
\section{Fairness Evaluation Framework}\label{sec:3}

\subsection{Background} 
\label{sec:background}
\rev{\marginpar{R1.D2}Given two sets of records $A$ and $B$, the EM problem is to identify all correspondences between record pairs in $A\times B$ that correspond to the same real-world entity.} A correspondence $c=(e_i, e_j, s)$ interrelates two records $e_i$ and $e_j$  with a confidence value $s\in[0,1]$ that indicates the similarity of  $e_i$ and $e_j$  or the  confidence of a matcher about $e_i$ and $e_j$  referring to the same entity ~\cite{kopcke2010frameworks}.
To decide whether the record pair of  $c=(e_i, e_j, s)$ is a {\em match} or {\em non-match}, matchers often apply a threshold on $s$~\cite{yu2016string,barlaug2021neural}. We decouple the choice of a threshold from the outcome of the matching and consider the outcome of an EM task as pairs of matching and non-matching records. Formally, we consider the following EM problem:

\submit{\vspace{-3mm}}
\begin{definition}[Entity Matching Problem]\label{entity-matching-def} 
\sloppy{
\rev{\marginpar{R1.D2}Consider two sets of records $A$ and $B$.}  
For every pair of records $(e_i,e_j)\in A\times B$, let $y_{ij}$ be the ground-truth label indicating if $e_i$ and $e_j$ refer to the same entity. 
Given all pairs $(e_i,e_j)\in A\times B$, the EM problem is
to predict $y_{ij}$ with a label $h_{ij}$. That is, $h_{ij}$ refers to the decision of the matcher about the label of $e_i$ and $e_j$ (match or non-match).}
\end{definition} 
\submit{\vspace{-2mm}}

\sloppy{In a fairness-sensitive setting, records are accompanied by sensitive attributes (e.g. {\tt \small gender}, {\tt \small country}, {\tt \small race}, etc.). Let $\mathcal{A}=\{A_1, \ldots, A_n\}$ be the sensitive  attributes, $dom(A_i)$ be the domain of $A_i$, and $\gee=\{g_1, \ldots, g_m\}$ be the set of all groups of interest, i.e. $\gee=\bigcup_{A_i\in\mathcal{A}}dom(A_i)$. 
The mapping $L(e_i)$ relates a record to its associated groups $G_{i}\subseteq\gee$. In other words, $G_i$ is the group that $e_i$ belongs to.}
\sloppy{\rev{\marginpar{R1.D2}Given two sets of records $A$ and $B$ 
and the set $[(e_i, e_j, G_{i}, G_{j}, h_{ij}, y_{ij})]_{\forall (e_i,e_j)\in A\times B}$, 
we would like to audit the fairness of a matcher with respect to groups.}} 

\submit{\vspace{-2mm}}
\subsection{Single and Pairwise Lens}\label{sec:single_pairwise_lens}

\subsubsection{Group Selection}

The first step in auditing an entity matcher for fairness is identifying 
meaningful (sub-)groups in sensitive attributes. An input dataset to a matcher  $\mathcal{M}$  includes record ids, the value ((sub-)group) of each record for sensitive attributes, the decisions of $\mathcal{M}$, as well as true labels for the record pairs.  
Depending on the type, cardinality, and the number of sensitive attributes, multiple fairness cases may happen that are presented in Table~\ref{tbl:attvalues}.

\begin{table*}
\caption{Fairness types based on the number and cardinality of sensitive attributes.}\submit{\vspace{-3mm}}
\small
\begin{tabular}{l l l}
\toprule
\thead{{\bf Type}} & \thead{{\bf Description}} & \thead{{\bf Example}} \\
\midrule
{\bf Single Attribute w/ Binary Values} & Each record belongs to one of & attribute: {\tt \footnotesize gender=\{male, female\}} \\
& two groups in the attribute domain. & {\tt \footnotesize group(e) = \{female\}}\\
\midrule
{\bf Single attribute w/ multiple} & Each record belongs to exactly & attribute: {\tt \footnotesize gender=\{male, female, transgender, non-binary, other\}}\\
{\bf exclusive values} & one group in the attribute domain. &{\tt \footnotesize group(e) = \{non-binary\}}\\
\midrule
{\bf Single setwise attribute} & Each record belongs to a subset of  & attribute: {\tt \footnotesize genre=\{Pop, Rock, Jazz\}}\\
& values in the attribute domain. &{\tt \footnotesize group(e) =\{Pop, Rock\}}\\
\midrule
{\bf Multiple attributes} & Groups could be either one or  & attributes: {\tt \footnotesize genre} and  {\tt \footnotesize  gender} \\
& a combination of the three cases above. &{\tt \footnotesize group(e) = \{male-Pop, male-Rock, male-Jazz\}}\\
 \bottomrule
\end{tabular}
\label{tbl:attvalues}
\submit{\vspace{-2mm}}
\end{table*}

\eat{
\squishlist    
    \item \textit{Single attribute with binary values} means fairness is studied on a single sensitive attribute. Each entity belongs to one of the two groups, e.g. {\tt \small gender=\{male, female\}}.
    \item \textit{Single attribute with multiple (exclusive) values} means  fairness is studied on a single sensitive attribute. Each entity belongs to one of the multiple demographic groups, e.g. {\tt \small gender=\{male, female, transgender, non-binary, other\}}.
    \item \textit{Single setwise attribute} means fairness is studied on a single sensitive attribute. Each entity can belong to a subset of the universe of possible values of an attribute. For example, assuming {\tt \small genre=\{Pop, Rock, Jazz\}}, an entity $e$ can have multiple genres: {\tt \small genre(e)=\{Pop, Rock\}}. Note that identified  groups can have overlapping values, e.g. we may have a pair of entities {\tt \small genre($e_1$)=\{Pop, Rock\}}, {\tt \small genre($e_2$)=\{Pop, Jazz\}}. 
    \item \textit{Multiple attributes} means fairness is studied on the intersection of multiple sensitive attributes. The values could be either one or a combination of the three cases discussed before. An example includes the   groups defined on intersection of (single setwise attribute) {\tt \small genre} and (binary attribute) {\tt \small gender}: {\tt \small \{male-Pop, male-Rock, male-Jazz, female-Pop, female-Rock, female-Jazz, male-Pop-Rock, male-Pop-Jazz, male-Rock-Jazz, female-Pop-Rock, female-Pop-Jazz, female-Rock-Jazz, male-Pop-Rock-Jazz, female-Pop-Rock-Jazz\}}. 
\squishend
}

The space of groups for a single attribute with binary or multiple values is the domain of the corresponding attribute. 
In multiple-attribute settings,  we can define intersectional subgroups,\rev{\marginpar{R1.D9} as the cartesian product of group values.
\submit{More details about intersectional subgroups are provided in the technical report~\cite{techrep}.}
}
\techrep{The subgroups can be presented in a hierarchical data structure, where the first level includes all groups (of all attributes), while the $k$-th level includes the set of non-overlapping groups created by combining groups from $k$ different attributes. When one attribute is of the single setwise type, level $k$ includes  $k-1$ groups from the setwise attribute with one group from a binary or multi-value attribute.   

\begin{example}\label{ex:hierarchy}
Figure~\ref{fig:lattice} shows the intersectional subgroup hierarchy  of sensitive attributes {\tt \small gender} and {\tt \small genre} for a dataset that matches songs of different artists. Note that {\tt \small gender} is a binary attribute  and {\tt \small genre} is a setwise attribute. The level-$2$ of this hierarchy includes all combinations of groups from {\tt \small gender} and {\tt \small genre} in level-$1$. Level-$3$ enumerates $2$-combinations of the domain of {\tt \small genre} with groups from {\tt \small gender}. 
\end{example}

Note that a subgroup hierarchy represents the space of groups and does not mean a data set must or does contain all these groups. 
In addition to enabling fairness audit on a particular group selected by a user, we allow batch auditing subgroups of each level. That is, a matcher's fairness is evaluated for all subgroups of a particular level selected by a user. 
}
\techrep{
\begin{figure}
    \centering
    \hbox{\includegraphics[scale=0.26]{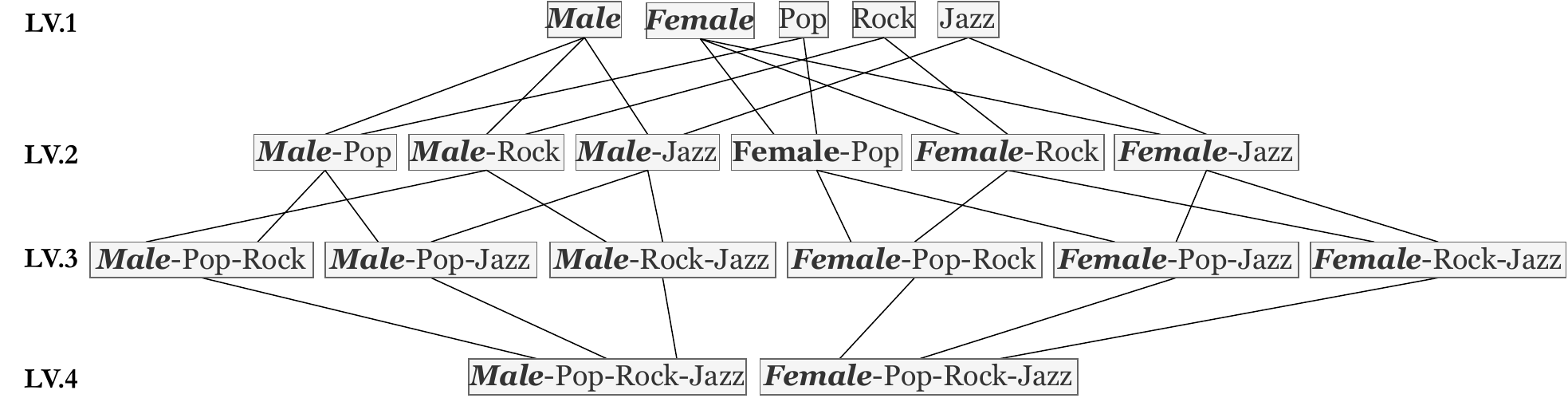}}
    \submit{\vspace{-3mm}}
    \caption{Intersectional subgroup hierarchy for single setwise and multiple attributes} 
    \label{fig:lattice}
    \submit{\vspace{-6mm}}
\end{figure}
}

\submit{\vspace{-1mm}}
\subsubsection{Single and Pairwise Fairness Evaluation}
Given the pairwise nature of EM tasks, there are two ways to audit entity matchers:
\begin{itemize}[leftmargin=*]
    \item{\textit{Single Fairness:}} The performance of a matcher is evaluated for one group 
    $s$ against either record in a pair. 
Given a correspondence $c = (e_i, e_j, h, y)$ and a group 
$s$ of interest,  $c$ is legitimate, if either $e_i$ or $e_j$ belong to 
group $s$. 
    \item{\textit{Pairwise Fairness:}}
The performance of a matcher is evaluated for a pair of 
groups $s,s^\prime$ against both records in a pair. 
Given a correspondence $c = (e_i, e_j, h, y)$ and a pair of 
groups $(s,s^\prime)$ of interest, $c$ is legitimate, if  $e_i$ belongs to $s$ and $e_j$ belongs to $s^\prime$, or vice versa.  From an encoding perspective, we concatenate the encodings of 
groups $s$ and $s^\prime$ into a vector $c$ and the encodings (explained in the technical report \cite{techrep}) of $e_i$ and $e_j$ into a vector $e$ and validates vector $e$ belongs to $c$ with both directions of $\langle s, s^\prime\rangle$ and $\langle s^\prime, s \rangle$. 
\end{itemize}

We consider the EM task to be symmetric in single and pairwise fairness definitions. We remark that these definitions can be extended to ordered single and ordered pairwise fairness where the 
groups are defined on left or right records. In this paper, we focus on non-directional single and pairwise fairness. 
\submit{\vspace{-2mm}}
\subsection{Correctness}
\label{sec:correctness}
The correctness of a matcher measures how well its matching predictions conform with the ground-truth. Given a test dataset with correspondences of  $t=(e_i, e_j, h, y)$, where $h$ is a binary variable indicating the result of EM (\textit{match} or \textit{non-match}) for records with encodings $e_i$ and $e_j$, and $y$ is a binary variable indicating the ground-truth for matching, we profile predictions of $h$  using the numbers of true positives (TP), true negatives (TN), false positives (FP), and false negatives (FN), respectively. Unlike a classification task, in the confusion matrix of a matching task, the result is counted both for the group(s) of $e_i$ and the group(s) of $e_j$. For further explanations, please refer to the technical report \cite{techrep}.

\subsection{Fairness Measures}
\label{sec:fairness}

\rev{\marginpar{R2.D1}\marginpar{M3}Similar to many of the existing works on algorithmic fairness and responsible data science \cite{shetiya2022fairness,nargesian2021tailoring,calmon2017optimized,asudeh2019designing,celis2019classification,feldman2015certifying,hardt2016equality,swift2022maximizing}, we use the terms bias and unfairness interchangeably to refer to when a matcher is not fair.} 
At a high level, fairness definitions can be viewed from three perspectives: group, subgroup, and individual fairness~\cite{fairmlbook}. 
The most granular notion of fairness is individual fairness that requires similar outcomes for similar individuals~\cite{dwork2012fairness}. 
The more popular perspective of fairness, (sub)group fairness, requires similar treatment for different (sub)groups. 
A model/algorithm satisfies some fairness constraints if it has equal or similar performance (according to some fairness measure) on different (sub)groups. 
{\em The focus of this paper is on (sub)group fairness}. 
Most of the group fairness measures belong to one of the following four categories~\cite{fairmlbook,asudeh2020fairly}. (1) {\em Independence} requires independence of analysis outcome from demographic groups. (2) {\em Separation} requires independence of the outcome from demographic groups conditioned on the target variable. (3) {\em Sufficiency} requires independence of the target variable from demographic groups conditioned on the outcome. (4) {\em Causation} requires that in a counterfactual world, the decision would not change had the individual belonged to a different demographic group. \techrep{Since we only assume access to a matcher's decisions and true labels, w}\submit{W}e do not consider Causal fairness in our audit. In Table~\ref{tbl:measures}, we present \rev{our suite of fairness measures, adapted from the} notions of fairness in classification~\cite{fairmlbook}, for auditing an entity matcher $\mathcal{M}$ for a set $\gee$ of (sub)groups. 
\begin{table*}
\caption{Fairness measures. $h(e,e^\prime)$ is the output of  a matcher $\mathcal{M}$  ({\em match} (`M') or {\em non-match} (`N')) and $y$ is the ground-truth\techrep{ on entities $e$ and $e^\prime$}.}
\submit{\vspace{-4mm}}
\small
\begin{tabular}{l@{\hskip 1mm} l l}
\toprule
\thead{{\bf Name}} & \thead{{\bf Description}} & \thead{{\bf Equation} ($\forall g_i\in \gee$)} \\
\midrule
{\bf Accuracy Parity (AP)} & requires the independence of matchers's accuracy from groups& $Pr(h(e,e^\prime)=y\vert g_i) \simeq Pr(h(e,e^\prime)=y)$\\ \midrule
{\bf Statistical Parity (SP)} & requires the independence of the matcher from  groups& $Pr(h(e,e^\prime)=`M\textrm'~\vert~g_i) \simeq Pr(h(e,e^\prime)=`M\textrm')$\\ \midrule
{\bf \footnotemark[1]True Positive Rate} & a.k.a \textit{Equal Opportunity}; in the group of true matches& $Pr(h(e,e^\prime) = `M\textrm'\vert g_i, y=`M\textrm') \simeq Pr(h(e,e^\prime) = `M\textrm' \vert y=`M\textrm')$\\ 
{\bf Parity (TPRP)} & requires the independence of match predictions from groups& \\ \midrule
{\bf False Positive Rate} &  in the group of true non-matches, requires& $Pr(h(e,e^\prime) = `M\textrm' \vert g_i, y=`N\textrm') \simeq Pr(h(e,e^\prime) = `M\textrm' \vert  y=`N\textrm')$\\ 
{\bf Parity (FPRP)} & the independence of match predictions from groups&\\ \midrule
{\bf \footnotemark[1]False Negative Rate} & in the group of true matches, requires& $Pr(h(e,e^\prime) = `N\textrm' \vert g_i, y=`M\textrm') \simeq Pr(h(e,e^\prime) = `N\textrm' \vert  y=`M\textrm') $\\ 
{\bf Parity (FNRP)} & the independence of non-match predictions from groups& \\ \midrule
{\bf True Negative Rate} & in the group of true non-matches, requires& $Pr(h(e,e^\prime) = `N\textrm' \vert g_i, y=`N\textrm') \simeq Pr(h(e,e^\prime) = `N\textrm' \vert  y=`N\textrm') $\\ 
{\bf Parity (TNRP)} & the independence of non-match predictions from groups&\\ \midrule
{\bf \footnotemark Equalized Odds (EO)} &in both groups of true matches and true non-matches& $Pr(h(e,e^\prime)=`M\textrm'\vert g_i, y=`M\textrm') \simeq	 Pr(h(e,e^\prime)=`M\textrm'\vert  y=`M\textrm')$\\ 
& requires the independence of match predictions from groups&$Pr(h(e,e^\prime)=`M\textrm'\vert g_i, y=`N\textrm') \simeq	 Pr(h(e,e^\prime)=`M\textrm'\vert  y=`N\textrm')$\\ \midrule
{\bf \footnotemark[1]Positive Predictive} &among the pairs predicted as match& $Pr(y=`M\textrm'\vert h(e,e^\prime)=`M\textrm', g_i) \simeq Pr(y=`M\textrm'\vert h(e,e^\prime)=`M\textrm')$\\ 
{\bf Value Parity (PPVP)}& requires the independence of true matches from groups&\\ \midrule
{\bf \footnotemark[1]Negative Predictive} & among the pairs predicted as non-match, requires& $Pr(y=`N\textrm'\vert h(e,e^\prime)=`N\textrm', g_i) \simeq Pr(y=`N\textrm'\vert h(e,e^\prime)=`N\textrm')$\\ 
{\bf Value Parity (NPVP)}& the independence of true non-matches from groups&\\ \midrule
{\bf \footnotemark[1]False Discovery Rate} & among the pairs predicted as match, requires& $Pr(y = `N\textrm' \vert g_i, h(e,e^\prime) = `M\textrm') \simeq Pr(y = `N\textrm' \vert  h(e,e^\prime) = `M\textrm')$\\ 
{\bf Parity (FDRP)}&the independence of true non-matches from groups&\\ \midrule
{\bf \footnotemark[1]False Omission Rate} &  among the pairs predicted as non-match, requires& $Pr(y = `M\textrm' \vert g_i, h(e,e^\prime) = `N\textrm') \simeq Pr(y = `M\textrm' \vert  h(e,e^\prime) = `N\textrm')$\\ 
{\bf Parity (FORP)}&the independence of true matches from groups&\\
 \bottomrule
\end{tabular}
\label{tbl:measures}
\vspace{-4mm}
\end{table*}

\subsection{Selecting Fairness Measures for EM}
\label{sec:measuresdiscussion}
\submit{\vspace{-1mm}}

Depending on the context of an EM task at hand, proper fairness measures should be employed.
Besides, a major difference between EM and regular classification tasks is that the input to EM tasks is a pair of records.
\rev{\marginpar{R2.D1 R2.D3 R2.W3 M1 M3}
Due to its pairwise matching nature, {\em class imbalance} is a distinguishing property of EM, compared to regular classification tasks.
To better explain this, let us consider a toy example, where two data sources $D$ and $D^\prime$ contain exactly the same set of $n$ records.
Each pair of records $e\in D$ and $e^\prime\in D^\prime$ is passed as an input to an entity matcher.
In this setting, only $n$ of the $n^2$ pairs are matches, and the others are non-matches. \marginpar{R3.W2 M2} 
In other words, the probability a random pair is a match is as low as $\frac{1}{n}$.
Class imbalance is also a challenge in some of classification problems but the degree of imbalance is often constant while in EM tasks, it is as high as $O({n})$.
Indeed, blocking techniques~\cite{papadakis2020blocking} can help in reducing the extreme class imbalance. 
Even after blocking, a high class imbalance is expected for EM tasks. 
}

\stitle{\rev{Guide for Practitioners}} 
\rev{\marginpar{R2.D1 R2.D3 R2.W3 R3.W3 M1 M5} 
Let us consider the fairness measures in Table~\ref{tbl:measures}.
Which measures to choose depends on the downstream task and the problem context. Therefore, practitioners should choose measures depending  on the importance of TPs (true match), FPs, FNs, and TNs in the problem context. 
For example, among the fairness measures,
{\em statistical parity} does not consider the ground-truth labels and  requires equal match ratios from different groups, independent of whether they really are a match or not. As a result, this measure \techrep{(and other measures in the independence category) }does not seem reasonable for deduplication tasks using EM.
On the other hand, it may be useful to ensure equal representation of different groups when using EM for joining tables.}

\rev{\marginpar{R2.D1 R2.D3 R2.W3 M1 M5}When the input to the EM task is imbalanced and most of the pairs are non-matches, some measures are more capable of revealing the unfairness of matchers. 
First, note that even a matcher that marks all pairs as non-matches has high accuracy in this setting. Subsequently, accuracy parity may not reveal the unfairness.
Similarly, measures such as FPRP and TNRP may fail to reveal unfairness in detecting true matches. In these settings, the fairness measure for successfully discovering these events is {\bf Positive Predictive Value Parity (PPVP)}. Another important measure in this context is {\bf True Positive Rate Parity (TPRP)}, a.k.a {\bf Equal Opportunity}, which focuses on correct match predictions among the (rare) true matches.
These recommendations are consistent with our comprehensive experiments on several data sets, in \S~\ref{sec:exp}, where PPVP and TPRP were the two measures that could reveal the unfairness of the matchers.}

\rev{We also note that some of the measures cannot be applied in pairwise fairness scenarios where conceptually, the equality of groups restricts matching results. 
 In some scenarios, two records with different groups can never be considered {\em match} in the ground-truth. 
For instance, in a matching task defined between {\em DBLP} and {\em ACM} publications, two records with different venues (after standardization) are never a true {\em match}. More concretely, when pairwise fairness is evaluated on 
non-overlapping groups, TPs and FNs are always zero; hence, measures based on TPs and FNs become inapplicable.
}

\footnotetext[1]{This measure is only meaningful for (a) \textit{single} fairness and (b) \textit{pairwise} fairness cases when groups are overlapping.}

\submit{\vspace{-3mm}}
\subsection{Measuring Unfairness}\label{sec:measuring_unfairness}
Consider a fairness notion and a 
group $g_i\in \gee$. In a perfect situation, the matcher should satisfy the parity (equality) between two probabilities in the following form: 
$\forall g_i\in \gee, Pr(\alpha~\vert~ \beta,g_i) = Pr(\alpha~\vert~ \beta)$, 
where $\alpha$ and $\beta$ are specified by the fairness measure.
\rev{\marginpar{R2.D1 R2.D3 R2.W3 M1 M5}For example, for Positive Predictive Parity, $\alpha$ is $y=`M\textrm'$ and $\beta$ is $h(e,e^\prime)=`M\textrm'$. 
On the other hand, due to the trade-offs~\cite{kleinberg2016inherent} between different fairness notions and the impossibilities theorems~\cite{chouldechova2017fair}, it is often not possible to satisfy complete parity on all fairness measures. As a result, the objective is to make sure that {\em disparity} (also known as {\em unfairness}) is less than a given threshold for a matcher to be fair. 
Given a fairness notion and a 
group $g_i\in \gee$, one way to compute disparity is to use  subtraction~\cite{bellamy2018ai}, as follows. }

\submit{\vspace{-4mm}}
\begin{equation}
    \label{equation:ds}
F^{(s)}_{\alpha,\beta}(g_i) = \max \Big(0~,~ Pr(\alpha~\vert~ \beta) - Pr(\alpha~\vert~ \beta,g_i)\Big)
\end{equation}
\submit{\vspace{-4mm}}

\rev{
For example, for accuracy parity ($\alpha$ is $h(e,e^\prime)=y$ and $\beta$ is null), 
the disparity can be computed as follows. 
}

\submit{\vspace{-4mm}}
\begin{align}F^{(s)}_{\mbox{AP}}(g_i) = \max \Big(0~,~ Pr(h(e,e^\prime)=y) - Pr(h(e,e^\prime)=y~\vert~ g_i)\Big)
\end{align}
\submit{\vspace{-4mm}}

\rev{Alternatively, given a fairness notion and a 
group $g_i\in \gee$, the disparity can be computed using division~\cite{feldman2015certifying} , as follows.}

\submit{\vspace{-3mm}}
\begin{equation}
    \label{equation:dd}
F^{(d)}_{\alpha,\beta}(g_i) = \max \Big(0~,1 - \frac{Pr(\alpha~\vert~ \beta,g_i)}{Pr(\alpha~\vert~ \beta)}\Big)
\end{equation}
\submit{\vspace{-4mm}}

\stitle{\rev{Guide for Practitioners}} \rev{When evaluating the  unfairness of a matcher, the objective is to determine whether {\em unfairness} is less than a given  threshold. For example, the 20\% rule~\cite{feldman2015certifying} suggests the threshold as $0.2$.} 
\rev{Note that if the accuracy for the 
group $g_i$ is higher than the average accuracy of the matcher, it is {\em not} considered as unfairness.
Also, note that Equation~\ref{equation:ds} considers the higher the probability, the better.
Depending on fairness measures (and application), the direction may be as the lower the probability, the better. For example, for FNRP, a lower probability of a false negative is preferred. For such cases, one should consider $Pr(h(e,e^\prime)=y~\vert~ g_i)-Pr(h(e,e^\prime)=y)$. 
As a result, for false negative rate ($\alpha$ is $h(e,e^\prime)=`N\textrm'$ and $\beta$ is $y=`M\textrm'$)
the disparity can be computed as}

\submit{\vspace{-3mm}}
\begin{equation}
\begin{split}
F^{(s)}_{\mbox{FNRP}}(g_i) = \max \Big(0~,~ &Pr(h(e,e^\prime)=0~\vert~ y=`M\textrm', g_i) \\&- Pr(h(e,e^\prime)=0~\vert~ y=`M\textrm')\Big)
\end{split}
\end{equation}
\submit{\vspace{-4mm}}

\rev{Similar to Equation~\ref{equation:ds}, Equation~\ref{equation:dd} also considers the higher the probabilities the better. For the cases (such as FNRP or FDRP) where the lower probabilities are better, one should swap the numerator and the denominator in the equation.
Therefore, for false discovery rate ($\alpha$ is $y=0$ and $\beta$ is $h(x)=1$)
the disparity can be computed as
}

\submit{\vspace{-4mm}}
\begin{align*}
F^{(d)}_{\mbox{FDRP}}(g_i) = \max \Big(0~,1 - \frac{Pr(y=`N\textrm'~\vert~ h(e,e^\prime)=`M\textrm')}{Pr(y=`N\textrm'~\vert~ h(e,e^\prime)=`M\textrm', g_i)}\Big)
\end{align*}
\submit{\vspace{-3mm}}

Our proposal in this paper is agnostic to the choice of operation for computing the disparities.
Still, in our experiments, without any preference, we use subtraction to compute the disparities.

\section{Entity Matching Approaches}
The existing techniques for EM fall into one of the following three  categories: 1) declarative rule-based, 2) ML-based, and 3) crowd-sourcing-based approaches. The last class of techniques relies on crowd-worker knowledge for EM tasks and we do not include them in our analysis. From each of the remaining categories, we select a few important matchers to be assessed for fairness. The specifications of the evaluated matchers are presented in Table~\ref{tbl:models}.   
\submit{\vspace{-3mm}}
\subsection{Rule-based Matchers}
\sloppy{Rule-based approaches perform EM based on the conjunction/disjunction of a few logical predicates, each specifying a matching condition. Each matching condition consists of a similarity measure (e.g., {\em Hamming, cosine, Levenshtein, Jaccard,} etc.) computed between record pair columns, a comparison operator (e.g., <, =, >), and a threshold value specifying the similarity value. Rule-based matchers are scalable to large settings and provide results that are explainable. However, they highly depend on human experts with relevant domain knowledge to assist with rule specification. \rev{\marginpar{R1.D4}For more information on rule specification in EM, we would like to direct the reader to \cite{wang2011entity,singh2017synthesizing,panahi2017towards,paganelli2019tuner}.}} 
\submit{\vspace{-2mm}}
\subsection{ML-based Matchers }
A crucial part of rule-based matching that affects the overall correctness of the task is the selection and configuration of the rules used for comparison. This task is difficult and laborious even for domain experts. ML-based supervised EM approaches reduce the associated manual labor by benefiting from the training data at hand. They significantly reduce the rule discovery efforts by extracting fitting parameters (e.g., model weights) from the data. However, preparing the training data itself imposes an additional cost. \rev{\marginpar{R2.D3}Furthermore, such techniques are computationally expensive (demanding a blocking phase to reduce the search space)~\cite{kopcke2010frameworks}  and are less explainable on account of using black-box classification methods~\cite{barlaug2022lemon, wang2022minun}}. Depending on the employed classification technique, ML-based matchers belong to one of the \textit{non-neural} or \textit{neural} categories.  
\subsubsection{Non-neural Matchers}
This category of matchers uses traditional ML algorithms such as decision tree, SVM, etc., to decide whether or not a pair of records is a match. Since the number of meaningful insights that can be extracted from data and fed as features to the learning algorithm are limited to word-level similarity metrics and TF-IDF scores, non-neural matchers may not perform well for cases where datasets are less structured, and column values are more in a textual format consisting of long spans of text.
\subsubsection{Neural Matchers}
Deep learning techniques have recently shown promising results in NLP applications. Due to the growing demand for matching textual data instances, it only makes sense to adopt such techniques where the other approaches usually fall short. Deep learning methods transform text into numerical values using character/word embeddings often through pre-trained embedding models such as word2vec \cite{mikolov2013efficient}, GloVe \cite{pennington2014glove}, fastText \cite{bojanowski2016enriching}. Due to the sequential nature of text, to better capture the semantics of the data, sequence models such as RNN and its variants (e.g., LSTM, GRU, etc.), where prior sequences of inputs can affect the current input and output, are utilized \cite{barlaug2021neural}. Further improvement mechanisms such as attention \cite{vaswani2017attention}, pre-trained language models \cite{devlin2018bert}, domain knowledge injection, data augmentation, summarization, etc., deliver further insights into the models to make better matching decisions. The superiority of neural matchers for textual and dirty data sets has been pointed out in the existing research \cite{mudgal2018deep}. However, there are associated challenges, such as high computation costs and large training data requirements, making them not suitable for every EM scenario.

\begin{table*}
\caption{List of EM approaches evaluated for fairness}\submit{\vspace{-3mm}}
\small
\scalebox{0.95}{
\begin{tabular}{l l l}
\toprule
\thead{{\bf Name}} & \thead{{\bf Type}} & \thead{{\bf Description}} \\
\midrule
{\sc BooleanRuleMatcher \cite{konda2018magellan}} & Rule-based & Conjunction of rules defined using a similarity measure, a comparison operator, and a threshold value  \\
&&between the record pair columns, part of Magellan framework  \\
\midrule
{\sc Dedupe \cite{forest2022dedupe}} & Non-neural & Uses regularized logistic regression for agglomerative hierarchical clustering of records\\
\midrule
{\sc DTMatcher \cite{konda2018magellan}} & Non-neural & Uses decision tree classifier for matching, part of Magellan framework  \\
\midrule
{\sc SVMMatcher \cite{konda2018magellan}} & Non-neural & Uses SVM classifier for matching, part of Magellan framework  \\
\midrule
{\sc RFMatcher \cite{konda2018magellan}} & Non-neural & Uses random forest classifier for matching, part of Magellan framework  \\
\midrule
{\sc LogRegMatcher \cite{konda2018magellan}} & Non-neural & Uses logistic regression classifier for matching, part of Magellan framework  \\
\midrule
{\sc LinRegMatcher \cite{konda2018magellan}} & Non-neural & Uses linear regression classifier for matching, part of Magellan framework  \\
\midrule
{\sc NBMatcher \cite{konda2018magellan}} & Non-neural & Uses naive bayes classifier for matching, part of Magellan framework \\
\midrule
{\sc DeepMatcher \cite{mudgal2018deep}} & Neural & Provides a variety of deep learning approaches such as aggregation-based, RNN-based, attention-based and,\\ &&hybrid (RNN+attention) to learn latent semantic features for a pair of records \\
\midrule
{\sc Ditto \cite{li2020deep}} & Neural & Deep learning approach utilizing pre-trained transformer-based language models and optimizing\\ 
&& performance using domain knowledge injection, text summarization, and data augmentation techniques \\
\midrule
{\sc GNEM \cite{chen2021gnem}} & Neural & One-to-set neural framework (unlike remaining pairwise solutions) benefiting from graph neural networks\\
\midrule
{\sc HierMatcher \cite{fu2021hierarchical}} & Neural & Deep learning approach based on RNN, attribute-aware attention mechanism and cross attribute\\
&& token alignment, built on top of {\sc DeepMatcher} framework\\
\midrule
{\sc MCAN \cite{zhang2020multi}} & Neural & Deep learning approach based on RNN and multi-context attention mechanisms such as self-attention,\\
&& pair-attention, global-attention, and gating mechanism, built on top of {\sc DeepMatcher} framework \\
 \bottomrule
\end{tabular}}
\label{tbl:models}
\submit{\vspace{-2mm}}
\end{table*}

\begin{table*}
\caption{Overview of the datasets used in our analysis}\submit{\vspace{-3mm}}
\small
\scalebox{0.94}{
\begin{tabular}{l l l l l l l l l l}
\toprule
\thead{{\bf Name}} & \thead{{\bf Repository}} &\thead{{\bf Domain}} & \thead{{\bf Type}} & \thead{{\bf Train}} & \thead{{\bf Test}} & \thead{{\bf \% Pos.}} & \thead{{\bf \# Attr.}} & \thead{{\bf Sens. Attr.}} & \thead{{\bf Sens. Attr. Type}} \\
\midrule
{\sc \fmatch} &&Population&Structured&271108&1084432&0.21\%&2&{\tt country}&Single attr. w/ binary values  \\
\midrule
{\sc \nofly} &&Population&Structured&20122&75459&0.63\%&3&{\tt race}&Single attr. w/ binary values  \\
\midrule
{\sc iTunes-Amazon} &Magellan&Music&Structured&321&109&24.7\%&8&{\tt genre}&Single setwise attr.  \\
\midrule
{\sc Dblp-Acm} &Magellan &Publications&Structured&7417&2473&17.9\%&4&{\tt venue}&Single attr. w/ multiple exclusive values  \\
\midrule
{\sc Dblp-Scholar} &Magellan&Publications&Dirty&225&100&19\%&10&{\tt entry type}&Single attr. w/ multiple exclusive values \\
\midrule
{\sc Cricket} &Magellan&Sports&Dirty&2277&1013&96.5\%&20&{\tt batting style}&Single attr. w/ binary values \\
\midrule
{\sc Shoes} &WDC&Products&Textual&24111&10717&10.3\%&1&{\tt company}&Single attr. w/ multiple exclusive values\\
\midrule
{\sc Cameras} &WDC&Products&Textual&5476&2434&17.2\%&1&{\tt company}&Single attr. w/ multiple exclusive values \\
\bottomrule
\end{tabular}}
\label{tbl:datasets}
\submit{\vspace{-4mm}}
\end{table*}

\submit{\vspace{-2mm}}
\section{Evaluation and Analysis}\label{sec:exp}
\submit{\vspace{-1mm}}
\subsection{Evaluation Plan}
To evaluate the matchers for fairness, we investigate the performance of matchers in terms of single and pairwise fairness for all valid groups in the datasets w.r.t. a variety of fairness definitions. To present a side-by-side comparison and visualization, we aggregate the results based on the dataset and the type of fairness (i.e., single and pairwise). Next, we look into some of the identified discriminated groups from different settings and investigate the reasoning behind the unfair behavior of matchers.

\submit{\vspace{-2mm}}
\subsubsection{Experimental Settings}
We conducted the experiments on a 3.5 GHz Intel Core i9 processor, 128 GB memory, running Ubuntu. The evaluation framework was implemented in Python. We accessed the source code of the entity matchers either through the authors' public GitHub or by directly contacting the authors.  
 
\submit{\vspace{-2mm}}
\subsubsection{Social Datasets.}\label{sec:social_datasets}
The concept of fairness holds significant societal implications and carries more significance when studied on the {\em individual records}. Unfortunately, public access to such data, especially demographic information, is restricted owing to privacy concerns. 
Therefore, we construct semi-synthetic datasets based on two publicly available real-world datasets {\sc CSRankings} and {\sc Compas} \cite{COMPAS}. 
We selected these datasets based on our airline security example discussed in the introduction.
Particularly, we want to evaluate the fairness of the matchers under two conditions: (a) {\em when two demographic groups have different degrees of similarities in their names}, and (b) {\em when there is an over-representation of some groups in the data}.
CSRankings\footnote{\url{csrankings.org}} is a global ranking system that evaluates computer science departments based on the scholarly research activities of their faculty members from universities across the world.
{\sc CSRankings} dataset is publicly available~\cite{CSRankings}. For each faculty, in addition to their names, the dataset contains other information such as affiliation country. Having observed various name similarities between different geographical regions, we found this dataset as a good candidate for evaluating (a).
{\sc Compas}, on the other hand, is a public dataset of criminal records that has been widely used in Fair ML research. In addition to names and other information, the dataset contains demographic information for each individual. The dataset over-represents {\tt Black/African-Americans}, which makes it a good candidate for evaluating (b). 

To create our first EM dataset \fmatch
 based on {\sc CSRankings}, we do the following steps\footnote{Semi-synthetic data generators are available for public access in \cite{synthetic_data}.}: 
 \submit{
 Using {\tt fullName} and {\tt country} for matching, we focus on two groups of faculties working in Germany {\tt de} and China {\tt cn}.
     Next, we perform a Cartesian product on the sample and label each pair as a match if left and right records have identical {\tt scholarID}s.
     Finally, we perturb the values of {\tt fullName} column for the right-side records which involve randomly adding, removing, or replacing a random character in the cell.
 }
\techrep{
 \begin{itemize}[leftmargin=*]
    \item 
    Using {\tt fullName} and {\tt country} for matching, we only focus on two demographic groups of faculties working in Germany {\tt de} and China {\tt cn}.
     \item We perform a Cartesian product on the sample and label each pair as a match if left and right records have identical {\tt scholarID}s.
     \item We perturb the values of {\tt fullName} column for the right-side records which involve randomly adding, removing, or replacing a random character in the cell.
 \end{itemize}
 }

Following our motivating example in the introduction, we create \nofly, a no-fly list scenario based on {\sc Compas}:
\submit{
First, using {\tt firstName}, {\tt lastName},  and {\tt race} for matching, we focus on individuals that are either {\tt Caucasian} or {\tt Black/African-American}.
We next create the no-fly list by taking a uniform sample from {\sc Compas} comprising of 48\% {\tt Caucasian} records and 52\% {\tt Black/African-American} (the distribution of the two groups in the {\sc Compas} dataset).
Then, in accordance with the racial distribution of the U.S. population, as reported by the Census Bureau Data \cite{usCensus}, we create a passenger table by taking a sample from {\sc Compas} that includes 80\% {\tt Caucasian} and 20\% {\tt Black/African-American} individuals.
Next, we perform a Cartesian product on the two tables and label each pair as a match if left and right records have identical {\tt personID}s.
Similar to the   process for \fmatch, the right records (that correspond to the no-fly list table) undergo perturbation in the {\tt firstName} and {\tt lastName} columns. 
}
\techrep{
 \begin{itemize}[leftmargin=*]
     \item Using {\tt firstName}, {\tt lastName},  and {\tt race} for matching, we focus only on individuals that are either {\tt Caucasian} or {\tt Black/African-American}.
     \item We create the no-fly list by taking a uniform sample from {\sc Compas} comprising of 48\% {\tt Caucasian} records and 52\% {\tt Black/African-American} (the distribution of the two groups in the {\sc Compas} dataset).
     \item In accordance with the racial distribution of the U.S. population as reported by the Census Bureau Data \cite{usCensus}, we create a passenger table by taking a sample from {\sc Compas} that includes 80\% {\tt Caucasian} and 20\% {\tt Black/African-American} individuals.
     \item We perform a Cartesian product on the two tables and label each pair as a match if left and right records have identical {\tt personID}s.
     \item Similar to the aforementioned process for \fmatch, the right records (that correspond to the no-fly list table) undergo perturbation in the {\tt firstName} and {\tt lastName} columns.  
 \end{itemize}
}

\submit{\vspace{-1mm}}
\subsubsection{Complementary Datasets.}
Data in the context of EM tasks usually fall into one of the following categories:

\begin{itemize}[leftmargin=*]
    \item \textit{Structured:} In this category of datasets, attribute values are atomic, meaning that they cannot be broken into multiple values. Furthermore, there are no missing values in the data.
    \item \textit{Dirty:} This category of datasets is similar to structured datasets; however, they include far too many random missing values in their columns. Therefore an attribute value may appear for a record while it does not exist for another one.
    \item \textit{Textual:} Textual datasets are made of a single attribute per record containing a textual description.
\end{itemize}

For the completeness of our experiments, we select several datasets from each category on which we evaluate the matchers. The complementary datasets are chosen from WDC~\cite{primpeli2019wdc} and Magellan~\cite{mudgal2018deep} repositories which are the standard benchmark corpora used in EM literature. Aside from the dataset type, we carefully handpicked the datasets w.r.t. domain, sensitive attribute type, and ground-truth class balance to cover a variety of possible settings. For the textual datasets {\sc Shoes} and {\sc Cameras}, we extract the manufacturer of the corresponding product from the description as the sensitive attribute.
Table \ref{tbl:datasets} shows the details of the selected datasets.
\submit{\vspace{-6mm}}
\subsubsection{Entity Matchers} To cover the breadth of existing methods in our experiments, we picked 13 EM tools from each of the discussed approaches (1 rule-based, 7 non-neural, and 5 neural). The selection criteria included the public availability and error-free execution of the source codes.
To ensure the satisfactory performance of the entity matchers, we took the following steps:  
\paragraph{BooleanRuleMatcher} We used the automatic feature generation tool provided in the Magellan library to extract features based on the similarity of the columns in the input table w.r.t. multiple distance measures. Next, we handpick some of the generated features based on which we declare matching conditions. \rev{\marginpar{R1.D4 M2} For each attribute, the generator creates multiple features based on different distance measures.} Depending on the attribute involved in the generated features, we either use the exact match of the attribute values (for attributes with short and atomic values, e.g., \textit{year}) or \rev{choose one of the distance-based features (e.g. {\em cosine} similarity between left and right attribute values) with a} similarity \rev{threshold} of greater than 0.5 (for attributes with longer values, e.g., \textit{paper title}).\footnote{\rev{Details on the specified rules for each dataset are provided in the GitHub repository.}}

\submit{\vspace{-2mm}}
\paragraph{Non-neural Matchers} For all non-neural matchers except for {\sc Dedupe}, we used the automatic feature generation tool in the Magellan library. Next, all of the generated features are fed to the models for training.
{\sc Dedupe}'s active learning component requires manual labeling of difficult record pairs, which is an uphill task. To bypass this step, we converted the training data into {\sc Dedupe}'s generated cache file format and utilized the entire training samples to keep the experiment consistent with the other matchers. Finally, {\sc Dedupe} did not scale for \fmatch, \nofly, {\sc Shoes} and {\sc Cameras}.

\submit{\vspace{-2mm}}
\paragraph{Neural Matchers}
We tuned the hyper-parameters of all the matchers
according to their results on the validation set.
For {\sc DeepMatcher}, {\sc HierMatcher}, and {\sc Mcan} we trained the models for 10 epochs with a batch size of 16 and used \textit{fastText}~\cite{bojanowski2016enriching} pre-trained word embeddings. We used the \textit{hybrid} model of {\sc DeepMatcher} that reportedly performs better than the other models. For {\sc HierMatcher}, we used the attribute-aware attention mechanism. For {\sc Mcan}, we utilized self-attention, pair-attention, global-attention, and gating mechanisms that reportedly would achieve the best results.
For {\sc Gnem}, we trained the GCN models for 10 epochs with a batch size of 2 and 768 nodes at each layer.
For {\sc Ditto}, we trained the models for 40 epochs with a batch size of 64 while using the DistilBERT language model and optimizations such as data augmentation, sequence summarization, and domain knowledge injection.

For all datasets except {\sc Cricket}, we declare a pair of records as a ``match'' if the similarity between the two is greater than 0.5. 
\rev{\marginpar{R1.D6}
Our choice of threshold value has a probabilistic interpretation of having a higher likelihood of being a match rather than being non-match. 
In other words, we follow a randomized rounding logic, where non-integer values in range $(0,1)$ are rounded to the closer integer.
Moreover, we empirically observed that a reasonably high accuracy occurs among all the matchers with a threshold over $0.5$, as shown in \S~\ref{sec:exp:sensitivity}.}
For the {\sc Cricket} dataset, however, due to the high similarity of all pairs, we had to choose a higher similarity threshold of 0.9 because otherwise, all of the models would predict all pairs as ``match'', which would affect the models' correctness. As for the fairness threshold, we follow EEOC's 80\% rule~\cite{EEOC}, that only 20\% disparity is tolerated.


\vspace{-2mm}
\subsection{Results for Social Cases}\label{sec:social_exp}
\subsubsection{\nofly}
We begin our experiments by evaluating matchers' fairness on our \nofly dataset.
Recall that \nofly dataset is the matching between the no-fly list and the passengers list, where the two lists have different distributions of the demographic groups. In particular, while in the U.S. population (passenger list) the White population (75\%) is significantly higher than  the Black (13\%), in the no-fly list Blacks are over-represented and the White and Black ratios are almost the same. 
It is common for a no-fly list to suffer from sampling bias. 
\rev{\marginpar{R1.D5}Table~\ref{tbl:compas_measures} includes the breakdown of experiment results. All the non-neural matchers had a perfect prediction performance, meaning that the TPR, FDR, and disparity values were 1.00, 0.00, and 0.00 respectively.}  
\techrep{
Figure~\ref{fig:compas_single} reports  the unfair matchers toward the African-Americans with respect to different measures.
}

Due to the disjoint nature of the binary-sensitive attribute {\tt race} in our comparison, single and pairwise fairness results are identical and therefore we only report the single fairness results. 
The first observation is the {\em superiority of non-neural matchers} over neural matchers for this task, \rev{both on fairness and the overall performance}. \rev{The higher performance of these models for structured datasets} has previously been reported in \cite{mudgal2018deep}, where the majority of non-neural matchers performed on par with or outperform the neural matchers. Next, by looking into the neural matchers in Table \ref{tbl:compas_measures}, we see a significant disparity against {\tt African-American} group. More specifically, in terms of {\bf FDR}
the {\tt African-American}  to {\tt Caucasian} ratio is between 
1.11 to {\bf 3.8} ({\bf 280\% larger}) across different matchers.
This translates to {\em a significantly higher chance of preventing an {\tt African-American} person to board a flight} or enter a country compared to a {\tt Caucasian} person.  

\begin{table}
\caption{\nofly results.\rev{All non-neural matchers had a perfect prediction performance, with {\em no FP or FN for any of the groups}. Unfair matchers in bold.}}\vspace{-3mm}
\small
\scalebox{0.9}{
\begin{tabular}{l l l l  l | l l l l} 
 \toprule
 & \multicolumn{2}{c}{\thead{{\bf TPR}}}& \multicolumn{2}{c}{\thead{{\bf Disparity}}}& \multicolumn{2}{c}{\thead{{\bf FDR}}}& \multicolumn{2}{c}{\thead{{\bf Disparity}}}\\
\thead{{\bf Matcher}}& \thead{{\tt Afr.}} &\thead{{\tt Cauc.}}& \thead{{\bf sub}} &\thead{{\bf div}}& \thead{{\tt Afr.}} &\thead{{\tt Cauc.}}& \thead{{\bf sub}} &\thead{{\bf div}} \\ \midrule
{\sc DeepMatcher}&0.89&0.86&-0.03&-0.03&0.20&0.18&0.02&0.11\\ \midrule
{\sc Ditto}&0.76&0.82&0.06&0.08&{\color{red}{0.31}}&{\color{red}{0.22}}&0.09&{\color{red}{\bf 0.41}}\\\midrule
{\sc GNEM}&0.84&0.84&0.00&0.00&0.17&0.09&0.08&{\color{red}{\bf 0.88}}\\\midrule
{\sc HierMatcher}&0.72&0.74&0.02&0.10&0.22&0.16&0.06&{\color{red}{\bf 0.38}}\\\midrule
{\sc MCAN}&{\color{red}{0.54}}&{\color{red}{0.57}}&0.03&0.05&0.19&0.05&{{0.14}}&{\color{red}{\bf 2.8}}\\
 \bottomrule
\end{tabular}}
\label{tbl:compas_measures}
\vspace{-5mm}
\end{table}


\noindent To better illustrate the root cause behind the disparity, let us highlight the following case that is falsely labeled as ``match'' by {\sc Ditto}:
\begin{center}
\fbox{\begin{minipage}{.94\columnwidth}\small \tt
    (left record) {\bf firstName:} James {\bf lastName:} Brown {\bf race:} African-American\\
    (right record) {\bf firstName:} Samanthai {\bf lastName:} Browne {\bf race:} African-American
\end{minipage}}
\end{center}
Some names are more common within certain demographic groups than others. For example, last names that are very common among black people include Brown, Jackson, Williams, Johnson, etc.
Since the no-fly list in our \nofly dataset over-represents the Black group, for an individual in this group there is a higher chance of getting falsely labeled as a match. 

\vspace{-1mm}
\subsubsection{\fmatch} \techrep{Having seen the impact of disproportionate representation of records in data w.r.t. the sensitive attributes, we turn our attention to another type of discrimination attributable to EM systems.}
Consider the \fmatch dataset described in \S~\ref{sec:social_datasets}. There are 2,061 Chinese {\tt cn} faculty members in this dataset compared to 1,595 German {\tt de} ones. Therefore, when we create the EM dataset by performing the Cartesian product, the group of Chinese faculty members has the larger population in the dataset. To increase the population gap even wider, we remove 80\% of the non-match pairs that have a German faculty member either on the right side or the left side. As a result, the number of Chinese pairs becomes more than 6 times the number of German pairs in the final sample ensuring proper representation. Next, using a variety of matchers, we conduct the matching task on the data and audit the matchers for fairness. 
\rev{\marginpar{R1.D5 M2}Table~\ref{tbl:csranking_measures} includes the breakdown of experiment results.}
\techrep{Figure~\ref{fig:csranking_single} reports  the unfair matchers toward Chinese {\tt cn} faculties with respect to measures.} 
\rev{Overall, non-neural matchers outperform neural matchers in terms of model performance and fairness.
Within the neural matchers we observe between} 9\% to 22\% more prone to make an erroneous positive prediction (match) for the {\tt cn} group. Further investigating the false-positives, we observed that those mostly include names that are very similar in the English transcription. An example of such cases (FP by {\sc Ditto}) is brought in the following:

\fbox{\begin{minipage}{.94\columnwidth}\small \tt
    (left record) {\bf fullName:} Qingming Huang {\bf country:} cn\\
    (right record) {\bf fullName:} Qing-Hu Huang {\bf country:} cn
\end{minipage}}

Furthermore, the models make somewhere between 44\% to 75\% more mistakes in terms of false-negative predictions for the {\tt cn} group. Due to the higher degree of similarities in Chinese names, models in general become more sensitive to minor differences and tend to mismatch. An example of such cases is the following:


\fbox{\begin{minipage}{.94\columnwidth}\small \tt
    (left record) {\bf fullName:} LinLin Shen {\bf country:} cn\\
    (right record) {\bf fullName:} Linlin phen {\bf country:} cn
\end{minipage}}

\rev{\marginpar{R1.D8}More extensive results on the overall performance of the matchers across social datasets are provided in~\cite{techrep}.} 

\begin{table}
\caption{\fmatch results. \rev{Unfair matchers in bold.}}\vspace{-3mm}
\small
\scalebox{0.9}{
\begin{tabular}{l l l l l | l l l l} 
 \toprule
 & \multicolumn{2}{c}{\thead{{\bf TPR}}}& \multicolumn{2}{c}{\thead{{\bf Disparity}}}& \multicolumn{2}{c}{\thead{{\bf PPV}}}& \multicolumn{2}{c}{\thead{{\bf Disparity}}}\\
\thead{{\bf Matcher}}& \thead{{\tt cn}} &\thead{{\tt de}}& \thead{{\bf sub}} &\thead{{\bf div}}& \thead{{\tt cn}} &\thead{{\tt de}}& \thead{{\bf sub}} &\thead{{\bf div}} \\ \midrule
{\sc DeepMatcher}&0.48&0.72&{\color{red}{\bf0.23}}&{\color{red}{\bf0.50}}&0.79&0.87&0.08&0.11\\ \midrule
{\sc Ditto}&0.59&0.85&{\color{red}{\bf0.26}}&{\color{red}{\bf0.44}}&0.77&0.94&0.17&{\color{red}{\bf 0.22}}\\\midrule
{\sc GNEM}&0.78&0.90&0.12&0.15&0.83&0.92&0.08&0.11\\\midrule
{\sc HierMatcher}&0.47&0.78&{\color{red}{\bf0.31}}&{\color{red}{\bf0.66}}&0.78&0.89&0.11&0.14\\\midrule
{\sc MCAN}&0.40&0.70&{\color{red}{\bf0.30}}&{\color{red}{\bf0.75}}&0.86&0.94&0.08&0.09\\\midrule\midrule
{\sc DTMatcher}&0.95&0.90&-0.05&-0.05&0.89&0.98&0.09&0.10\\\midrule
{\sc LinRegMatcher}&{\color{red}{0.33}}&{\color{red}{0.23}}&-0.09&-0.43&0.44&0.96&{\color{red}{\bf 0.52}}&{\color{red}{\bf 1.18}}\\\midrule
{\sc LogRegMatcher}&0.95&0.88&-0.07&-0.08&0.93&1.0&0.07&0.07\\\midrule
{\sc NbMatcher}&{ 0.99}&{ 0.99}&{ 0.00}&{0.00}&{\color{red}{0.03}}&{\color{red}{0.58}}&{\color{red}{\bf 0.55}}&{\color{red}{\bf 18.3}}\\\midrule
{\sc RfMatcher}&0.96&0.89&-0.06&-0.08&{0.98}&{0.99}&{0.01}&{0.01}\\\midrule
{\sc SvmMatcher}&0.95&0.87&-0.07&-0.09&{0.94}&{0.99}&{0.05}&{0.05}\\
\bottomrule
\end{tabular}}
\label{tbl:csranking_measures}
\vspace{-5mm}
\end{table}

\techrep{
\begin{figure}[t]
  \centering
    \begin{minipage}{.48\columnwidth}
    \centering
    \includegraphics[width=\linewidth]{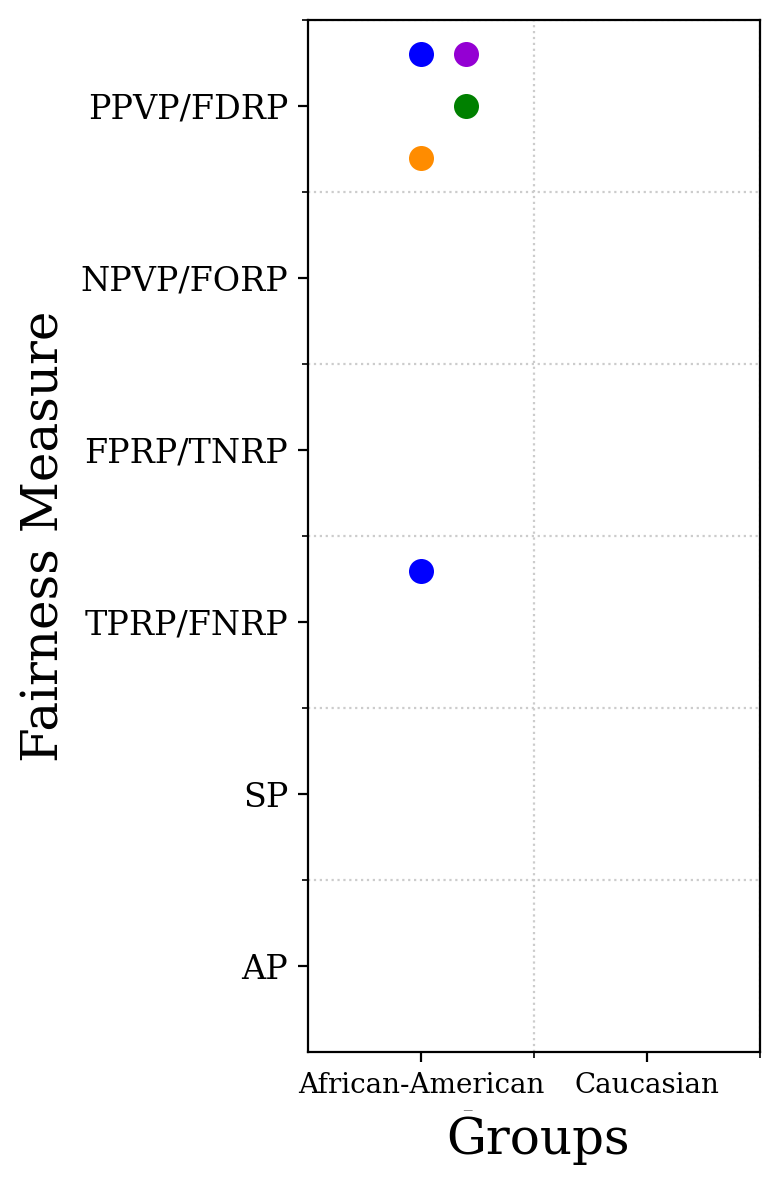}
    \vspace{-8mm}
    \caption{\rev{\fmatch: Unfair matchers towards African-Americans  w.r.t. single fairness;} plot markers provided in Figure~\ref{tbl:markers}.}
    \label{fig:compas_single}
  \end{minipage}
   \hfill
  \begin{minipage}{.48\columnwidth}
    \centering
    \vspace{-4mm}
    \includegraphics[width=\linewidth]{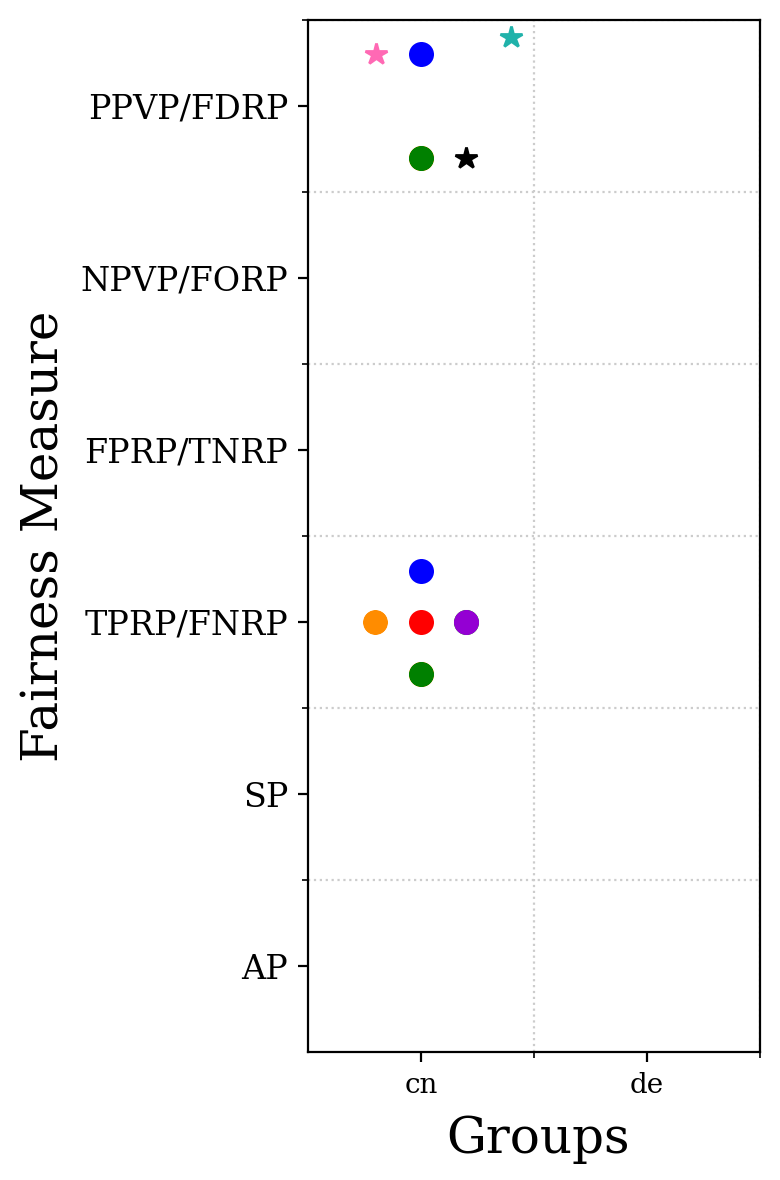}
    \vspace{-8mm}
    \caption{\rev{\nofly: Unfair matchers towards Chinese faculty  w.r.t. single fairness;}}
    \label{fig:csranking_single}
  \end{minipage}
  \vspace{-5mm}
\end{figure}
}

\submit{\vspace{-3mm}}
\subsection{Comprehensive  Results}\label{sec:comprehensive_results}


\rev{\marginpar{R1.D8} 
This section provides a comprehensive evaluation of the matchers' fairness and correctness using the benchmark datasets.} 

\stitle{\rev{Summary of result}}
\rev{ \marginpar{R3.D2 R1.W1 M4}
In summary, our results confirm the higher accuracy (\S~\ref{sec:evalcorrecteness}) and fairness (\S~\ref{sec:evalmatchers}) of neural matchers for textual and dirty data (Figure~\ref{fig:cameras-single}), and non-neural matchers for structured data (Figures~\ref{fig:itunes-amazon-single} and~\ref{fig:DBLP-Sch-single}).
While heavily relying on problematic proxies hurts the fairness of non-neural matchers, 
not fully considering the dataset structure and heavily relying on semantic similarities and (biased) pretrained models hurts the fairness of neural matchers.
TPRP and PPVP were more capable of revealing matching unfairness (\S~\ref{sec:evalmeasures}). Finally, we observed a higher fairness sensitivity of neural matchers on matching thresholds (\S~\ref{sec:exp:sensitivity}).
}

\submit{\vspace{-2mm}}
\subsubsection{\underline{Correctness}}
\label{sec:evalcorrecteness} 
\rev{\marginpar{R1.D8}Due to space limitations, here we present a summary of our correctness results. More extensive results on the overall performance of the matchers across the datasets, fairness and accuracy synergies, and detailed discussions can be found in the technical report~\cite{techrep}.} 
In summary, \rev{aligned with \cite{mudgal2018deep}}, throughout our extensive experiments, we observed that {\em neural matchers are more accurate than non-neural matchers on textual and dirty data}. Modern neural matchers draw on external knowledge by incorporating language models, which helps a matcher to learn the relevance of records despite the lack of structure and syntactic similarity in text records. This result is consistent with what is reported by the state-of-art matchers.
On the other hand, our results corroborate that {\em non-neural matchers are more accurate than neural matchers
on structured data}. Various combinations of correctness and fairness exist in EM as some matchers have low accuracy and F-1 score, while no unfairness issue is observed. This can be explained by the low accuracy of these matchers for all groups across the board which makes the disparity a low value.
\techrep{
In Figure ~\ref{tbl:accfairness}, we present a selective overview of the unfairness and accuracy of matchers across all datasets. 
The general message is that similar to accuracy, unfairness is dataset and measure-dependent. 
No matcher is unfair across all datasets and no matcher is unfair across all measures. 

\begin{figure}[!tb]
    \centering \small
    \scalebox{0.9}{
    \begin{tabular}{l l l} 
         \toprule
\thead{{\bf Accurate}} & \thead{{\bf Fair}} & \thead{{\bf Evidence}}  \\ \midrule
 $\times$ & $\times$& {\sc RfMatcher}: {\sc Cameras}: \{TPRP,PPVP\}\\
 & & {\sc BooleanRuleMatcher}: {\sc iTunes-Amazon}: \{AP,SP,PPVP\}\\
  & & {\sc Gnem}: {\sc iTunes-Amazon}: \{AP,PPVP,...\}\\\midrule
 $\times$ & \checkmark& {\sc LinRegMatcher}: {\sc Shoes}; {\sc Gnem}: {\sc Dblp-Acm} \\
 & &  {\sc BooleanRuleMatcher}: {\sc Cricket}\\\midrule
 \checkmark & $\times$& {\sc HierMatcher}: {\sc iTunes-Amazon}: \{AP,PPVP,...\} \\
 & & {\sc SvmMatcher}: {\sc Dblp-Acm}: PPVP\\
 & & {\sc Mcan}: {\sc Cameras}: TPRP; {\sc Ditto}: {\sc Dblp-Scholar}: \{AP,TPRP,...\}\\
 & & {\sc DeepMatcher}: {\sc Camera}: \{PPVP,TPRP\} \\\midrule
\checkmark & \checkmark&  {\sc Mcan}: {\sc Dblp-Acm}; {\sc Ditto}: {\sc Cricket}\\
& &  {\sc NbMatcher}: {\sc Dblp-Scholar}\\ 
\bottomrule
    \end{tabular}}
    \submit{\vspace{-4mm}}
    \caption{Fairness and accuracy synergies}
    \label{tbl:accfairness}
    \submit{\vspace{-6mm}}
\end{figure}
}

\begin{figure*}[!tb] 
    \begin{minipage}[t]{0.23\linewidth}
        \centering
        \footnotesize
        \vspace{-48mm}
        \begin{tabular}{@{}l@{}c@{}}
        \toprule
        \thead{{\bf Model}} & \thead{{\bf Marker}} \\
        \midrule
        {\sc DeepMatcher} & {\textcolor[RGB]{255, 0, 0}{\faCircle}} \\
        {\sc Ditto} & {\textcolor[RGB]{0,0,255}{\faCircle}} \\  
        {\sc Gnem} & {\textcolor[RGB]{255,140,0}{\faCircle}} \\  
        {\sc HierMatcher} & {\textcolor[RGB]{0,128,0}{\faCircle}} \\  
        {\sc Mcan} & {\textcolor[RGB]{148,0,211}{\faCircle}} \\
        {\sc SvmMatcher} & {\textcolor[RGB]{0,0,0}{\faStar}} \\
        {\sc RfMatcher} & {\textcolor[RGB]{255,105,180}{\faStar}} \\  
        {\sc NbMatcher} & {\textcolor[RGB]{128,128,128}{\faStar}} \\  
        {\sc LogRegMatcher} & {\textcolor[RGB]{32,177,170}{\faStar}} \\  
        {\sc LinRegMatcher} & {\textcolor[RGB]{144,238,144}{\faStar}} \\  
        {\sc DtMatcher} & {\textcolor[RGB]{159,82,45}{\faStar}} \\  
        {\sc Dedupe} & {\textcolor[RGB]{1,191,255}{\faStar}} \\  
        {\sc BooleanRuleMatcher} & {\textcolor[RGB]{255,215,0}{\faStar}} \\ 
        \bottomrule
        \end{tabular}
        \vspace{-2mm}\caption{plot markers\techrep{ for the entity matchers}}
        \label{tbl:markers}
    \end{minipage}
    \begin{minipage}[t]{0.32\linewidth}
        \centering 
        \includegraphics[width=\textwidth]{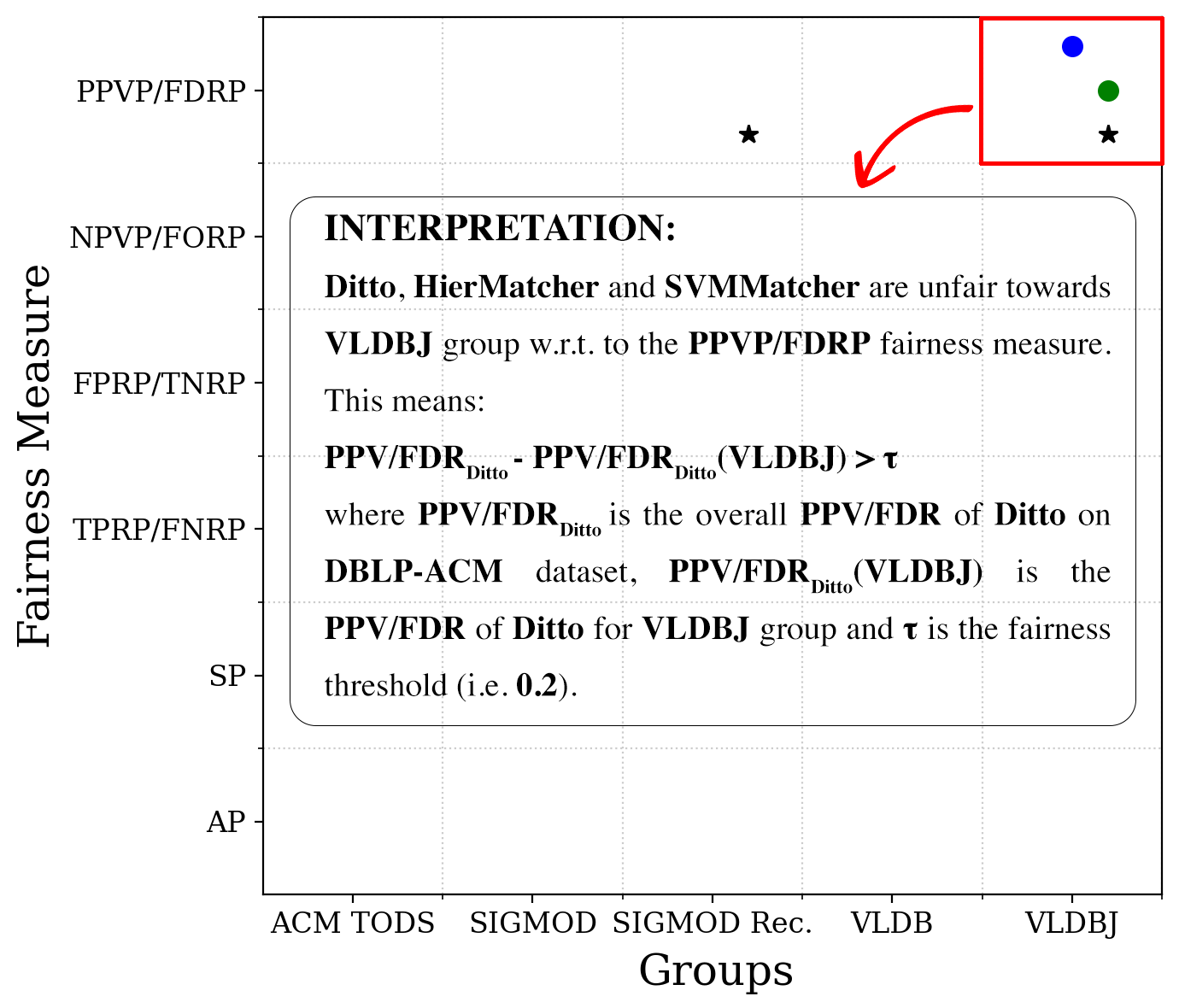}
        \vspace{-8mm}\caption{\rev{{\sc Dblp-Acm}: Single Fairness\protect\footnotemark[6]}}
        \label{fig:DBLP-ACM-single}
    \end{minipage} 
    \hfill
    \begin{minipage}[t]{0.39\linewidth}
        \centering
        \includegraphics[width=\textwidth]{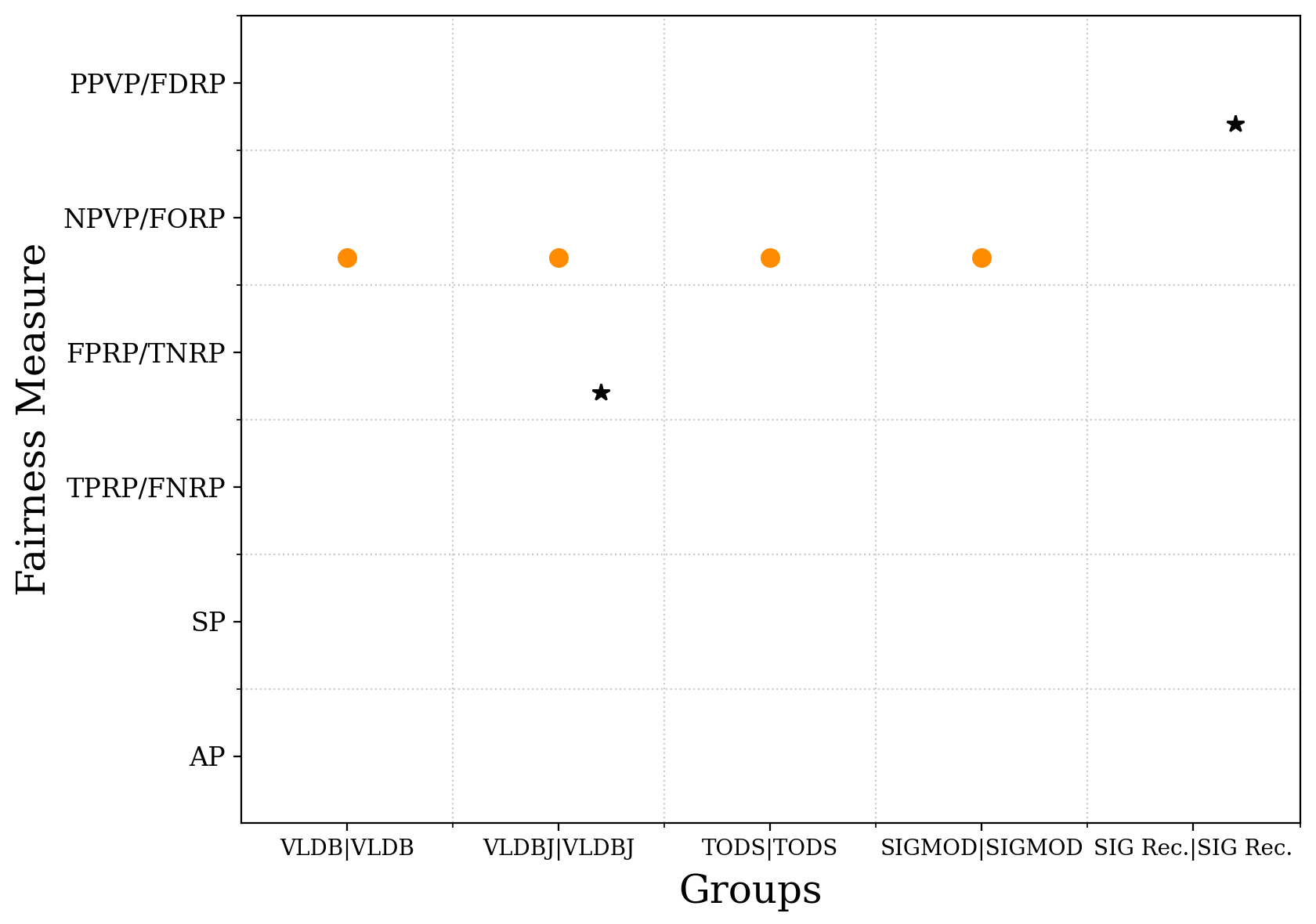}
        \vspace{-8mm}\caption{{\sc Dblp-Acm}: Pairwise Fairness}
        \label{fig:DBLP-ACM-pairwise}
    \end{minipage}
    \vspace{-4mm}
\end{figure*}

\begin{figure*}[!tb]
\centering 
    \includegraphics[width=.85\textwidth]{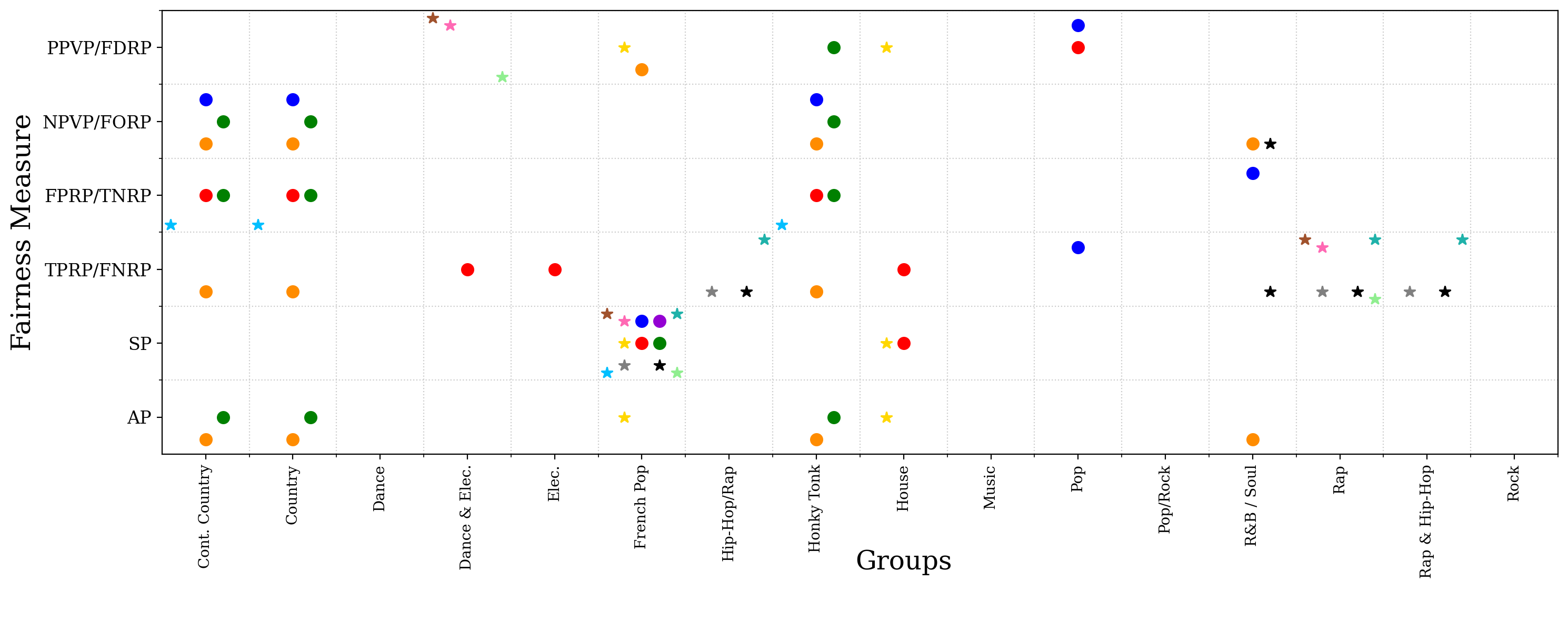}
    \submit{\vspace{-9mm}}\caption{{\sc iTunes-Amazon}: Single Fairness}
\label{fig:itunes-amazon-single}
\submit{\vspace{-4mm}}
\end{figure*}

\begin{figure*}[!tb] 
    \begin{minipage}[t]{0.29\linewidth}
        \centering
        \includegraphics[width=\textwidth]{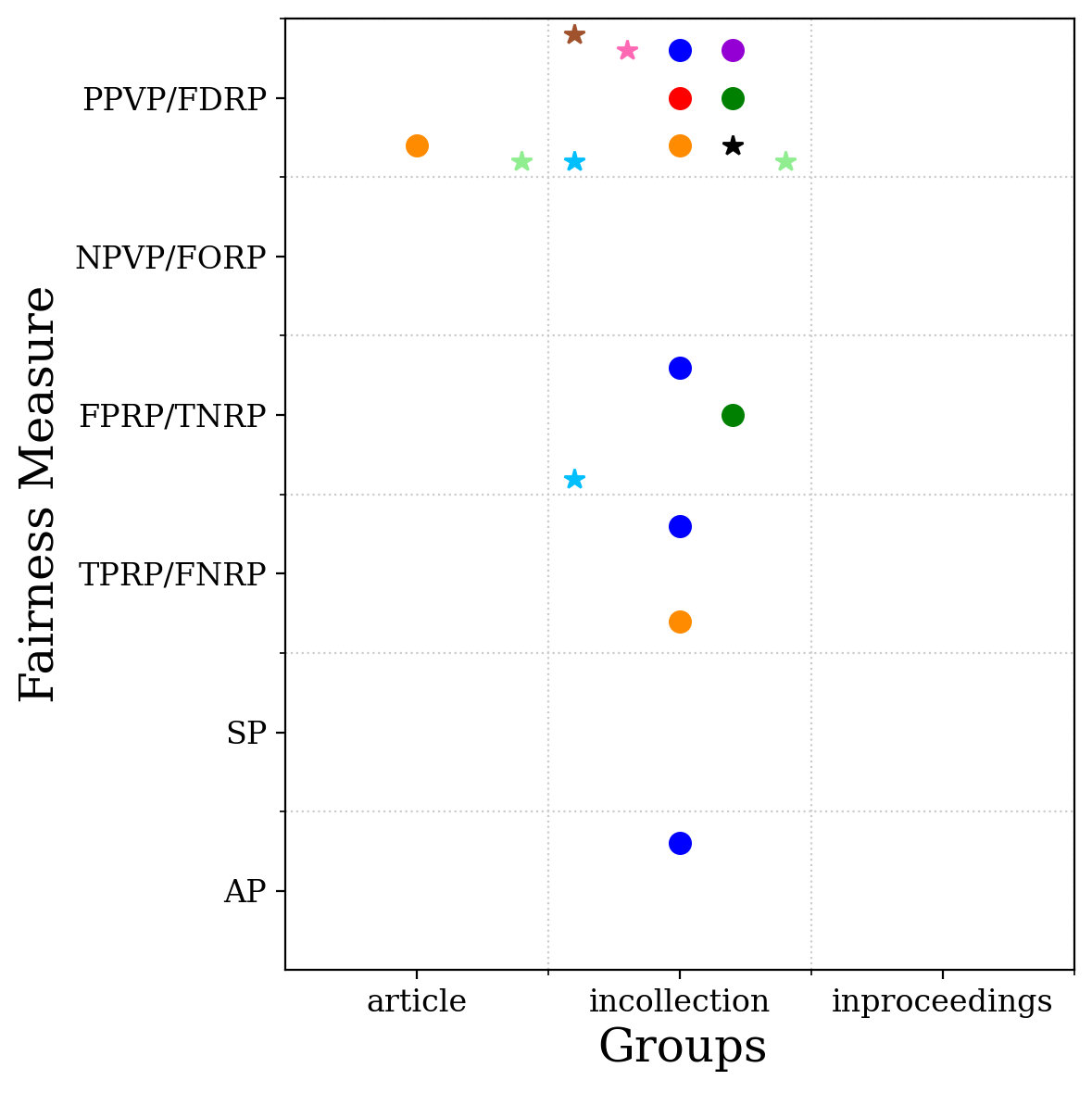}
        \vspace{-8mm}\caption{{\sc Dblp-Scholar}: Single}
        \label{fig:DBLP-Sch-single}
    \end{minipage}
    \hfill
    \begin{minipage}[t]{0.22\linewidth}
        \centering
        \includegraphics[width=\textwidth]{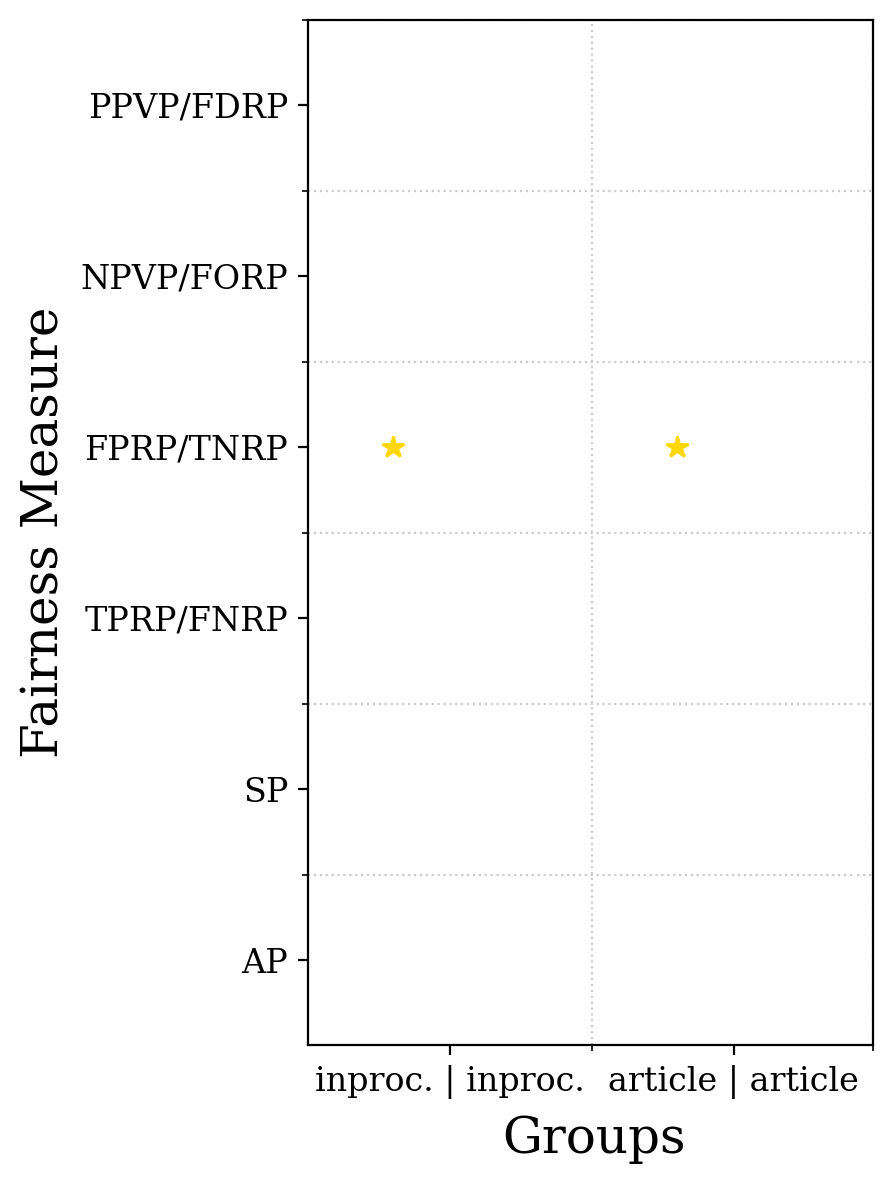}
        \vspace{-9mm}\caption{{\sc Dblp-Scholar}: Pairwise}
        \label{fig:DBLP-Sch-pairwise}
    \end{minipage}  
    \hfill
    \begin{minipage}[t]{0.19\linewidth}
        \centering
        \includegraphics[width=\textwidth]{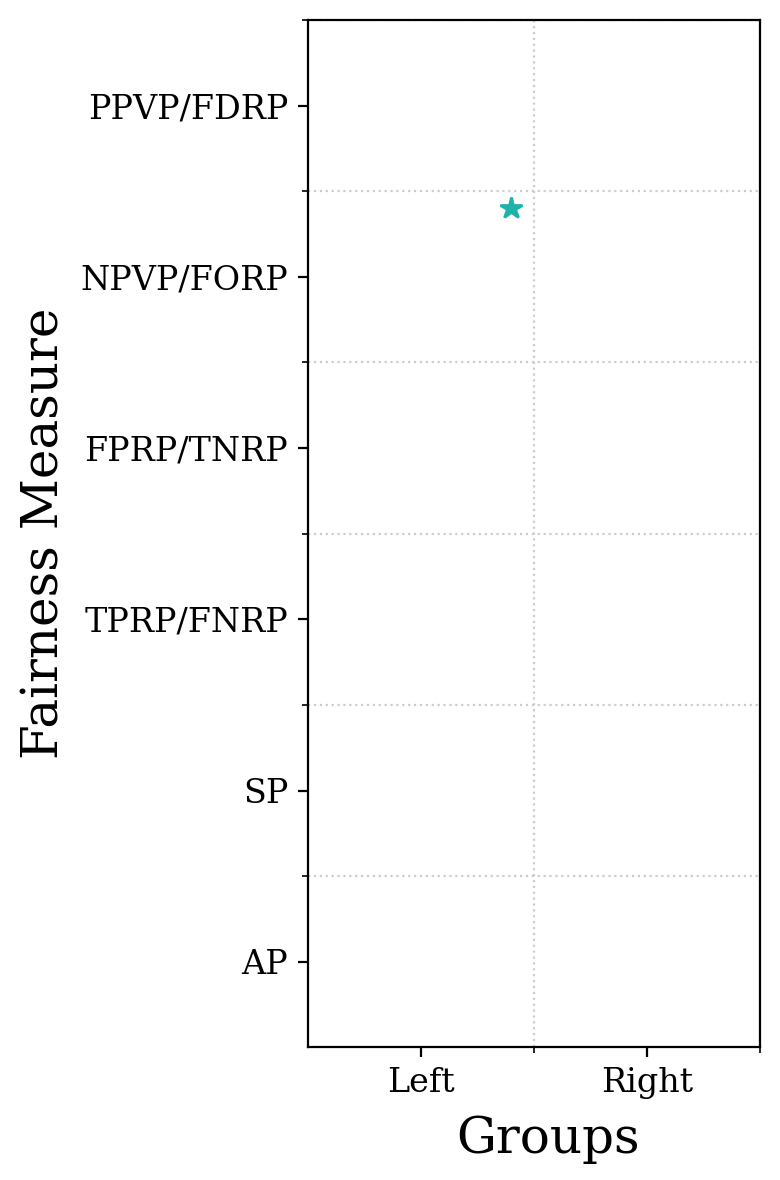}
        \vspace{-9mm}\caption{{\sc Cricket}: Single}
        \label{fig:cricket-single}
    \end{minipage}
    \hfill
    \begin{minipage}[t]{0.21\linewidth}
\centering
\includegraphics[width=.9\textwidth]{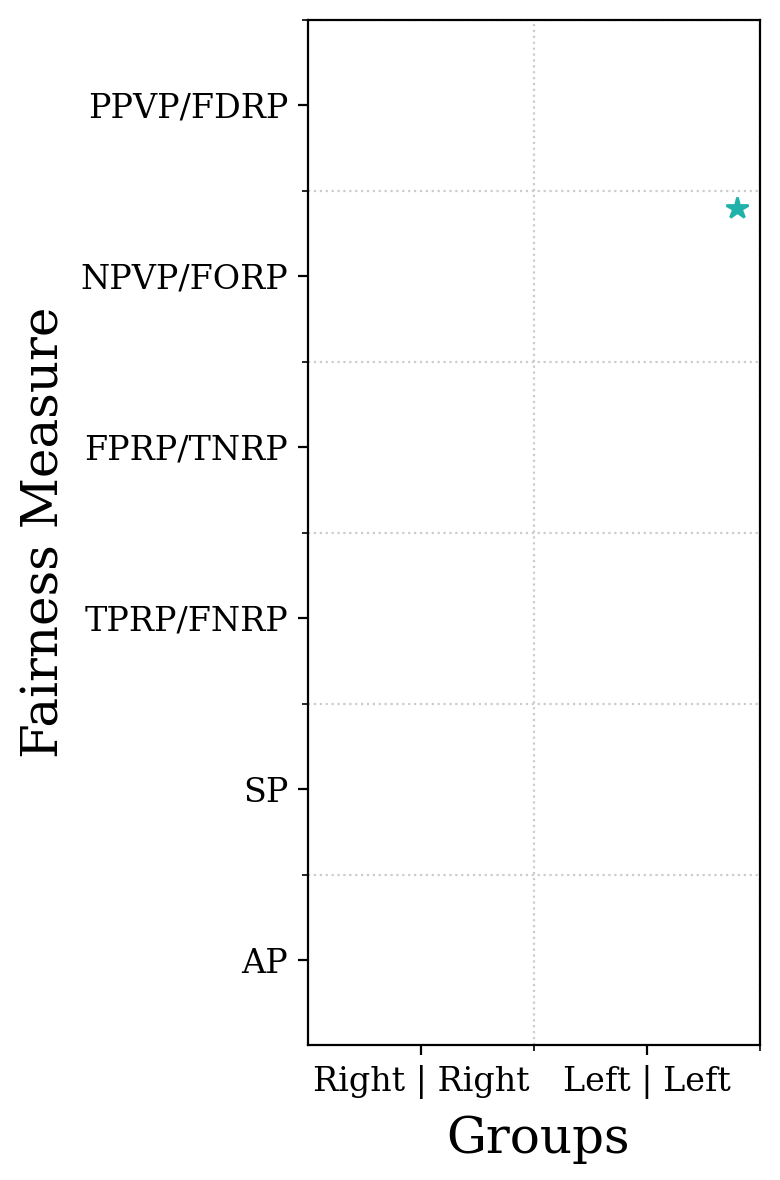}
        \vspace{-5mm}\caption{{\sc Cricket}: Pairwise}
        \label{fig:cricket-pairwise}
\end{minipage} 
\vspace{-5mm}
\end{figure*}

\begin{figure*}[!tb]
\centering 
    \includegraphics[width=\textwidth]{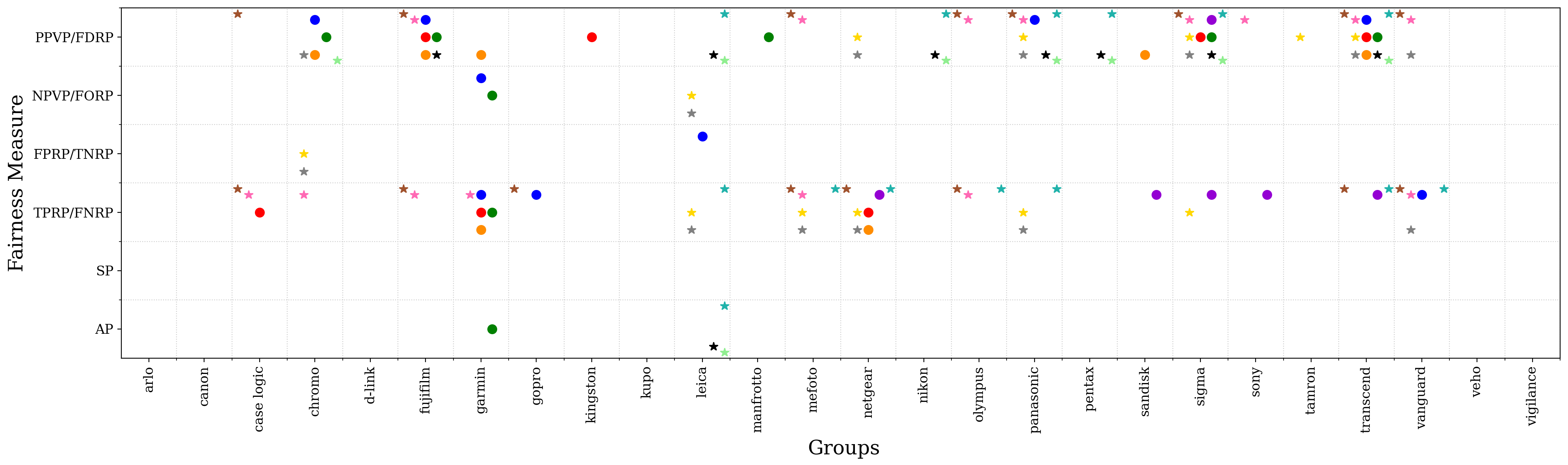}
    \submit{\vspace{-9mm}}\caption{{\sc Cameras}: Single Fairness}
\label{fig:cameras-single}
\submit{\vspace{-5mm}}
\end{figure*}

\subsubsection{\underline{Fairness: Measure Types}}
\label{sec:evalmeasures}
\footnotetext[5]{For the evaluation of ML-based matchers, we used random train/test splits from the datasets published by Magellan~\cite{konda2018magellan}. To be consistent, all matchers are  evaluated in a standard framework  against the same datasets. We acknowledge that these results may not exactly match the accuracy results reported by matchers' papers.}
\footnotetext[6]{
Across all plots, Equalized Odds (EO) is the union of FPRP and TPRP rows. A matcher that is appears either in row 3 or row 4 of any column is unfair from EO perspective.
}

\techrep{
{\em Some fairness measures are not applicable to EM, and some are more capable of revealing the unfairness of matcher.} We have extensively discussed this point in  \S~\ref{sec:measuresdiscussion}. Here, we bring some empirical evidence of such scenarios.
}
\rev{\marginpar{R1.W1 R1.D8}
Figures~\ref{fig:DBLP-ACM-single} to \ref{fig:cameras-single} show our fairness evaluation results for all matchers across the benchmark datasets.
In each plot, the x-axis shows the groups (single or pairwise), while the fairness measures are provided in the y-axis.
The corresponding marker of a matcher is placed in a specific cell, if that matcher is unfair for the group shown in the cell's column based on the measure in its row.
}
In the majority of our experiments, PPVP and TPRP were the measures that discover unfairness the most across all datasets and matchers. 
Nevertheless, it is not the case that one measure fits all settings. When data has match/non-match negative imbalance, i.e., the number of matching pairs is much higher than non-matching pairs in the ground-truth, 
NPVP and FPRP are the most appropriate measures. 
This is because while the majority of pairs are positive instances, the failure of a matcher in identifying non-matches makes it unfair to certain groups. Consider the {\sc Cricket} dataset that contains a larger number of pairs of matching cricket batters than non-matching batters. As shown in Figure~\ref{fig:cricket-single}, NPVP allows us to detect the unfairness of a matcher such as {\sc LogRegMatcher} to left-handed batters due to the large number of FNs generated by this matcher. 
SP does not consider the ground-truth labels and requires the independence of the matching prediction from the groups. In other words, SP requires equal match ratios from different groups, independent of whether they really are a match or not. Then, when the ground-truth has (non-)match imbalance for a group, that is, the ratio of matched pairs to unmatched ones, is low, the SP measure falsely identifies a matcher as unfair for that group. 
An example of this phenomenon can be observed in Figure~\ref{fig:itunes-amazon-single}, for {\tt\small French-Pop} group in the {\sc iTunes-Amazon} dataset, where SP unfairness is indeed due to the fact that the ground-truth only contains TNs. 

{\em Some measures can be explained by others.} 
For example, let us consider the AP unfairness of {\sc Gnem} on {\sc iTunes-Amazon} for the group of country genres, including {\tt\small Country}, {\tt\small Cont. Country}, and {\tt\small Honky Tonk}, reported in Figure~\ref{fig:itunes-amazon-single}. This matcher has low accuracy for this group of genres because 
it identifies a small number of true matches (i.e., has a low number of TPs, thus, suffers from TPRP). Instead, the matchers falsely identify many pairs as non-match (i.e., have a high number of FPs, thus, suffer from NPVP). 
Similarly, we observe that {\sc HierMatcher} demonstrates AP unfairness on {\sc iTunes-Amazon} for the group of country genres because it incurs a large number of FPs, thus, suffers from FPRP unfairness. 

{\em Single unfairness can potentially propagate to pairwise fairness.} 
In Figure~\ref{fig:cricket-single} and~\ref{fig:cricket-pairwise}, 
we observe that the unfairness of {\sc LogRegMatcher} for the single {\tt\small Left Handed} group incurs its unfairness for the pairwise {\tt\small Left Handed-Left Handed} groups because most likely only a left-handed batter can be matched with another left-handed batter. 

\submit{\vspace{-5mm}}
\subsubsection{\underline{Fairness: Matcher Types}}\label{sec:evalmatchers}
\stitle{Neural Matchers}
{\em Neural matchers demonstrate more  unfairness on structured datasets than non-neural matchers}, as shown in Figures~\ref{fig:itunes-amazon-single} and~\ref{fig:DBLP-Sch-single}.  
One reason is that matchers such as {\sc Ditto} merge the content of different attributes as a single block and use token similarity as a signal for matching. 
However, for structured data, this technique may lose the important information specified by the structure. In particular, in the following example from {\sc Dblp-Acm} dataset, the two records have similar titles and are predicted as match despite the fact that they are (i) written by different authors, (ii) published in different venues, and (iii) published in different years. 
\fbox{\begin{minipage}{.98\columnwidth} \small \tt
(left record) {\bf title:} lineage tracing for general data warehouse transformations; {\bf author:} jennifer widom , yingwei cui; {\bf venue:} VLDBJ; {\bf year:} 2003 \\
     (right record) {\bf title:} data extraction and transformation for the data warehouse; {\bf author:} case squire; {\bf venue:} SIGMOD; {\bf year:} 1995
\end{minipage}}\\
One of the reasons {\sc Ditto} was unfair for {\tt\small VLDBJ} is that, similar to the following example, it is common to publish extended versions of previously published papers in this venue. As a result,
after merging different attributes as a block of text for each record, 
similar titles and authors may cause enough similarity between the two phrases that the {\sc Ditto} mistakenly predicts them as a match.\\
\fbox{\begin{minipage}{.98\columnwidth} \small \tt
    (left record) {\bf title:} efficient schemes for managing multiversionxml documents; {\bf author:} shu-yao chien , carlo zaniolo , vassilis j. tsotras; {\bf venue:} VLDBJ; {\bf year:} 2002\\
    (right record) {\bf title:} efficient management of multiversion documents by object referencing; {\bf author:} shu-yao chien , vassilis j. tsotras , carlo zaniolo; {\bf venue:} VLDB; {\bf year:} 2001
\end{minipage}}

{\em External bias could be injected into neural matchers through the use of language models and word embeddings.} 
For example, {\sc Hiermatcher} uses language models and {\em word embeddings} to compare the attribute similarities of records. 
As a result, it may mistakenly match articles with similar titles. Below is an FP example for {\sc HierMatcher}. Both articles are published in the same year. But they appear in different venues and are written by different authors. Still, language models find sufficient similarity between titles to persuade the matcher to label the records as a match. Perhaps this is because of the similarity of words like ``efficient'' and ``effective'' in the embedding space.\\
\fbox{\begin{minipage}{.94\columnwidth}\small \tt
    (left record) {\bf title:} efficient and cost-effective techniques for browsing and indexing large video databases; {\bf author:} kien a. hua , jung-hwan oh; {\bf venue:} SIGMOD; {\bf year:} 2000\\
    (right record) {\bf title:} effective timestamping in databases; {\bf author:} kristian torp , christian s. jensen , richard thomas snodgrass; {\bf venue:} VLDBJ; {\bf year:} 2000
\end{minipage}}

Another example we bring is from {\sc iTunes-Amazon} dataset. The following pair records is an FP by {\sc Ditto}. First, both songs are by {\tt Kenny Chesney}.
But more importantly, using a pre-trained language model, {\tt \small Likes Me} and {\tt \small Loves Me} are considered (almost) identical. 
As a result, the model mistakenly labeled the left and right songs as a match. 
Interestingly, such cases happen to be more frequent in genres such {\tt Country}, resulting in FPRP unfairness for those groups, as shown in Figure~\ref{fig:itunes-amazon-single}. \\
\fbox{\begin{minipage}{.94\columnwidth}\small \tt
    (left record) {\bf song:} Tequila Loves Me; {\bf artist:} K. Chesney\\
    (right record) {\bf song:} Likes Me; {\bf artist:} K. Chesney
\end{minipage}}

Our fourth example is from the {\sc Cameras} dataset, where camera records are matched based on their descriptions. 
A successful matcher on a dataset that includes descriptions in many languages requires extensive coverage of language models on various languages. 
For example, {\sc Mcan} returns the following pair of records as an FN, although the model and the brand match, and {\em Prijzen} is the Dutch translation of word {\em Prices}. We suspect that this is due to the poor coverage of word embeddings on the Dutch language.  \\
\fbox{\begin{minipage}{.94\columnwidth}\small \tt
    (left record) {\bf title:} Sony Cyber-shot RX100@en RX100 Prices - CNET@en\\
    (right record) {\bf tile:} Sony Cyber-shot RX100 Zwart - Prijzen @NL Tweakers@NL
\end{minipage}}


{\em One model does not fit all.} 
In {\sc iTunes-Amazon} dataset, an interesting observation is that
neural matchers perform poorly for the class of country (because a neural matcher creates a curvy decision boundary for all groups and fails for easy groups), while non-neural matchers perform poorly for the class of rap (because non-neural matchers make simple decision boundaries which may not work for a difficult group such as the class of rap genres).

{\em For setwise attributes, matchers demonstrate similar unfair behavior on groups with overlapping semantics.}  In practice, we observe that, in single setwise sensitive attributes, different sets of groups highly overlap. This is sometimes due to the existence of a semantic hierarchy of groups. For example, in the {\sc iTunes-Amazon} dataset, {\tt\small Honky Tonk} and {\tt\small Cont. Country} are subclasses of {\tt\small Country} in the semantic taxonomy of Wikipedia. As a result, we observe similar behavior of matchers across these groups. For instance, Figure~\ref{fig:itunes-amazon-single} shows extensive unfair behavior of neural matchers on country music groups: {\tt\small Honky Tonk}, {\tt\small Cont. Country}, and {\tt\small Country}. Following the same trend, non-neural matchers perform poorly on  groups {\tt\small Hip-hop/Rap} and  {\tt\small Rap} and {\tt\small Rap \& Hip-Hop}, suggesting these matchers are unfair to rap and hip-hop singers. 

\vspace{-2mm}
\stitle{Non-neural Matchers}
\label{sec:evalnonneurals}
The non-neural matchers universally failed for the textual datasets ({\sc Camera} and {\sc Shoes}), with F-1 measures as low as zero in several cases. This underscores that these matchers are not fit for unstructured data. Still, in some settings, these matchers were both inaccurate and unfair for different groups, as shown in Figure~
~\ref{fig:cameras-single}.
Note that a matcher being fair in these cases simply means that it equally {\em failed} for all groups, not that it is a good choice.
\techrep{For example, {\sc LinRegMatcher} was fair for the {\sc Shoes} dataset.
However, looking at its overall performance, it turns out it did not correctly find {\em any} of the true matches for any of the groups, hence was equally bad for all groups.}
On the other hand, non-neural matchers performed well for the structured datasets. Still, similar to the neural matchers, all of them showed unfairness in multiple cases.
Further investigating this unfairness, we realized that by minimizing the overall error, these models put high weights on attributes that often indicate a match.
In other words, overall, those attributes are good {\em proxies} for the ground-truth labels. However, when it comes to certain groups, they may not be as good proxies, causing the model to underperform for those groups.
For example, consider {\sc SvmMatcher} for the {\sc Dblp-Acm} dataset, which was unfair for {\tt SIGMOD Rec.} and {\tt VLDBJ}. 
First, we realized that both these groups frequently publish reports or editorial articles with the same title but different years and authors. Being trained to perform for all groups, the {\sc SvmMatcher} model {\em assigned a high weight to the title}, assuming that different articles have different titles. Therefore, for examples like the one below, it matched them, although different authors wrote those in different years. This caused a higher ratio of false match detection (FP) compared to the other groups resulting in PPVP unfairness.\\
\fbox{\begin{minipage}{.94\columnwidth} \small \tt
    (left record) {\bf title:} guest editorial; {\bf author:} alon y. halevy; {\bf venue:} VLDBJ; {\bf year:} 2002 \\
    (right record) {\bf title:} guest editorial ; {\bf author:} vijay atluri , anupam joshi , yelena yesha; {\bf venue:} VLDBJ; {\bf year:} 2003
\end{minipage}}\\
Besides, in Figure~\ref{fig:DBLP-ACM-pairwise} the unfairness due to the high FP for {\tt SIGMOD Rec.} and {\tt VLDBJ}, caused pairwise unfairness for these two groups as well.
Note that this issue is not necessarily limited to the non-neural matchers. For example, \cite{di2019interpreting} also reports that an RNN-based matcher heavily relied on the ``time'' attribute when matching songs in the {\sc iTunes-Amazon} dataset.
{\em Lack of proper coverage~\cite{asudeh2019assessing,shahbazi2022survey,asudeh2021identifying} from some groups} is the reason the models do not get well-trained for those. For example, in the {\sc Dblp-Acm} case, the training data did not include enough non-match cases with (almost) identical titles to reduce the correlation of the title with the ground-truth label.

\begin{figure}[!tb] 
  \begin{minipage}[b]{\linewidth}
    \centering
    \vspace{-10mm}
    \includegraphics[scale=0.36]{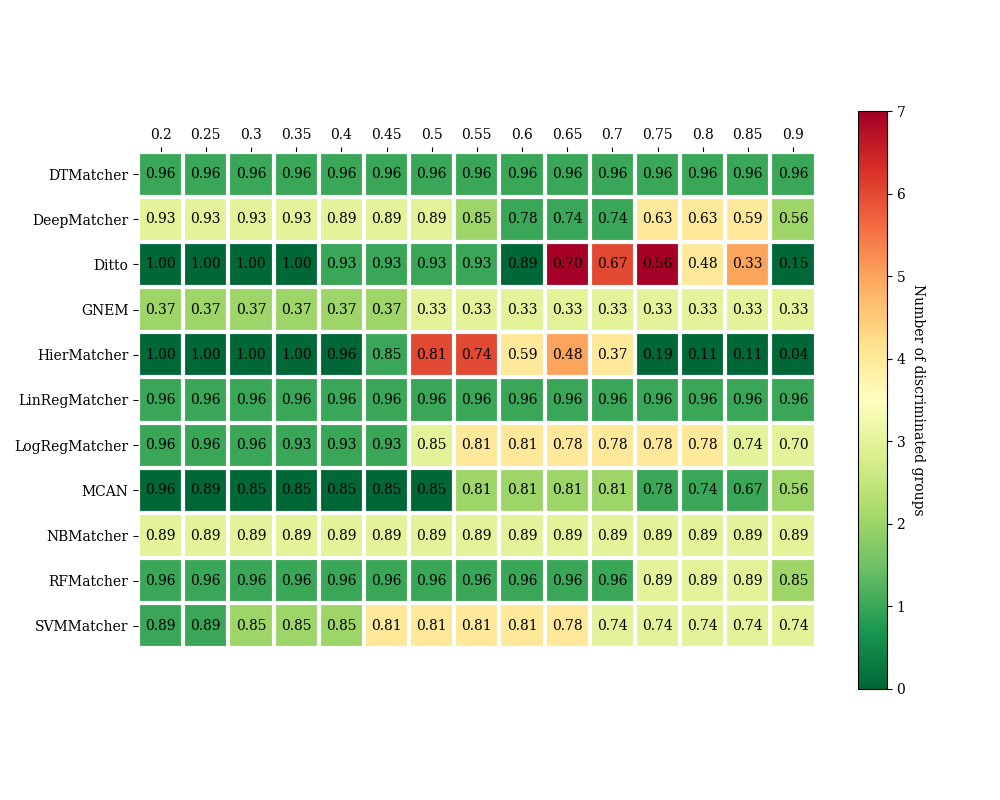} 
    \vspace{-19mm}
    \caption{\rev{The effect of matching threshold on fairness (TPRP) and accuracy (TPR) of the matchers on {\sc iTunes-Amazon} dataset. Cell color specifies the unfairness and the cell value indicates the accuracy.}} 
    \label{fig:heatmap_tprp_itunes_amazon}
  \end{minipage}
  \submit{\vspace{-8mm}}
\end{figure}

\rev{
\submit{\vspace{-5mm}}
\subsubsection{Matching Threshold vs. Fairness and Accuracy}\label{sec:exp:sensitivity} 
In this experiment, we study the sensitivity of the models' fairness to the matching threshold. Based on our previous results, we only focus on the two measures of TPRP and PPVP. 
\marginpar{R1.W2 R1.D6} 
Figure \ref{fig:heatmap_tprp_itunes_amazon} shows the number of discriminated 
groups with respect to TPRP (as the color code) and the overall TPR values (written in every cell) for different threshold values on the {\sc iTunes-Amazon} dataset. 
\submit{The complementary results on other datasets are provided in \cite{techrep}.}
It is evident in Figure \ref{fig:heatmap_tprp_itunes_amazon} that neural matchers are more sensitive to the choice of thresholds. For example, while {\sc Ditto} is completely fair at threshold 0.6, with a small increase in the matching threshold (to 0.65) it becomes unfair for 7 groups. 
To further investigate this empirical observation over various datasets and both TPRP and PPVP,
we define the threshold sensitivity of each matcher on a dataset 
as the $\ell_2$ distance on the number of groups a matcher is being unfair for between the adjacent matching thresholds.
The results are provided in Table \ref{fig:threshold_sensitivity}.
Larger values indicate more sensitivity to the matching threshold.
Aligned with our observation in Figure~\ref{fig:heatmap_tprp_itunes_amazon}, the table shows higher sensitivity (less robustness) for the neural matchers.
Some of the non-neural matchers have high sensitivity values for {\sc Cameras} dataset.
However, the model accuracy was universally bad for non-neural matchers on this dataset and regardless of fairness those matchers are not reliable.
}

\begin{table}[!tb]
    \centering \footnotesize
    \submit{\vspace{-1mm}}
    \caption{\rev{Sensitivity of fairness measures w.r.t. varying matching threshold. Cells with gray values show low-accuracy models that did not perform well regardless of fairness. Values highlighted in orange and red show moderate and high degrees of sensitivity.}}
    \label{fig:threshold_sensitivity}
    \submit{\vspace{-5mm}}
    \begin{tabular}{p{0.01\textwidth}|p{0.1\textwidth}|p{0.01\textwidth}p{0.01\textwidth}p{0.01\textwidth}p{0.01\textwidth}p{0.01\textwidth}p{0.01\textwidth}|p{0.01\textwidth}p{0.01\textwidth}p{0.01\textwidth}p{0.01\textwidth}p{0.01\textwidth}} 
         & & \multicolumn{6}{c|}{\thead{{\bf Non-neural}}}&  \multicolumn{5}{c}{\thead{{\bf Neural}}}\\ 
& \thead{{\bf Dataset}}& \thead{\rotatebox{90}{\sc DtMatcher}} &\thead{\rotatebox{90}{\sc LinRegMatcher}}& \thead{\rotatebox{90}{\sc LogRegMatcher}} &\thead{\rotatebox{90}{\sc NbMatcher}}& \thead{\rotatebox{90}{\sc RfMatcher}} &\thead{\rotatebox{90}{\sc SvmMatcher}}& \thead{\rotatebox{90}{\sc DeepMatcher}} &\thead{\rotatebox{90}{\sc Ditto}}& \thead{\rotatebox{90}{\sc Gnem}} &\thead{\rotatebox{90}{\sc HierMatcher }}& \thead{\rotatebox{90}{\sc Mcan}} \\ \hline
\multirow{4}{*}{\rotatebox{90}{\bf TPRP}}&{\sc iTunes-Amazon}&0&0&{\vspace{-2mm}\color{orange}{\bf2.4}}&0&{\vspace{-2mm}\color{orange}{\bf2.2}}&{\vspace{-2mm}\color{orange}{\bf2.4}}&{\vspace{-2mm}\color{red}{\bf3.9}}&{\vspace{-2mm}\color{red}{\bf9.3}}&{\vspace{-2mm}\color{gray}{1}}&{\vspace{-2mm}\color{red}{\bf6.9}}&{\vspace{-2mm}\color{orange}{\bf2.4}}\\\cline{2-13}
&{\sc Cameras}&{\vspace{-2mm}\color{gray}{1}}&{\vspace{-2mm}\color{gray}{0}}&{\vspace{-2mm}\color{gray}{8.4}}&{\vspace{-2mm}\color{gray}{2.8}}&{\vspace{-2mm}\color{gray}{8.7}}&{\vspace{-2mm}\color{gray}{7.1}}&{\vspace{-2mm}\color{red}{\bf3.3}}&{\vspace{-2mm}\color{orange}{\bf2.8}}&1&{\vspace{-2mm}\color{orange}{\bf2.6}}&{\vspace{-2mm}\color{red}{\bf3.6}}\\\cline{2-13}
&{\sc Dblp-Acm}&0&0&0&0&0&0&0&{\vspace{-2mm}\color{orange}{\bf2}}&0&0&0\\\cline{2-13}
&{\sc Dblp-Scholar}&0&0&0&0&0&1&{\vspace{-2mm}\color{orange}{\bf2.4}}&{\vspace{-2mm}\color{gray}{2}}&{\vspace{-2mm}\color{gray}{0}}&{\vspace{-2mm}\color{gray}{2.2}}&{\vspace{-2mm}\color{orange}{\bf2.4}}\\\hline
\multirow{4}{*}{\rotatebox{90}{\bf PPVP}}&{\sc iTunes-Amazon}&0&0&0&0&{\vspace{-2mm}\color{orange}{\bf2}}&0&1.7&{\vspace{-2mm}\color{red}{\bf5.2}}&{\vspace{-2mm}\color{gray}{0}}&{\vspace{-2mm}\color{orange}{\bf2}}&1.4\\\cline{2-13} 
&{\sc Cameras}&{\vspace{-2mm}\color{gray}{1}}&{\vspace{-2mm}\color{gray}{0}}&{\vspace{-2mm}\color{gray}{5.8}}&{\vspace{-2mm}\color{gray}{4.5}}&{\vspace{-2mm}\color{gray}{4.6}}&{\vspace{-2mm}\color{gray}{3.7}}&{\vspace{-2mm}\color{red}{\bf3.4}}&{\vspace{-2mm}\color{orange}{\bf2.4}}&1.7&{\vspace{-2mm}\color{red}{\bf4.6}}&{\vspace{-2mm}\color{red}{\bf3.6}}\\\cline{2-13}
&{\sc Dblp-Acm}&0&0&0&0&0&{\vspace{-2mm}\color{orange}{\bf2.6}}&0&0&0&0&0\\\cline{2-13}
&{\sc Dblp-Scholar}&0&0&1&1&1&1.4&1&{\vspace{-2mm}\color{gray}{1.4}}&{\vspace{-2mm}\color{gray}{0}}&{\vspace{-2mm}\color{gray}{2.4}}&1\\
\end{tabular}
\submit{\vspace{-7mm}}
\end{table}

\section{Lessons and Discussion}\label{sec:discussion}
Some of the lessons learned in this study include:

\vspace{1mm}\noindent
(i) \rev{\marginpar{R2.W2 R3.W3 M4}
\em Call for action to collect EM benchmarks on societal applications:} Perhaps the most challenging burden when auditing EM techniques from the fairness perspective is {\em lack of proper benchmark} datasets.
\techrep{
Fairness is a societal issue and is meaningful when the EM task is on {\em individual records}.
In real-world, EM is frequently used for tasks such as terrorist watch list screening~\cite{krouse2009terrorist}, record linkage for data integration on medical data~\cite{acheson1967medical,nargesian2022responsible}, individual records deduplication~\cite{scanlon2016battling}, and many more.
On the other hand, due to reasons such as privacy, such data are not publicly available.}
Although the EM community already has some benchmarks~\cite{Wang0H21,konda2018magellan}, a thorough audit of existing and future EM techniques requires benchmark entity-matching data for societal applications. 
In this paper, we took the first steps\techrep{ towards addressing this need} by creating and publishing two semi-synthetic social datasets using publicly available real datasets. \nofly and \fmatch are generated for auditing the fairness of EM techniques when some groups are over-represented in data and when two demographic groups have different degrees of similarity in their names.

\vspace{1mm}\noindent
(ii) {\em Over-representation and name similarity in social data:}
Group over-representation and higher similarity degrees for specific groups are common in social data.
Experiment results on our social datasets confirm the general unfairness of entity matchers under these conditions.
Interestingly, under the group over-representation, we observed the superiority of non-neural matchers in terms of model performance and accuracy.
Over-representation in general can increase the chance of finding similar non-matches for an entity, which can be falsely labeled as a match.
Likewise, when names in one group are more similar, there is a higher chance of mistakenly labeling non-matching tuples from that group as a match. 
\rev{Considering more (unbiased) attributes can help in such situations.}

\noindent
(iii) \rev{\em Unbiased and representative training data:}
Responsible training of EM techniques requires access to unbiased data with proper coverage of different groups and possible cases.
Insufficient coverage of different groups can bias the models in favor of some of the groups, making the model unfair.
\rev{
In particular, given the class imbalance nature of EM tasks, it is important to ensure enough representatives from different groups in both (match/unmatch) classes.
}

\vspace{1mm}\noindent
(iv) {\em Proper fairness measures for EM:}
Different fairness definitions are valuable for different settings.
Still, due to its pairwise matching nature, class imbalance, with most of the records being non-match, is a distinguishing property of EM.
In this setting, {\em positive predictive value parity} and {\em true positive rate parity} a.k.a. equal opportunity is more capable of revealing the matchers' unfairness.
Finally, some of the unfairness of a matcher, such as AP, could be explained using other measures, such as TPRP.

\vspace{1mm}\noindent
(v) {\em Proper matching techniques for different settings:}
Different matching techniques perform differently for different dataset types. At a high level, non-neural matchers fail for textual datasets while performing well for structured data. Lack of proper coverage in training data can bias these models to significantly rely on attributes (such as name) that are highly correlated with the ground-truth label but may bias their performance for the minority groups.
Neural matchers, on the other hand, generally perform well for different dataset types. Still, (a) using pre-trained language models and embeddings, (b) relying less on the structure of data caused these matchers to be unfair for different settings.
\rev{
The fairness of neural matchers is more sensitive to the choice of the similarity threshold, as small changes in the threshold value can significantly change their fairness.
Therefore, it is important to identify the right threshold to find the most fair/accurate one.
}

\vspace{1mm}\noindent
(vi) {\em Ensemble learning for fair EM:}
We observed that, in a fixed dataset, 
some groups needed matchers with more complex decision boundaries, while others required matchers with simpler decision boundaries. As a result, adapting either of the neural/non-neural matchers would show unfairness for some groups. 
This observation underscores the need for techniques such as {\em ensemble learning} to consider a range of matchers with different properties to assure similar performance across different groups.
\rev{\marginpar{R1.D10} 
Specifically, for cases with a single sensitive attribute with exclusive values, we recommend to train {\em a set} of matchers, for each group identify the one that performs the best, and use it for that group.
This helps to maximize the performance for the under-performing groups.
Note that this is different from the ensemble-learning-based approaches, since their goal is to improve the overall EM accuracy~\cite{liu2022ensemble}, not reducing unfairness. For instance, Jurek et al. apply ensemble learning based on feature selection~\cite{jurek2017novel}, while Yu et al.  create an ensemble of models based on different similarity metric schemes~\cite{yi2017method}.
We leave designing ensemble EM techniques for fairness as future work.
} 

\submit{\vspace{-1mm}}
\stitle{\rev{Rule of Thumb}} 
\rev{ \marginpar{R2.D2 R2.W2} \marginpar{R3.W3} 
Following our findings and the lessons learned in this study, we would like to conclude the paper by Table~\ref{tab:Rules}, which provides some actionable insights and rules of thumb for responsible entity matching.
}

\begin{table}
    \centering \footnotesize
    \color{black}
    \caption{\rev{Rules of Thumb for Responsible Entity Matching.}}
    \label{tab:Rules}
    \submit{\vspace{-4mm}}
    \begin{tabular}{|@{}c|l@{}|} 
    \hline
    &{\bf Rules of Thumb} \\ \hline
    \multirow{5}{*}{\spheading{\bf Structured datasets}}& -- Non-neural matchers are preferred\\
    & -- Obtain attributes with min correlation with sensitive attributes\\
    & -- Minimize Representation bias in training data\\
    & -- Make sure the model is not putting high weights on only a few attributes\\
    &\\
    \hline
    \multirow{5}{*}{\spheadingp{\bf Textual\& dirty datasets}}& -- Neural matchers are preferred\\
    & -- Obtain additional (unbiased) features\\
    & -- Use unbiased pretrained models\\
    & -- Minimize Representation bias in training data\\
    & -- Considering their sensitivity, try out different matching thresholds\\
    & \hspace{2mm}and select the most fair/accurate one\\
    \hline
    \multicolumn{2}{|l|}{
    {\bf Fairness measure:} TPRP and PPVP are usually preferred (see \S~\ref{sec:measuresdiscussion} and \S~\ref{sec:evalmeasures})
    }\\
    \hline
    \multicolumn{2}{|l@{}|}{
    {\bf Use an ensemble of matchers}
    (for single sensitive attributes with exclusive}\\
    \multicolumn{2}{|l@{}|}{
    values): construct a set of matchers; for each group use the matcher with best}\\
    \multicolumn{2}{|l|}{performance on it (using separate test sets for each group)}\\
    \hline
\end{tabular}
\submit{\vspace{-7mm}}
\end{table}

\section*{Acknowledgement}
The authors would like to thank 
Dongxiang Zhang and Zepeng Li for providing  {\sc Mcan} matcher's code.
The work by Nima Shahbazi and Abolfazl Asudeh was supported in part by the National Science Foundation (Grant No. 2107290). Fatemeh Nargesian was supported in part by the National Science Foundation (Grant No. 2107050). 
\submit{\newpage}
\balance
\bibliographystyle{ACM-Reference-Format}
\bibliography{ref}


\begin{thebibliography}{68}


\ifx \showCODEN    \undefined \def \showCODEN     #1{\unskip}     \fi
\ifx \showDOI      \undefined \def \showDOI       #1{#1}\fi
\ifx \showISBNx    \undefined \def \showISBNx     #1{\unskip}     \fi
\ifx \showISBNxiii \undefined \def \showISBNxiii  #1{\unskip}     \fi
\ifx \showISSN     \undefined \def \showISSN      #1{\unskip}     \fi
\ifx \showLCCN     \undefined \def \showLCCN      #1{\unskip}     \fi
\ifx \shownote     \undefined \def \shownote      #1{#1}          \fi
\ifx \showarticletitle \undefined \def \showarticletitle #1{#1}   \fi
\ifx \showURL      \undefined \def \showURL       {\relax}        \fi
\providecommand\bibfield[2]{#2}
\providecommand\bibinfo[2]{#2}
\providecommand\natexlab[1]{#1}
\providecommand\showeprint[2][]{arXiv:#2}

\bibitem[\protect\citeauthoryear{??}{usC}{[n.d.]}]%
        {usCensus}
 \bibinfo{year}{[n.d.]}\natexlab{}.
\newblock \bibinfo{title}{u.s. census bureau quickfacts: united states}.
\newblock
\newblock
\urldef\tempurl%
\url{https://www.census.gov/quickfacts/fact/table/US/PST045221}
\showURL{%
\tempurl}


\bibitem[\protect\citeauthoryear{??}{COM}{2015}]%
        {COMPAS}
 \bibinfo{year}{2015}\natexlab{}.
\newblock \bibinfo{title}{COMPAS Recidivism Risk Score Data and Analysis}.
\newblock
  \bibinfo{howpublished}{\url{www.propublica.org/datastore/dataset/compas-recidivism-risk-score-data-and-analysis}}.
\newblock


\bibitem[\protect\citeauthoryear{??}{CSR}{2023}]%
        {CSRankings}
 \bibinfo{year}{[visited: 2023]}\natexlab{}.
\newblock \bibinfo{title}{CSRankings GitHub Repository}.
\newblock
  \bibinfo{howpublished}{\url{https://github.com/emeryberger/CSRankings}}.
\newblock


\bibitem[\protect\citeauthoryear{Acheson et~al\mbox{.}}{Acheson
  et~al\mbox{.}}{1967}]%
        {acheson1967medical}
\bibfield{author}{\bibinfo{person}{Ernest~Donald Acheson} {et~al\mbox{.}}}
  \bibinfo{year}{1967}\natexlab{}.
\newblock \showarticletitle{Medical record linkage.}
\newblock \bibinfo{journal}{\emph{Medical record linkage.}}
  (\bibinfo{year}{1967}).
\newblock


\bibitem[\protect\citeauthoryear{Advertising}{Advertising}{2022}]%
        {AdBiasIBM}
\bibfield{author}{\bibinfo{person}{IBM~Watson Advertising}.}
  \bibinfo{year}{2022}\natexlab{}.
\newblock \bibinfo{title}{Bias in Advertising: Confronting \& Addressing the
  Challenge}.
\newblock
  \bibinfo{howpublished}{\url{https://www.ibm.com/watson-advertising/thought-leadership/bias-in-advertising}}.
\newblock


\bibitem[\protect\citeauthoryear{Asudeh, Jagadish, Stoyanovich, and Das}{Asudeh
  et~al\mbox{.}}{2019a}]%
        {asudeh2019designing}
\bibfield{author}{\bibinfo{person}{Abolfazl Asudeh}, \bibinfo{person}{HV
  Jagadish}, \bibinfo{person}{Julia Stoyanovich}, {and} \bibinfo{person}{Gautam
  Das}.} \bibinfo{year}{2019}\natexlab{a}.
\newblock \showarticletitle{Designing fair ranking schemes}. In
  \bibinfo{booktitle}{\emph{Proceedings of the 2019 international conference on
  management of data}}. \bibinfo{pages}{1259--1276}.
\newblock


\bibitem[\protect\citeauthoryear{Asudeh and Jagadish}{Asudeh and
  Jagadish}{2020}]%
        {asudeh2020fairly}
\bibfield{author}{\bibinfo{person}{Abolfazl Asudeh} {and}
  \bibinfo{person}{H.~V. Jagadish}.} \bibinfo{year}{2020}\natexlab{}.
\newblock \showarticletitle{Fairly evaluating and scoring items in a data set}.
\newblock \bibinfo{journal}{\emph{PVLDB}} \bibinfo{volume}{13},
  \bibinfo{number}{12} (\bibinfo{year}{2020}), \bibinfo{pages}{3445--3448}.
\newblock


\bibitem[\protect\citeauthoryear{Asudeh, Jin, and Jagadish}{Asudeh
  et~al\mbox{.}}{2019b}]%
        {asudeh2019assessing}
\bibfield{author}{\bibinfo{person}{Abolfazl Asudeh}, \bibinfo{person}{Zhongjun
  Jin}, {and} \bibinfo{person}{HV Jagadish}.} \bibinfo{year}{2019}\natexlab{b}.
\newblock \showarticletitle{Assessing and remedying coverage for a given
  dataset}. In \bibinfo{booktitle}{\emph{ICDE}}. IEEE,
  \bibinfo{pages}{554--565}.
\newblock


\bibitem[\protect\citeauthoryear{Asudeh, Shahbazi, Jin, and Jagadish}{Asudeh
  et~al\mbox{.}}{2021}]%
        {asudeh2021identifying}
\bibfield{author}{\bibinfo{person}{Abolfazl Asudeh}, \bibinfo{person}{Nima
  Shahbazi}, \bibinfo{person}{Zhongjun Jin}, {and} \bibinfo{person}{HV
  Jagadish}.} \bibinfo{year}{2021}\natexlab{}.
\newblock \showarticletitle{Identifying insufficient data coverage for ordinal
  continuous-valued attributes}. In \bibinfo{booktitle}{\emph{Proceedings of
  the 2021 international conference on management of data}}.
  \bibinfo{pages}{129--141}.
\newblock


\bibitem[\protect\citeauthoryear{Bach and Bernat}{Bach and Bernat}{2022}]%
        {AdBias1}
\bibfield{author}{\bibinfo{person}{Tho Bach} {and} \bibinfo{person}{Kenny
  Bernat}.} \bibinfo{year}{2022}\natexlab{}.
\newblock \bibinfo{title}{The Business Impact of Biased Advertising (and How to
  Fix It)}.
\newblock
  \bibinfo{howpublished}{\url{https://www.wpromote.com/blog/digital-marketing/biased-advertising}}.
\newblock


\bibitem[\protect\citeauthoryear{Barlaug}{Barlaug}{2022}]%
        {barlaug2022lemon}
\bibfield{author}{\bibinfo{person}{Nils Barlaug}.}
  \bibinfo{year}{2022}\natexlab{}.
\newblock \showarticletitle{LEMON: explainable entity matching}.
\newblock \bibinfo{journal}{\emph{IEEE Transactions on Knowledge and Data
  Engineering}} (\bibinfo{year}{2022}).
\newblock


\bibitem[\protect\citeauthoryear{Barlaug and Gulla}{Barlaug and Gulla}{2021}]%
        {barlaug2021neural}
\bibfield{author}{\bibinfo{person}{Nils Barlaug} {and}
  \bibinfo{person}{Jon~Atle Gulla}.} \bibinfo{year}{2021}\natexlab{}.
\newblock \showarticletitle{Neural networks for entity matching: A survey}.
\newblock \bibinfo{journal}{\emph{ACM Transactions on Knowledge Discovery from
  Data (TKDD)}} \bibinfo{volume}{15}, \bibinfo{number}{3}
  (\bibinfo{year}{2021}), \bibinfo{pages}{1--37}.
\newblock


\bibitem[\protect\citeauthoryear{Barocas, Hardt, and Narayanan}{Barocas
  et~al\mbox{.}}{2019}]%
        {fairmlbook}
\bibfield{author}{\bibinfo{person}{Solon Barocas}, \bibinfo{person}{Moritz
  Hardt}, {and} \bibinfo{person}{Arvind Narayanan}.}
  \bibinfo{year}{2019}\natexlab{}.
\newblock \bibinfo{title}{Fairness and machine learning: Limitations and
  opportunities}.
\newblock \bibinfo{howpublished}{\url{fairmlbook.org}}.
\newblock


\bibitem[\protect\citeauthoryear{Bellamy, Dey, Hind, Hoffman, Houde, Kannan,
  Lohia, Martino, Mehta, Mojsilovic, et~al\mbox{.}}{Bellamy
  et~al\mbox{.}}{2018}]%
        {bellamy2018ai}
\bibfield{author}{\bibinfo{person}{Rachel~KE Bellamy}, \bibinfo{person}{Kuntal
  Dey}, \bibinfo{person}{Michael Hind}, \bibinfo{person}{Samuel~C Hoffman},
  \bibinfo{person}{Stephanie Houde}, \bibinfo{person}{Kalapriya Kannan},
  \bibinfo{person}{Pranay Lohia}, \bibinfo{person}{Jacquelyn Martino},
  \bibinfo{person}{Sameep Mehta}, \bibinfo{person}{Aleksandra Mojsilovic},
  {et~al\mbox{.}}} \bibinfo{year}{2018}\natexlab{}.
\newblock \showarticletitle{AI Fairness 360: An extensible toolkit for
  detecting, understanding, and mitigating unwanted algorithmic bias}.
\newblock \bibinfo{journal}{\emph{arXiv preprint arXiv:1810.01943}}
  (\bibinfo{year}{2018}).
\newblock


\bibitem[\protect\citeauthoryear{Bojanowski, Grave, Joulin, and
  Mikolov}{Bojanowski et~al\mbox{.}}{2016}]%
        {bojanowski2016enriching}
\bibfield{author}{\bibinfo{person}{Piotr Bojanowski}, \bibinfo{person}{Edouard
  Grave}, \bibinfo{person}{Armand Joulin}, {and} \bibinfo{person}{Tomas
  Mikolov}.} \bibinfo{year}{2016}\natexlab{}.
\newblock \showarticletitle{Enriching Word Vectors with Subword Information}.
\newblock \bibinfo{journal}{\emph{arXiv preprint arXiv:1607.04606}}
  (\bibinfo{year}{2016}).
\newblock


\bibitem[\protect\citeauthoryear{Calmon, Wei, Vinzamuri, Natesan~Ramamurthy,
  and Varshney}{Calmon et~al\mbox{.}}{2017}]%
        {calmon2017optimized}
\bibfield{author}{\bibinfo{person}{Flavio Calmon}, \bibinfo{person}{Dennis
  Wei}, \bibinfo{person}{Bhanukiran Vinzamuri}, \bibinfo{person}{Karthikeyan
  Natesan~Ramamurthy}, {and} \bibinfo{person}{Kush~R Varshney}.}
  \bibinfo{year}{2017}\natexlab{}.
\newblock \showarticletitle{Optimized pre-processing for discrimination
  prevention}.
\newblock \bibinfo{journal}{\emph{Advances in neural information processing
  systems}}  \bibinfo{volume}{30} (\bibinfo{year}{2017}).
\newblock


\bibitem[\protect\citeauthoryear{Celis, Huang, Keswani, and Vishnoi}{Celis
  et~al\mbox{.}}{2019}]%
        {celis2019classification}
\bibfield{author}{\bibinfo{person}{L~Elisa Celis}, \bibinfo{person}{Lingxiao
  Huang}, \bibinfo{person}{Vijay Keswani}, {and} \bibinfo{person}{Nisheeth~K
  Vishnoi}.} \bibinfo{year}{2019}\natexlab{}.
\newblock \showarticletitle{Classification with fairness constraints: A
  meta-algorithm with provable guarantees}. In
  \bibinfo{booktitle}{\emph{Proceedings of the conference on fairness,
  accountability, and transparency}}. \bibinfo{pages}{319--328}.
\newblock


\bibitem[\protect\citeauthoryear{Chen, Shen, and Zhang}{Chen
  et~al\mbox{.}}{2021}]%
        {chen2021gnem}
\bibfield{author}{\bibinfo{person}{Runjin Chen}, \bibinfo{person}{Yanyan Shen},
  {and} \bibinfo{person}{Dongxiang Zhang}.} \bibinfo{year}{2021}\natexlab{}.
\newblock \showarticletitle{GNEM: a generic one-to-set neural entity matching
  framework}. In \bibinfo{booktitle}{\emph{Proceedings of the Web Conference
  2021}}. \bibinfo{pages}{1686--1694}.
\newblock


\bibitem[\protect\citeauthoryear{Chouldechova}{Chouldechova}{2017}]%
        {chouldechova2017fair}
\bibfield{author}{\bibinfo{person}{Alexandra Chouldechova}.}
  \bibinfo{year}{2017}\natexlab{}.
\newblock \showarticletitle{Fair prediction with disparate impact: A study of
  bias in recidivism prediction instruments}.
\newblock \bibinfo{journal}{\emph{Big data}} \bibinfo{volume}{5},
  \bibinfo{number}{2} (\bibinfo{year}{2017}), \bibinfo{pages}{153--163}.
\newblock


\bibitem[\protect\citeauthoryear{Christophides, Efthymiou, Palpanas, Papadakis,
  and Stefanidis}{Christophides et~al\mbox{.}}{2020}]%
        {christophides2020overview}
\bibfield{author}{\bibinfo{person}{Vassilis Christophides},
  \bibinfo{person}{Vasilis Efthymiou}, \bibinfo{person}{Themis Palpanas},
  \bibinfo{person}{George Papadakis}, {and} \bibinfo{person}{Kostas
  Stefanidis}.} \bibinfo{year}{2020}\natexlab{}.
\newblock \showarticletitle{An overview of end-to-end entity resolution for big
  data}.
\newblock \bibinfo{journal}{\emph{ACM Computing Surveys (CSUR)}}
  \bibinfo{volume}{53}, \bibinfo{number}{6} (\bibinfo{year}{2020}),
  \bibinfo{pages}{1--42}.
\newblock


\bibitem[\protect\citeauthoryear{Commission}{Commission}{1979}]%
        {EEOC}
\bibfield{author}{\bibinfo{person}{Equal Employment~Opportunity Commission}.}
  \bibinfo{year}{1979}\natexlab{}.
\newblock \bibinfo{title}{The U.S. Uniform guidelines on employee selection
  procedures}.
\newblock
\newblock


\bibitem[\protect\citeauthoryear{Devlin, Chang, Lee, and Toutanova}{Devlin
  et~al\mbox{.}}{2018}]%
        {devlin2018bert}
\bibfield{author}{\bibinfo{person}{Jacob Devlin}, \bibinfo{person}{Ming-Wei
  Chang}, \bibinfo{person}{Kenton Lee}, {and} \bibinfo{person}{Kristina
  Toutanova}.} \bibinfo{year}{2018}\natexlab{}.
\newblock \showarticletitle{Bert: Pre-training of deep bidirectional
  transformers for language understanding}.
\newblock \bibinfo{journal}{\emph{arXiv preprint arXiv:1810.04805}}
  (\bibinfo{year}{2018}).
\newblock


\bibitem[\protect\citeauthoryear{Di~Cicco, Firmani, Koudas, Merialdo, and
  Srivastava}{Di~Cicco et~al\mbox{.}}{2019}]%
        {di2019interpreting}
\bibfield{author}{\bibinfo{person}{Vincenzo Di~Cicco},
  \bibinfo{person}{Donatella Firmani}, \bibinfo{person}{Nick Koudas},
  \bibinfo{person}{Paolo Merialdo}, {and} \bibinfo{person}{Divesh Srivastava}.}
  \bibinfo{year}{2019}\natexlab{}.
\newblock \showarticletitle{Interpreting deep learning models for entity
  resolution: an experience report using LIME}. In
  \bibinfo{booktitle}{\emph{Proceedings of the Second International Workshop on
  Exploiting Artificial Intelligence Techniques for Data Management}}.
  \bibinfo{pages}{1--4}.
\newblock


\bibitem[\protect\citeauthoryear{Dwork, Hardt, Pitassi, Reingold, and
  Zemel}{Dwork et~al\mbox{.}}{2012}]%
        {dwork2012fairness}
\bibfield{author}{\bibinfo{person}{Cynthia Dwork}, \bibinfo{person}{Moritz
  Hardt}, \bibinfo{person}{Toniann Pitassi}, \bibinfo{person}{Omer Reingold},
  {and} \bibinfo{person}{Richard Zemel}.} \bibinfo{year}{2012}\natexlab{}.
\newblock \showarticletitle{Fairness through awareness}. In
  \bibinfo{booktitle}{\emph{Proceedings of the 3rd innovations in theoretical
  computer science conference}}. \bibinfo{pages}{214--226}.
\newblock


\bibitem[\protect\citeauthoryear{Efthymiou, Stefanidis, Pitoura, and
  Christophides}{Efthymiou et~al\mbox{.}}{2021}]%
        {efthymiou2021fairer}
\bibfield{author}{\bibinfo{person}{Vasilis Efthymiou}, \bibinfo{person}{Kostas
  Stefanidis}, \bibinfo{person}{Evaggelia Pitoura}, {and}
  \bibinfo{person}{Vassilis Christophides}.} \bibinfo{year}{2021}\natexlab{}.
\newblock \showarticletitle{FairER: Entity Resolution With Fairness
  Constraints}. In \bibinfo{booktitle}{\emph{Proceedings of the 30th ACM
  International Conference on Information \& Knowledge Management}}.
  \bibinfo{pages}{3004--3008}.
\newblock


\bibitem[\protect\citeauthoryear{Feldman, Friedler, Moeller, Scheidegger, and
  Venkatasubramanian}{Feldman et~al\mbox{.}}{2015}]%
        {feldman2015certifying}
\bibfield{author}{\bibinfo{person}{Michael Feldman}, \bibinfo{person}{Sorelle~A
  Friedler}, \bibinfo{person}{John Moeller}, \bibinfo{person}{Carlos
  Scheidegger}, {and} \bibinfo{person}{Suresh Venkatasubramanian}.}
  \bibinfo{year}{2015}\natexlab{}.
\newblock \showarticletitle{Certifying and removing disparate impact}. In
  \bibinfo{booktitle}{\emph{proceedings of the 21th ACM SIGKDD international
  conference on knowledge discovery and data mining}}.
  \bibinfo{pages}{259--268}.
\newblock


\bibitem[\protect\citeauthoryear{Fu, Han, He, and Sun}{Fu
  et~al\mbox{.}}{2021}]%
        {fu2021hierarchical}
\bibfield{author}{\bibinfo{person}{Cheng Fu}, \bibinfo{person}{Xianpei Han},
  \bibinfo{person}{Jiaming He}, {and} \bibinfo{person}{Le Sun}.}
  \bibinfo{year}{2021}\natexlab{}.
\newblock \showarticletitle{Hierarchical matching network for heterogeneous
  entity resolution}. In \bibinfo{booktitle}{\emph{Proceedings of the
  Twenty-Ninth International Conference on International Joint Conferences on
  Artificial Intelligence}}. \bibinfo{pages}{3665--3671}.
\newblock


\bibitem[\protect\citeauthoryear{Gregg and Eder}{Gregg and Eder}{2022}]%
        {forest2022dedupe}
\bibfield{author}{\bibinfo{person}{Forest Gregg} {and} \bibinfo{person}{Derek
  Eder}.} \bibinfo{year}{2022}\natexlab{}.
\newblock \bibinfo{title}{Dedupe}.
\newblock \bibinfo{howpublished}{\url{https://github.com/dedupeio/dedupe}}.
\newblock


\bibitem[\protect\citeauthoryear{Hardt, Price, and Srebro}{Hardt
  et~al\mbox{.}}{2016}]%
        {hardt2016equality}
\bibfield{author}{\bibinfo{person}{Moritz Hardt}, \bibinfo{person}{Eric Price},
  {and} \bibinfo{person}{Nati Srebro}.} \bibinfo{year}{2016}\natexlab{}.
\newblock \showarticletitle{Equality of opportunity in supervised learning}.
\newblock \bibinfo{journal}{\emph{Advances in neural information processing
  systems}}  \bibinfo{volume}{29} (\bibinfo{year}{2016}).
\newblock


\bibitem[\protect\citeauthoryear{InDeXLab}{InDeXLab}{2023}]%
        {synthetic_data}
\bibfield{author}{\bibinfo{person}{InDeXLab}.} \bibinfo{year}{2023}\natexlab{}.
\newblock \bibinfo{title}{Fair Entity Matching}.
\newblock
  \bibinfo{howpublished}{\url{github.com/UIC-InDeXLab/fair_entity_matching/tree/main/synthetic\%20dataset\%20generator}}.
\newblock


\bibitem[\protect\citeauthoryear{Jurek, Hong, Chi, and Liu}{Jurek
  et~al\mbox{.}}{2017}]%
        {jurek2017novel}
\bibfield{author}{\bibinfo{person}{Anna Jurek}, \bibinfo{person}{Jun Hong},
  \bibinfo{person}{Yuan Chi}, {and} \bibinfo{person}{Weiru Liu}.}
  \bibinfo{year}{2017}\natexlab{}.
\newblock \showarticletitle{A novel ensemble learning approach to unsupervised
  record linkage}.
\newblock \bibinfo{journal}{\emph{Information Systems}}  \bibinfo{volume}{71}
  (\bibinfo{year}{2017}), \bibinfo{pages}{40--54}.
\newblock


\bibitem[\protect\citeauthoryear{Kleinberg, Mullainathan, and
  Raghavan}{Kleinberg et~al\mbox{.}}{2016}]%
        {kleinberg2016inherent}
\bibfield{author}{\bibinfo{person}{Jon Kleinberg}, \bibinfo{person}{Sendhil
  Mullainathan}, {and} \bibinfo{person}{Manish Raghavan}.}
  \bibinfo{year}{2016}\natexlab{}.
\newblock \showarticletitle{Inherent trade-offs in the fair determination of
  risk scores}.
\newblock \bibinfo{journal}{\emph{arXiv preprint arXiv:1609.05807}}
  (\bibinfo{year}{2016}).
\newblock


\bibitem[\protect\citeauthoryear{Konda}{Konda}{2018}]%
        {konda2018magellan}
\bibfield{author}{\bibinfo{person}{Pradap~Venkatramanan Konda}.}
  \bibinfo{year}{2018}\natexlab{}.
\newblock \bibinfo{booktitle}{\emph{Magellan: Toward building entity matching
  management systems}}.
\newblock \bibinfo{publisher}{The University of Wisconsin-Madison}.
\newblock


\bibitem[\protect\citeauthoryear{K{\"o}pcke and Rahm}{K{\"o}pcke and
  Rahm}{2010}]%
        {kopcke2010frameworks}
\bibfield{author}{\bibinfo{person}{Hanna K{\"o}pcke} {and}
  \bibinfo{person}{Erhard Rahm}.} \bibinfo{year}{2010}\natexlab{}.
\newblock \showarticletitle{Frameworks for entity matching: A comparison}.
\newblock \bibinfo{journal}{\emph{Data \& Knowledge Engineering}}
  \bibinfo{volume}{69}, \bibinfo{number}{2} (\bibinfo{year}{2010}),
  \bibinfo{pages}{197--210}.
\newblock


\bibitem[\protect\citeauthoryear{Krouse and Elias}{Krouse and Elias}{2009}]%
        {krouse2009terrorist}
\bibfield{author}{\bibinfo{person}{William~J Krouse} {and}
  \bibinfo{person}{Bart Elias}.} \bibinfo{year}{2009}\natexlab{}.
\newblock \showarticletitle{Terrorist watchlist checks and air passenger
  prescreening}. LIBRARY OF CONGRESS WASHINGTON DC CONGRESSIONAL RESEARCH
  SERVICE.
\newblock


\bibitem[\protect\citeauthoryear{Kusner, Loftus, Russell, and Silva}{Kusner
  et~al\mbox{.}}{2017}]%
        {kusner2017counterfactual}
\bibfield{author}{\bibinfo{person}{Matt~J Kusner}, \bibinfo{person}{Joshua
  Loftus}, \bibinfo{person}{Chris Russell}, {and} \bibinfo{person}{Ricardo
  Silva}.} \bibinfo{year}{2017}\natexlab{}.
\newblock \showarticletitle{Counterfactual fairness}.
\newblock \bibinfo{journal}{\emph{Advances in neural information processing
  systems}}  \bibinfo{volume}{30} (\bibinfo{year}{2017}).
\newblock


\bibitem[\protect\citeauthoryear{Li, Liu, Zhang, Wang, and Wan}{Li
  et~al\mbox{.}}{2020b}]%
        {li2020survey}
\bibfield{author}{\bibinfo{person}{Bo-Han Li}, \bibinfo{person}{Yi Liu},
  \bibinfo{person}{An-Man Zhang}, \bibinfo{person}{Wen-Huan Wang}, {and}
  \bibinfo{person}{Shuo Wan}.} \bibinfo{year}{2020}\natexlab{b}.
\newblock \showarticletitle{A survey on blocking technology of entity
  resolution}.
\newblock \bibinfo{journal}{\emph{Journal of Computer Science and Technology}}
  \bibinfo{volume}{35} (\bibinfo{year}{2020}), \bibinfo{pages}{769--793}.
\newblock


\bibitem[\protect\citeauthoryear{Li, Li, Suhara, Doan, and Tan}{Li
  et~al\mbox{.}}{2020a}]%
        {li2020deep}
\bibfield{author}{\bibinfo{person}{Yuliang Li}, \bibinfo{person}{Jinfeng Li},
  \bibinfo{person}{Yoshihiko Suhara}, \bibinfo{person}{AnHai Doan}, {and}
  \bibinfo{person}{Wang-Chiew Tan}.} \bibinfo{year}{2020}\natexlab{a}.
\newblock \showarticletitle{Deep entity matching with pre-trained language
  models}.
\newblock \bibinfo{journal}{\emph{arXiv preprint arXiv:2004.00584}}
  (\bibinfo{year}{2020}).
\newblock


\bibitem[\protect\citeauthoryear{Liu}{Liu}{2022}]%
        {liu2022ensemble}
\bibfield{author}{\bibinfo{person}{Ling Liu}.} \bibinfo{year}{2022}\natexlab{}.
\newblock \showarticletitle{Ensemble Learning Methods for Dirty Data}. In
  \bibinfo{booktitle}{\emph{CIKM, Keynote}}.
\newblock


\bibitem[\protect\citeauthoryear{Makri, Karakasidis, and Pitoura}{Makri
  et~al\mbox{.}}{2022}]%
        {makri2022towards}
\bibfield{author}{\bibinfo{person}{Christina Makri},
  \bibinfo{person}{Alexandros Karakasidis}, {and} \bibinfo{person}{Evaggelia
  Pitoura}.} \bibinfo{year}{2022}\natexlab{}.
\newblock \showarticletitle{Towards a more Accurate and Fair SVM-based Record
  Linkage}. In \bibinfo{booktitle}{\emph{2022 IEEE International Conference on
  Big Data (Big Data)}}. IEEE, \bibinfo{pages}{4691--4699}.
\newblock


\bibitem[\protect\citeauthoryear{Mikolov, Chen, Corrado, and Dean}{Mikolov
  et~al\mbox{.}}{2013}]%
        {mikolov2013efficient}
\bibfield{author}{\bibinfo{person}{Tomas Mikolov}, \bibinfo{person}{Kai Chen},
  \bibinfo{person}{Greg Corrado}, {and} \bibinfo{person}{Jeffrey Dean}.}
  \bibinfo{year}{2013}\natexlab{}.
\newblock \showarticletitle{Efficient estimation of word representations in
  vector space}.
\newblock \bibinfo{journal}{\emph{arXiv preprint arXiv:1301.3781}}
  (\bibinfo{year}{2013}).
\newblock


\bibitem[\protect\citeauthoryear{Miller and Hosanagar}{Miller and
  Hosanagar}{2019}]%
        {miller2019targeted}
\bibfield{author}{\bibinfo{person}{Alex~P Miller} {and} \bibinfo{person}{Kartik
  Hosanagar}.} \bibinfo{year}{2019}\natexlab{}.
\newblock \showarticletitle{How targeted ads and dynamic pricing can perpetuate
  bias}.
\newblock \bibinfo{journal}{\emph{Harvard Business Review}}
  (\bibinfo{year}{2019}).
\newblock


\bibitem[\protect\citeauthoryear{Mudgal, Li, Rekatsinas, Doan, Park, Krishnan,
  Deep, Arcaute, and Raghavendra}{Mudgal et~al\mbox{.}}{2018}]%
        {mudgal2018deep}
\bibfield{author}{\bibinfo{person}{Sidharth Mudgal}, \bibinfo{person}{Han Li},
  \bibinfo{person}{Theodoros Rekatsinas}, \bibinfo{person}{AnHai Doan},
  \bibinfo{person}{Youngchoon Park}, \bibinfo{person}{Ganesh Krishnan},
  \bibinfo{person}{Rohit Deep}, \bibinfo{person}{Esteban Arcaute}, {and}
  \bibinfo{person}{Vijay Raghavendra}.} \bibinfo{year}{2018}\natexlab{}.
\newblock \showarticletitle{Deep learning for entity matching: A design space
  exploration}. In \bibinfo{booktitle}{\emph{Proceedings of the 2018
  International Conference on Management of Data}}. \bibinfo{pages}{19--34}.
\newblock


\bibitem[\protect\citeauthoryear{Nargesian, Asudeh, and Jagadish}{Nargesian
  et~al\mbox{.}}{2021}]%
        {nargesian2021tailoring}
\bibfield{author}{\bibinfo{person}{Fatemeh Nargesian},
  \bibinfo{person}{Abolfazl Asudeh}, {and} \bibinfo{person}{HV Jagadish}.}
  \bibinfo{year}{2021}\natexlab{}.
\newblock \showarticletitle{Tailoring data source distributions for
  fairness-aware data integration}.
\newblock \bibinfo{journal}{\emph{PVLDB}} \bibinfo{volume}{14},
  \bibinfo{number}{11} (\bibinfo{year}{2021}), \bibinfo{pages}{2519--2532}.
\newblock


\bibitem[\protect\citeauthoryear{Nargesian, Asudeh, and Jagadish}{Nargesian
  et~al\mbox{.}}{2022}]%
        {nargesian2022responsible}
\bibfield{author}{\bibinfo{person}{Fatemeh Nargesian},
  \bibinfo{person}{Abolfazl Asudeh}, {and} \bibinfo{person}{HV Jagadish}.}
  \bibinfo{year}{2022}\natexlab{}.
\newblock \showarticletitle{Responsible Data Integration: Next-generation
  Challenges}. In \bibinfo{booktitle}{\emph{Proceedings of the 2022
  International Conference on Management of Data}}.
  \bibinfo{pages}{2458--2464}.
\newblock


\bibitem[\protect\citeauthoryear{Nilforoushan, Wu, and Milani}{Nilforoushan
  et~al\mbox{.}}{2022}]%
        {nilforoushan2022entity}
\bibfield{author}{\bibinfo{person}{Soudeh Nilforoushan},
  \bibinfo{person}{Qianfan Wu}, {and} \bibinfo{person}{Mostafa Milani}.}
  \bibinfo{year}{2022}\natexlab{}.
\newblock \showarticletitle{Entity Matching with AUC-Based Fairness}. In
  \bibinfo{booktitle}{\emph{2022 IEEE International Conference on Big Data (Big
  Data)}}. IEEE, \bibinfo{pages}{5068--5075}.
\newblock


\bibitem[\protect\citeauthoryear{Paganelli, Sottovia, Guerra, and
  Velegrakis}{Paganelli et~al\mbox{.}}{2019}]%
        {paganelli2019tuner}
\bibfield{author}{\bibinfo{person}{Matteo Paganelli}, \bibinfo{person}{Paolo
  Sottovia}, \bibinfo{person}{Francesco Guerra}, {and} \bibinfo{person}{Yannis
  Velegrakis}.} \bibinfo{year}{2019}\natexlab{}.
\newblock \showarticletitle{Tuner: Fine tuning of rule-based entity matchers}.
  In \bibinfo{booktitle}{\emph{Proceedings of the 28th ACM International
  Conference on Information and Knowledge Management}}.
  \bibinfo{pages}{2945--2948}.
\newblock


\bibitem[\protect\citeauthoryear{Panahi, Wu, Doan, and Naughton}{Panahi
  et~al\mbox{.}}{2017}]%
        {panahi2017towards}
\bibfield{author}{\bibinfo{person}{Fatemah Panahi}, \bibinfo{person}{Wentao
  Wu}, \bibinfo{person}{AnHai Doan}, {and} \bibinfo{person}{Jeffrey~F
  Naughton}.} \bibinfo{year}{2017}\natexlab{}.
\newblock \showarticletitle{Towards Interactive Debugging of Rule-based Entity
  Matching.}. In \bibinfo{booktitle}{\emph{EDBT}}. \bibinfo{pages}{354--365}.
\newblock


\bibitem[\protect\citeauthoryear{Papadakis, Skoutas, Thanos, and
  Palpanas}{Papadakis et~al\mbox{.}}{2020}]%
        {papadakis2020blocking}
\bibfield{author}{\bibinfo{person}{George Papadakis},
  \bibinfo{person}{Dimitrios Skoutas}, \bibinfo{person}{Emmanouil Thanos},
  {and} \bibinfo{person}{Themis Palpanas}.} \bibinfo{year}{2020}\natexlab{}.
\newblock \showarticletitle{Blocking and filtering techniques for entity
  resolution: A survey}.
\newblock \bibinfo{journal}{\emph{ACM Computing Surveys (CSUR)}}
  \bibinfo{volume}{53}, \bibinfo{number}{2} (\bibinfo{year}{2020}),
  \bibinfo{pages}{1--42}.
\newblock


\bibitem[\protect\citeauthoryear{Papadakis, Svirsky, Gal, and
  Palpanas}{Papadakis et~al\mbox{.}}{2016}]%
        {papadakis2016comparative}
\bibfield{author}{\bibinfo{person}{George Papadakis}, \bibinfo{person}{Jonathan
  Svirsky}, \bibinfo{person}{Avigdor Gal}, {and} \bibinfo{person}{Themis
  Palpanas}.} \bibinfo{year}{2016}\natexlab{}.
\newblock \showarticletitle{Comparative analysis of approximate blocking
  techniques for entity resolution}.
\newblock \bibinfo{journal}{\emph{Proceedings of the VLDB Endowment}}
  \bibinfo{volume}{9}, \bibinfo{number}{9} (\bibinfo{year}{2016}),
  \bibinfo{pages}{684--695}.
\newblock


\bibitem[\protect\citeauthoryear{Pennington, Socher, and Manning}{Pennington
  et~al\mbox{.}}{2014}]%
        {pennington2014glove}
\bibfield{author}{\bibinfo{person}{Jeffrey Pennington},
  \bibinfo{person}{Richard Socher}, {and} \bibinfo{person}{Christopher~D
  Manning}.} \bibinfo{year}{2014}\natexlab{}.
\newblock \showarticletitle{Glove: Global vectors for word representation}. In
  \bibinfo{booktitle}{\emph{Proceedings of the 2014 conference on empirical
  methods in natural language processing (EMNLP)}}.
  \bibinfo{pages}{1532--1543}.
\newblock


\bibitem[\protect\citeauthoryear{Primpeli, Peeters, and Bizer}{Primpeli
  et~al\mbox{.}}{2019}]%
        {primpeli2019wdc}
\bibfield{author}{\bibinfo{person}{Anna Primpeli}, \bibinfo{person}{Ralph
  Peeters}, {and} \bibinfo{person}{Christian Bizer}.}
  \bibinfo{year}{2019}\natexlab{}.
\newblock \showarticletitle{The WDC training dataset and gold standard for
  large-scale product matching}. In \bibinfo{booktitle}{\emph{Companion
  Proceedings of The 2019 World Wide Web Conference}}.
  \bibinfo{pages}{381--386}.
\newblock


\bibitem[\protect\citeauthoryear{Scanlon}{Scanlon}{2016}]%
        {scanlon2016battling}
\bibfield{author}{\bibinfo{person}{Mark Scanlon}.}
  \bibinfo{year}{2016}\natexlab{}.
\newblock \showarticletitle{Battling the digital forensic backlog through data
  deduplication}. In \bibinfo{booktitle}{\emph{2016 sixth international
  conference on innovative computing technology (INTECH)}}. IEEE,
  \bibinfo{pages}{10--14}.
\newblock


\bibitem[\protect\citeauthoryear{Shahbazi, Danevski, Nargesian, Asudeh, and
  Srivastava}{Shahbazi et~al\mbox{.}}{2023a}]%
        {techrep}
\bibfield{author}{\bibinfo{person}{Nima Shahbazi}, \bibinfo{person}{Nikola
  Danevski}, \bibinfo{person}{Fatemeh Nargesian}, \bibinfo{person}{Abolfazl
  Asudeh}, {and} \bibinfo{person}{Divesh Srivastava}.}
  \bibinfo{year}{2023}\natexlab{a}.
\newblock \bibinfo{title}{Through the Fairness Lens: Experimental Analysis and
  Evaluation of Entity Matching}.
\newblock
  \bibinfo{howpublished}{\url{https://github.com/UIC-InDeXLab/fair_entity_matching/blob/main/techrep.pdf}}.
\newblock


\bibitem[\protect\citeauthoryear{Shahbazi, Lin, Asudeh, and Jagadish}{Shahbazi
  et~al\mbox{.}}{2023b}]%
        {shahbazi2022survey}
\bibfield{author}{\bibinfo{person}{Nima Shahbazi}, \bibinfo{person}{Yin Lin},
  \bibinfo{person}{Abolfazl Asudeh}, {and} \bibinfo{person}{HV Jagadish}.}
  \bibinfo{year}{2023}\natexlab{b}.
\newblock \showarticletitle{Representation Bias in Data: A Survey on
  Identification and Resolution Techniques}.
\newblock \bibinfo{journal}{\emph{{ACM Computing Surveys}}}
  (\bibinfo{year}{2023}).
\newblock


\bibitem[\protect\citeauthoryear{Shetiya, Swift, Asudeh, and Das}{Shetiya
  et~al\mbox{.}}{2022}]%
        {shetiya2022fairness}
\bibfield{author}{\bibinfo{person}{Suraj Shetiya}, \bibinfo{person}{Ian~P.
  Swift}, \bibinfo{person}{Abolfazl Asudeh}, {and} \bibinfo{person}{Gautam
  Das}.} \bibinfo{year}{2022}\natexlab{}.
\newblock \showarticletitle{Fairness-Aware Range Queries for Selecting Unbiased
  Data}. In \bibinfo{booktitle}{\emph{2022 IEEE 38th International Conference
  on Data Engineering (ICDE)}}. IEEE.
\newblock


\bibitem[\protect\citeauthoryear{Singh, Meduri, Elmagarmid, Madden, Papotti,
  Quian{\'e}-Ruiz, Solar-Lezama, and Tang}{Singh et~al\mbox{.}}{2017}]%
        {singh2017synthesizing}
\bibfield{author}{\bibinfo{person}{Rohit Singh},
  \bibinfo{person}{Venkata~Vamsikrishna Meduri}, \bibinfo{person}{Ahmed
  Elmagarmid}, \bibinfo{person}{Samuel Madden}, \bibinfo{person}{Paolo
  Papotti}, \bibinfo{person}{Jorge-Arnulfo Quian{\'e}-Ruiz},
  \bibinfo{person}{Armando Solar-Lezama}, {and} \bibinfo{person}{Nan Tang}.}
  \bibinfo{year}{2017}\natexlab{}.
\newblock \showarticletitle{Synthesizing entity matching rules by examples}.
\newblock \bibinfo{journal}{\emph{Proceedings of the VLDB Endowment}}
  \bibinfo{volume}{11}, \bibinfo{number}{2} (\bibinfo{year}{2017}),
  \bibinfo{pages}{189--202}.
\newblock


\bibitem[\protect\citeauthoryear{Swift, Ebrahimi, Nova, and Asudeh}{Swift
  et~al\mbox{.}}{2022}]%
        {swift2022maximizing}
\bibfield{author}{\bibinfo{person}{Ian~P Swift}, \bibinfo{person}{Sana
  Ebrahimi}, \bibinfo{person}{Azade Nova}, {and} \bibinfo{person}{Abolfazl
  Asudeh}.} \bibinfo{year}{2022}\natexlab{}.
\newblock \showarticletitle{Maximizing Fair Content Spread via Edge Suggestion
  in Social Networks}.
\newblock \bibinfo{journal}{\emph{Proceedings of the VLDB Endowment}}
  \bibinfo{volume}{15}, \bibinfo{number}{11} (\bibinfo{year}{2022}).
\newblock


\bibitem[\protect\citeauthoryear{Vaswani, Shazeer, Parmar, Uszkoreit, Jones,
  Gomez, Kaiser, and Polosukhin}{Vaswani et~al\mbox{.}}{2017}]%
        {vaswani2017attention}
\bibfield{author}{\bibinfo{person}{Ashish Vaswani}, \bibinfo{person}{Noam
  Shazeer}, \bibinfo{person}{Niki Parmar}, \bibinfo{person}{Jakob Uszkoreit},
  \bibinfo{person}{Llion Jones}, \bibinfo{person}{Aidan~N Gomez},
  \bibinfo{person}{{\L}ukasz Kaiser}, {and} \bibinfo{person}{Illia
  Polosukhin}.} \bibinfo{year}{2017}\natexlab{}.
\newblock \showarticletitle{Attention is all you need}.
\newblock \bibinfo{journal}{\emph{Advances in neural information processing
  systems}}  \bibinfo{volume}{30} (\bibinfo{year}{2017}).
\newblock


\bibitem[\protect\citeauthoryear{Wang, Li, Yu, and Feng}{Wang
  et~al\mbox{.}}{2011}]%
        {wang2011entity}
\bibfield{author}{\bibinfo{person}{Jiannan Wang}, \bibinfo{person}{Guoliang
  Li}, \bibinfo{person}{Jeffrey~Xu Yu}, {and} \bibinfo{person}{Jianhua Feng}.}
  \bibinfo{year}{2011}\natexlab{}.
\newblock \showarticletitle{Entity matching: How similar is similar}.
\newblock \bibinfo{journal}{\emph{Proceedings of the VLDB Endowment}}
  \bibinfo{volume}{4}, \bibinfo{number}{10} (\bibinfo{year}{2011}),
  \bibinfo{pages}{622--633}.
\newblock


\bibitem[\protect\citeauthoryear{Wang and Li}{Wang and Li}{2022}]%
        {wang2022minun}
\bibfield{author}{\bibinfo{person}{Jin Wang} {and} \bibinfo{person}{Yuliang
  Li}.} \bibinfo{year}{2022}\natexlab{}.
\newblock \showarticletitle{Minun: evaluating counterfactual explanations for
  entity matching}. In \bibinfo{booktitle}{\emph{Proceedings of the Sixth
  Workshop on Data Management for End-To-End Machine Learning}}.
  \bibinfo{pages}{1--11}.
\newblock


\bibitem[\protect\citeauthoryear{Wang, Li, and Hirota}{Wang
  et~al\mbox{.}}{2021}]%
        {Wang0H21}
\bibfield{author}{\bibinfo{person}{Jin Wang}, \bibinfo{person}{Yuliang Li},
  {and} \bibinfo{person}{Wataru Hirota}.} \bibinfo{year}{2021}\natexlab{}.
\newblock \showarticletitle{Machamp: {A} Generalized Entity Matching
  Benchmark}. In \bibinfo{booktitle}{\emph{{CIKM}}}.
  \bibinfo{publisher}{{ACM}}, \bibinfo{pages}{4633--4642}.
\newblock


\bibitem[\protect\citeauthoryear{Yi, Xing-Chun, Jian-Jun, Xing, and Yu-Ling}{Yi
  et~al\mbox{.}}{2017}]%
        {yi2017method}
\bibfield{author}{\bibinfo{person}{Liu Yi}, \bibinfo{person}{Diao Xing-Chun},
  \bibinfo{person}{Cao Jian-Jun}, \bibinfo{person}{Zhou Xing}, {and}
  \bibinfo{person}{Shang Yu-Ling}.} \bibinfo{year}{2017}\natexlab{}.
\newblock \showarticletitle{A method for entity resolution in high dimensional
  data using ensemble classifiers}.
\newblock \bibinfo{journal}{\emph{Mathematical Problems in Engineering}}
  \bibinfo{volume}{2017} (\bibinfo{year}{2017}).
\newblock


\bibitem[\protect\citeauthoryear{Yu, Li, Deng, and Feng}{Yu
  et~al\mbox{.}}{2016}]%
        {yu2016string}
\bibfield{author}{\bibinfo{person}{Minghe Yu}, \bibinfo{person}{Guoliang Li},
  \bibinfo{person}{Dong Deng}, {and} \bibinfo{person}{Jianhua Feng}.}
  \bibinfo{year}{2016}\natexlab{}.
\newblock \showarticletitle{String similarity search and join: a survey}.
\newblock \bibinfo{journal}{\emph{Frontiers of Computer Science}}
  \bibinfo{volume}{10}, \bibinfo{number}{3} (\bibinfo{year}{2016}),
  \bibinfo{pages}{399--417}.
\newblock


\bibitem[\protect\citeauthoryear{Zafar, Valera, Rogriguez, and Gummadi}{Zafar
  et~al\mbox{.}}{2017}]%
        {zafar2017fairness}
\bibfield{author}{\bibinfo{person}{Muhammad~Bilal Zafar},
  \bibinfo{person}{Isabel Valera}, \bibinfo{person}{Manuel~Gomez Rogriguez},
  {and} \bibinfo{person}{Krishna~P Gummadi}.} \bibinfo{year}{2017}\natexlab{}.
\newblock \showarticletitle{Fairness constraints: Mechanisms for fair
  classification}. In \bibinfo{booktitle}{\emph{Artificial intelligence and
  statistics}}. PMLR, \bibinfo{pages}{962--970}.
\newblock


\bibitem[\protect\citeauthoryear{Zemel, Wu, Swersky, Pitassi, and Dwork}{Zemel
  et~al\mbox{.}}{2013}]%
        {zemel2013learning}
\bibfield{author}{\bibinfo{person}{Rich Zemel}, \bibinfo{person}{Yu Wu},
  \bibinfo{person}{Kevin Swersky}, \bibinfo{person}{Toni Pitassi}, {and}
  \bibinfo{person}{Cynthia Dwork}.} \bibinfo{year}{2013}\natexlab{}.
\newblock \showarticletitle{Learning fair representations}. In
  \bibinfo{booktitle}{\emph{International conference on machine learning}}.
  PMLR, \bibinfo{pages}{325--333}.
\newblock


\bibitem[\protect\citeauthoryear{Zhang, Nie, Wu, Shen, and Tan}{Zhang
  et~al\mbox{.}}{2020}]%
        {zhang2020multi}
\bibfield{author}{\bibinfo{person}{Dongxiang Zhang}, \bibinfo{person}{Yuyang
  Nie}, \bibinfo{person}{Sai Wu}, \bibinfo{person}{Yanyan Shen}, {and}
  \bibinfo{person}{Kian-Lee Tan}.} \bibinfo{year}{2020}\natexlab{}.
\newblock \showarticletitle{Multi-context attention for entity matching}. In
  \bibinfo{booktitle}{\emph{Proceedings of The Web Conference 2020}}.
  \bibinfo{pages}{2634--2640}.
\newblock


\bibitem[\protect\citeauthoryear{Zhang, Shahbazi, Chu, and Asudeh}{Zhang
  et~al\mbox{.}}{2021}]%
        {zhang2021fairrover}
\bibfield{author}{\bibinfo{person}{Hantian Zhang}, \bibinfo{person}{Nima
  Shahbazi}, \bibinfo{person}{Xu Chu}, {and} \bibinfo{person}{Abolfazl
  Asudeh}.} \bibinfo{year}{2021}\natexlab{}.
\newblock \showarticletitle{FairRover: explorative model building for fair and
  responsible machine learning}. In \bibinfo{booktitle}{\emph{Proceedings of
  the Fifth Workshop on Data Management for End-To-End Machine Learning}}.
  \bibinfo{pages}{1--10}.
\newblock


\end{thebibliography}

\techrep{
\newpage
\section*{Appendix} \label{apx}
\appendix
\section{Group Encoding} 

To unify all attribute-value types, we  summarize subgroups in an encoding  and use this encoding to represent individual entities and entity pairs. Given a set of sensitive attributes $\mathcal{A}=\{A_1, \ldots, A_n\}$  and value domains $dom(A_i)$ for attributes $A_i$,  $\gee=\{g_1, \ldots, g_m\}$ denotes the set of all level-1 groups, i.e. $\gee=\bigcup_{A_i\in\mathcal{A}}dom(A_i)$. 
We represent a subgroup $s$ of level $k$ ($k$-combination) consisting of  groups  $s=\{g_1,\ldots,g_k\}$, with a binary encoding $s=\langle a_1, \ldots, a_m \rangle$, where $m=|dom(A_1)|\times\ldots|dom(A_i)|)$ and  $a_i$ is one if $g_i\in s$ and is zero otherwise. Note that for a $k$-combination subgroup, exactly $k$ entries of $s$ get the value one. 
We represent an entity $e$ associated with groups $G\subseteq\gee$ with a binary encoding $\langle b_1, \ldots, b_m \rangle$, where  $b_i$ is one if $g_i\in G$ and is zero otherwise. An entity $e$ with groups $G$ belongs to subgroup $s$ if $s\subseteq G$. 
Given an entity encoding $e=\langle b_1, \ldots, b_m \rangle$ and a subgroup encoding $s=\langle a_1, \ldots, a_m \rangle$, 
we say $e$ belongs to subgroup $s$ if $s~AND~e~==~s$, i.e. the entity belongs to every group that defines the subgroup $s$. The encoding of an  entity pair $e_i, e_j$ is the concatenation of the encodings of  $e_i$ and $e_j$. 

\begin{example} Consider attributes {\tt \small genre} and {\tt \small gender} of Figure~\ref{fig:lattice}. Assuming a  lexicographical order on all groups,  the encoding of entity $e$ with associated groups  $G$=\{{\tt \small Female}, {\tt \small Pop}, {\tt \small Rock}\} is $\langle 1, 0, 0, 1, 1 \rangle$. The encoding of a level-$2$ subgroup $s$=\{{\tt \small Female}, {\tt \small Pop}\} is $\langle 1, 0, 0, 1, 0 \rangle$. 
\end{example}

\section{Evaluating the Correctness of Entity Matchers}
\label{sec:apx:correctness}
The correctness of a matcher measures how well its matching predictions match the ground-truth. Given a test dataset with correspondences of  $t=(e_i, e_j, h, y)$, where $h$ is a binary variable indicating the result of EM (\textit{match} or \textit{non-match}) for entities with encodings $e_i$ and $e_j$, and $y$ is a binary variable indicating the ground-truth for matching, we profile predictions of $h$  using the numbers of true positives (TP), true negatives (TN), false positives (FP), and false negatives (FN), respectively. Unlike a classification task, in the confusion matrix of a matching task, the result is counted both for the group(s) of $e_i$ and the group(s) of $e_j$. 

\begin{example}\label{ex:apx:confusion} 
Consider a test dataset, shown in Table~\ref{fig:ex-1}, for a matcher $\mathcal{M}$, where columns {\tt $id_1$} and {\tt $id_2$} contain entity encodings, column $h$ is the output decision of $\mathcal{M}$, and column $y$ is the ground-truth. 
Comparing columns $h$ and $y$, we add and populate column $Result$ for each entity pair. Consider the simple case of having two groups $\gee=\{g_1,g_2\}$. 
Suppose we would like to evaluate single fairness for $g_1$ and $g_2$. We describe how the confusion matrices of these groups are created. 
Consider the first row in Table~\ref{fig:ex:1:2}, which happens to be an FP. Since $e_1$ and $e_2$ both belong to group $g_1$, the value $2$ will be added to the count of FPs in the confusion matrix of $g_1$. However, in the second row which happens to be a TN, $e_3$ belongs to $g_2$ while $e_4$ belongs to $g_1$. Thereby, we will add one to both TN values of the confusion matrix corresponding to group $g_1$ and $g_2$. We repeat the same procedure for rows three and four. The completed confusion matrices are shown in Figures~\ref{fig:ex:1:3} and~\ref{fig:ex:1:4}.
\end{example}

We measure correctness for single and pairwise fairness  through well-studied metrics in
literature, including precision, recall, and F-1 score.

\begin{figure}[H] 
\centering
    \begin{subfigure}[t]{0.48\textwidth}
    \centering
    \small
    \begin{tabular}{||c|c|@{}c@{}|@{}c@{}|c|c|c||}
            \hline
            id$_1$&id$_2$&group($id_1$)&group($id_2$)&$h$&$y$&Result \\ [0.5ex] 
            \hline \hline 
            $e_1$&$e_2$&$g_1$&$g_1$&\textit{`M'}&\textit{`N'}&FP \\ \hline
            $e_3$&$e_4$&$g_2$&$g_1$&\textit{`N'}&\textit{`N'}&TN \\ \hline
            $e_1$&$e_4$&$g_1$&$g_1$&\textit{`M'}&\textit{`M'}&TP \\ \hline
            $e_2$&$e_3$&$g_1$&$g_2$&\textit{`N'}&\textit{`M'}&FN \\ \hline
        \end{tabular}
        \caption{}
        \label{fig:ex:1:2}
    \end{subfigure}
    
    \hspace{-13mm}\begin{subfigure}[t]{0.245\linewidth}
        \small
        \begin{tabular}{cc|c|c|c|}
            &\multicolumn{1}{c}{}&\multicolumn{2}{c}{\textbf{Actual}}\\
            &\multicolumn{1}{c}{}&\multicolumn{1}{c}{\textbf{y=`M'}}
            &\multicolumn{1}{c}{\textbf{y=`N'}}\\
            \cline{3-4}
            \multicolumn{1}{c}{\multirow{2}{*}{\rotatebox{90}{\textbf{Predicted}}}}
            &\textbf{h=`M'} &TP=2 &FP=2\\
            \cline{3-4}
            &\textbf{h=`N'} &FN=1 &TN=1\\
            \cline{3-4}
            \end{tabular}
            \vspace{2mm}
        \caption{}
        \label{fig:ex:1:3}
    \end{subfigure}
    \hspace{20mm}
    \begin{subfigure}[t]{0.245\linewidth}
        \small
        \begin{tabular}{cc|c|c|c|}
            &\multicolumn{1}{c}{}&\multicolumn{2}{c}{\textbf{Actual}}\\
            &\multicolumn{1}{c}{}&\multicolumn{1}{c}{\textbf{y=`M'}}
            &\multicolumn{1}{c}{\textbf{y=`N'}}\\
            \cline{3-4}
            \multicolumn{1}{c}{\multirow{2}{*}{\rotatebox{90}{\textbf{Predicted}}}}
            &\textbf{h=`M'} &TP=0 &FP=0\\
            \cline{3-4}
            &\textbf{h=`N'} &FN=1 &TN=1\\
            \cline{3-4}
            \end{tabular}
            \vspace{2mm}
        \caption{}
        \label{fig:ex:1:4}
    \end{subfigure}
    \vspace{-3mm}
    \caption{(a) Matching Results (b) Confusion Matrix of $g_1$ (c) Confusion Matrix of $g_2$.} 
    \label{fig:ex-1}
\end{figure}

\section{Deploying Fair Entity Matching Framework}
In this section, we provide a step-by-step instruction set on how to use our proposed fair entity matching framework:

\begin{algorithm}[!tb]
    \label{alg:pseudo-code}
    \caption{Fair Entity Matching Evaluation}
    \begin{algorithmic}[1] \small
    \Require{Dataset $\dee$, Sensitive attribute $\gee$, Fairness threshold $\tau$, Matching ground-truth $y$}
    \Ensure{Lists of discriminated groups $g_{\mathit{single}}$, $g_{\mathit{pairwise}}$ w.r.t {\em single} and {\em pairwise} fairness definitions}
    \State {\em train}, {\em test}, {\em valid} $\gets$ {\sc split}($\dee$, $y$)
    \State $\mathcal{M} \gets$ {\sc matchingModel}()
    \State $\mathcal{M}$.{\sc fit}({\em train}, {\em valid})
    \State $h \gets \mathcal{M}$.{\sc predict}({\em test})
    \State Let {\em utility} be a matching quality metric value such as TPR, PPV, etc.
    \State {\em utility} $\gets \mathcal{M}$.{\sc evaluate}($y_{\mathit{test}}$, $h$)
    \State
    \State {\tt\footnotesize ~~//single fairness}
    \State $g_{\mathit{single}} \gets$ empty list
    \For{g$_i \in \gee$}:
    \State Let $h_{\text{g}_i}$ be the prediction of $\mathcal{M}$ on tuples where either left or right entity belong to g$_i$
    \State Let $y_{\text{g}_i}$ be the ground-truth of tuples where either left or right entity belongs to g$_i$
    \State $\mathit{utility_{\text{g}_i}} \gets \mathcal{M}$.{\sc evaluate}($y_{\text{g}_i}$, $h_{\text{g}_i}$)
    \If{{\sc disparity}($\mathit{utility},\mathit{utility_{\text{g}_i}})>\tau$}:
        \State $g_{\mathit{single}}$.{\sc add}(g$_i$)
    \EndIf
    \EndFor
    \State
    \State {\tt\footnotesize ~~//pairwise fairness}
    \State $g_{\mathit{paiwise}} \gets$ empty list
    \For{g$_i$, g$_j \in \gee$}:
    \State Let $h_{\text{g}_{ij}}$ be the prediction of $\mathcal{M}$ on tuples where left entity belongs to g$_i$ and the right entity belongs to g$_j$
    \State Let $y_{\text{g}_{ij}}$ be the ground-truth of tuples where left entity belongs to g$_i$ and the right entity belongs to g$_j$
    \State $\mathit{utility_{\text{g}_{ij}}} \gets \mathcal{M}$.{\sc evaluate}($y_{\text{g}_{ij}}$, $h_{\text{g}_{ij}}$)
    \If{{\sc disparity}($\mathit{utility},\mathit{utility_{\text{g}_{ij}}})>\tau$}:
        \State $g_{\mathit{pairwise}}$.{\sc add}($\langle$ g$_i$, g$_j\rangle$)
    \EndIf
    \EndFor
    \State
    \State {\bf return} $g_{\mathit{single}}, g_{\mathit{pairwise}}$
    \end{algorithmic}
\end{algorithm}

\section{Additional Experiments}
\subsection{Correctness}
\label{sec:apx:evalcorrecteness}

\begin{table*}
    \centering \small
    \begin{tabular}{l l l l l l l l l l l l l l l l l} 
         \toprule
         &  \multicolumn{2}{c}{\thead{{\nofly}}}&  \multicolumn{2}{c}{\thead{{\fmatch}}}&  \multicolumn{2}{c}
         {\thead{{\sc iTunes-Amazon}}}&  \multicolumn{2}{c}{\thead{{\sc Dblp-Acm}}}&  \multicolumn{2}{c}{\thead{{\sc Dblp-Scholar}}}&  \multicolumn{2}{c}{\thead{{\sc Cricket}}}&  \multicolumn{2}{c}{\thead{{\sc Shoes}}}&  \multicolumn{2}{c}{\thead{{\sc Camera}}} \\ 
\thead{{\bf Matcher}}& \thead{{\bf Acc}} &\thead{{\bf F-1}}& \thead{{\bf Acc}} &\thead{{\bf F-1}}& \thead{{\bf Acc}} &\thead{{\bf F-1}}& \thead{{\bf Acc}} &\thead{{\bf F-1}}& \thead{{\bf Acc}} &\thead{{\bf F-1}}& \thead{{\bf Acc}} &\thead{{\bf F-1}}& \thead{{\bf Acc}} &\thead{{\bf F-1}}& \thead{{\bf Acc}} &\thead{{\bf F-1}} \\ \midrule
{\sc BooleanRuleMatcher} &0.99&\textcolor{red}{0.14}& 0.99&\textcolor{red}{0.37}&\textcolor{red}{0.29}&\textcolor{red}{0.41}&\textcolor{red}{0.41}& \textcolor{red}{0.38}&\textcolor{red}{0.38}& \textcolor{red}{0.38}&\textcolor{red}{0.03}& \textcolor{red}{0.0}&0.82&\textcolor{red}{0.28}&0.81&\textcolor{red}{0.4} \\ \midrule
\midrule
{\sc DeepMatcher}&0.99&0.84& 0.99&0.70&0.94& 0.88& 0.99& 0.98&0.97& 0.92&0.87&  0.92&0.96& 0.82 &0.93& 0.81 \\ \midrule
{\sc Ditto}&0.99&0.79& 0.99&0.78&0.91& 0.84&0.99& 0.98&0.95&0.87&0.96&  0.98&0.95& 0.78&0.91& 0.76 \\ \midrule
{\sc Gnem}&0.99&0.86&0.99&0.85&0.64&\textcolor{red}{0.31}&0.70&\textcolor{red}{0.18}&0.83& 0.58& 0.96&  0.98&0.96&0.80&0.97&0.91 \\ \midrule
{\sc HierMatcher}&0.99&0.77&0.99&0.72&0.93&0.87&0.95&0.88&0.96& 0.9&0.81&  0.89& 0.96&0.81&0.94&0.83 \\ \midrule
{\sc Mcan}&0.99&0.69&0.99&0.68&0.97&0.94&0.99&0.99&0.97& 0.92&0.95&  0.97&0.95&0.73&0.93&0.78 \\ \midrule
\midrule
{\sc SvmMatcher}&0.99&0.99&0.99&0.94 & 0.92&0.84& 0.96&0.90&0.94& 0.86&0.96&  0.98&0.89&\textcolor{red}{0.0}&0.84&\textcolor{red}{0.27} \\ \midrule
{\sc RfMatcher}&1.00&1.00&0.99&0.95& 0.94&0.89&0.99&0.97&0.98& 0.94&0.96&  0.98&0.88&\textcolor{red}{0.29}&0.82&\textcolor{red}{0.38} \\ \midrule
{\sc NbMatcher} &0.99&0.98&0.96&\textcolor{red}{0.11}& 0.88&0.78&0.98&0.97&0.99& 0.97&0.96& 931& 0.86&\textcolor{red}{0.26}&0.82&\textcolor{red}{0.38} \\ \midrule
{\sc LogRegMatcher} &1.00&1.00&0.99&0.93& 0.91 &0.83&0.99 &0.97&0.99& 0.97& 0.96&  0.98& 0.89 &\textcolor{red}{0.04}& 0.84 &\textcolor{red}{0.31} \\ \midrule
{\sc LinRegMatcher} &0.99&0.95&0.99&\textcolor{red}{0.38}& 0.97&0.94& 0.97&0.93&0.95& 0.88&0.96&  0.97& 0.89&\textcolor{red}{0.0}&0.84&\textcolor{red}{0.30} \\ \midrule
{\sc Dedupe} &-&-&-&-& 0.94&0.89&0.95&0.85&0.95& 0.87& 0.96& 0.98 & - & - & - & -\\ \midrule
{\sc DtMatcher} &1.00&1.00&0.99&0.93& 0.94&0.89&0.99&0.97&0.98& 0.94& 0.93&  0.96&0.85&\textcolor{red}{0.30}&0.84&\textcolor{red}{0.31} \\
\bottomrule
    \end{tabular}
    \caption{Overall performance of matchers across different datasets (Acc: Accuracy)}
    \label{tbl:correctness}
\end{table*}

\begin{figure}[H]
    \centering \small
    \scalebox{0.9}{
    \begin{tabular}{l l l} 
         \toprule
\thead{{\bf Accurate}} & \thead{{\bf Fair}} & \thead{{\bf Evidence}}  \\ \midrule
 $\times$ & $\times$& {\sc RfMatcher}: {\sc Cameras}: \{TPRP,PPVP\}\\
 & & {\sc BooleanRuleMatcher}: {\sc iTunes-Amazon}: \{AP,SP,PPVP\}\\
  & & {\sc Gnem}: {\sc iTunes-Amazon}: \{AP,PPVP,...\}\\\midrule
 $\times$ & \checkmark& {\sc LinRegMatcher}: {\sc Shoes} \\
  & & {\sc Gnem}: {\sc Dblp-Acm}\\
 & &  {\sc BooleanRuleMatcher}: {\sc Cricket}\\\midrule
 \checkmark & $\times$& {\sc HierMatcher}: {\sc iTunes-Amazon}: \{AP,PPVP,...\} \\
 & & {\sc SvmMatcher}: {\sc Dblp-Acm}: PPVP\\
 & & {\sc Mcan}: {\sc Cameras}: TPRP\\
 & & {\sc Ditto}: {\sc Dblp-Scholar}: \{AP,TPRP,...\}\\
 & & {\sc DeepMatcher}: {\sc Camera}: \{PPVP,TPRP\} \\\midrule
\checkmark & \checkmark&  {\sc Mcan}: {\sc Dblp-Acm}\\
& &  {\sc Ditto}: {\sc Cricket}\\
& &  {\sc NbMatcher}: {\sc Dblp-Scholar}\\ 
\bottomrule
    \end{tabular}}
    \caption{Fairness and accuracy synergies}
    \label{tbl:acc-fairness}
\end{figure}

{\em Neural matchers are more accurate than non-neural matchers on textual and dirty data.}  The correctness results of the textual datasets: {\sc Shoes} and {\sc Camera},  can be found in  
Table~\ref{tbl:correctness}. Non-neural matchers extensively suffer in F-1 score, compared to neural matchers that have higher ranges of F-1 score. Modern neural matchers such as {\sc Ditto} and {\sc DeepMatcher} draw on external knowledge by incorporating language models, which helps a matcher to learn the relevance of entities despite the lack of structure and syntactic similarity in text entities. This result is consistent with what is reported by the state-of-art matchers. 

{\em Non-neural matchers are more accurate than neural matchers on structured data.}  
Considering the structured datasets: {\sc iTunes-Amazon} and {\sc Dblp-Acm}, although all matchers, with the exception of {\sc BooleanRuleMatcher}, perform quite well, the non-neural matchers have slightly higher F-1 scores overall.  

The main job of a matcher is to find matching entities. A failure in doing so results in a low number of TPs,  which reflects in not only a low F-1 score but also unfairness with respect to TPRP and PPVP measures for many groups across the board, as we observe in Figures \ref{fig:cameras-single} and~\ref{fig:shoes-single}. Recall that these measures verify how well a matcher performs in identifying true positives. Similar observations can be made in both datasets regarding the unfairness of the {\sc BooleanRuleMatcher}. 
One interesting observation in {\sc Gnem}, which is the only neural matcher with a low F-1 score for the {\sc Dblp-Acm} dataset, is that the high number of FNs of {\sc Gnem} results in the pairwise  unfairness for $g_i|g_i$ pairs  (e.g., {\tt\small SIGMOD|SIGMOD} and {\tt\small ACM TODS|ACM TODS}). {\sc Gnem} does not demonstrate  NPVP unfairness on  $g_i|g_j$ pairs, particularly in the {\sc Dblp-Acm} dataset, because often two  entities with $g_i|g_j$ are not a match (e.g. it is rare that a {\tt\small ACM TODS} publication is matched with a publication {\tt\small SIGMOD}).  

The flip side of correctness and fairness also exists in EM. For example, the {\sc BooleanRuleMatcher} and {\sc Gnem} have low accuracy and F-1 score on {\sc Dblp-Acm}, while no unfairness issue is reported for these matchers, in Figure~\ref{fig:DBLP-ACM-single}. This can be explained by the low accuracy of these matchers for all groups across the board which makes the disparity a low value. 
In Figure ~\ref{tbl:acc-fairness}, we present a selective overview of the unfairness and accuracy of matchers across all datasets. 
The general message is that similar to accuracy, unfairness is dataset and measure dependent. 
First, no matcher is unfair across all datasets. 
For example, {\sc Gnem} is unfair for {\sc iTunes-Amazon} and {\sc Dblp-Acm} but is not unfair for the {\sc Cricket} dataset. No matcher is unfair across all measures. For example, on {\sc Dblp-Scholar}  dataset, {\sc HierMatcher} is only unfair with respect to PPVP and FPRP and not other measures. We will have a more detailed discussion on the behavior of matcher w.r.t measures in \S~\ref{sec:evalmeasures}.

\subsection{{\sc Cameras} Dataset: Pairwise Fairness}

\begin{figure*}[h]
\begin{minipage}[t]{0.19\linewidth}
        \vspace{-72mm} \footnotesize
\begin{tabular}{@{}l@{}c@{}}
        \toprule
        \thead{{\bf Model}} & \thead{{\bf Marker}} \\
        \midrule
        {\sc DeepMatcher} & {\textcolor[RGB]{255, 0, 0}{\faCircle}} \\
        {\sc Ditto} & {\textcolor[RGB]{0,0,255}{\faCircle}} \\  
        {\sc Gnem} & {\textcolor[RGB]{255,140,0}{\faCircle}} \\  
        {\sc HierMatcher} & {\textcolor[RGB]{0,128,0}{\faCircle}} \\  
        {\sc Mcan} & {\textcolor[RGB]{148,0,211}{\faCircle}} \\
        {\sc SvmMatcher} & {\textcolor[RGB]{0,0,0}{\faStar}} \\
        {\sc RfMatcher} & {\textcolor[RGB]{255,105,180}{\faStar}} \\  
        {\sc NbMatcher} & {\textcolor[RGB]{128,128,128}{\faStar}} \\  
        {\sc LogRegMatcher} & {\textcolor[RGB]{32,177,170}{\faStar}} \\  
        {\sc LinRegMatcher} & {\textcolor[RGB]{144,238,144}{\faStar}} \\  
        {\sc DtMatcher} & {\textcolor[RGB]{159,82,45}{\faStar}} \\  
        {\sc Dedupe} & {\textcolor[RGB]{1,191,255}{\faStar}} \\  
        {\sc BooleanRuleMatcher} & {\textcolor[RGB]{255,215,0}{\faStar}} \\ 
        \bottomrule
        \end{tabular}
    \end{minipage}
    \hfill
\begin{minipage}[t]{0.8\linewidth}
\centering 
    \includegraphics[width=\textwidth]{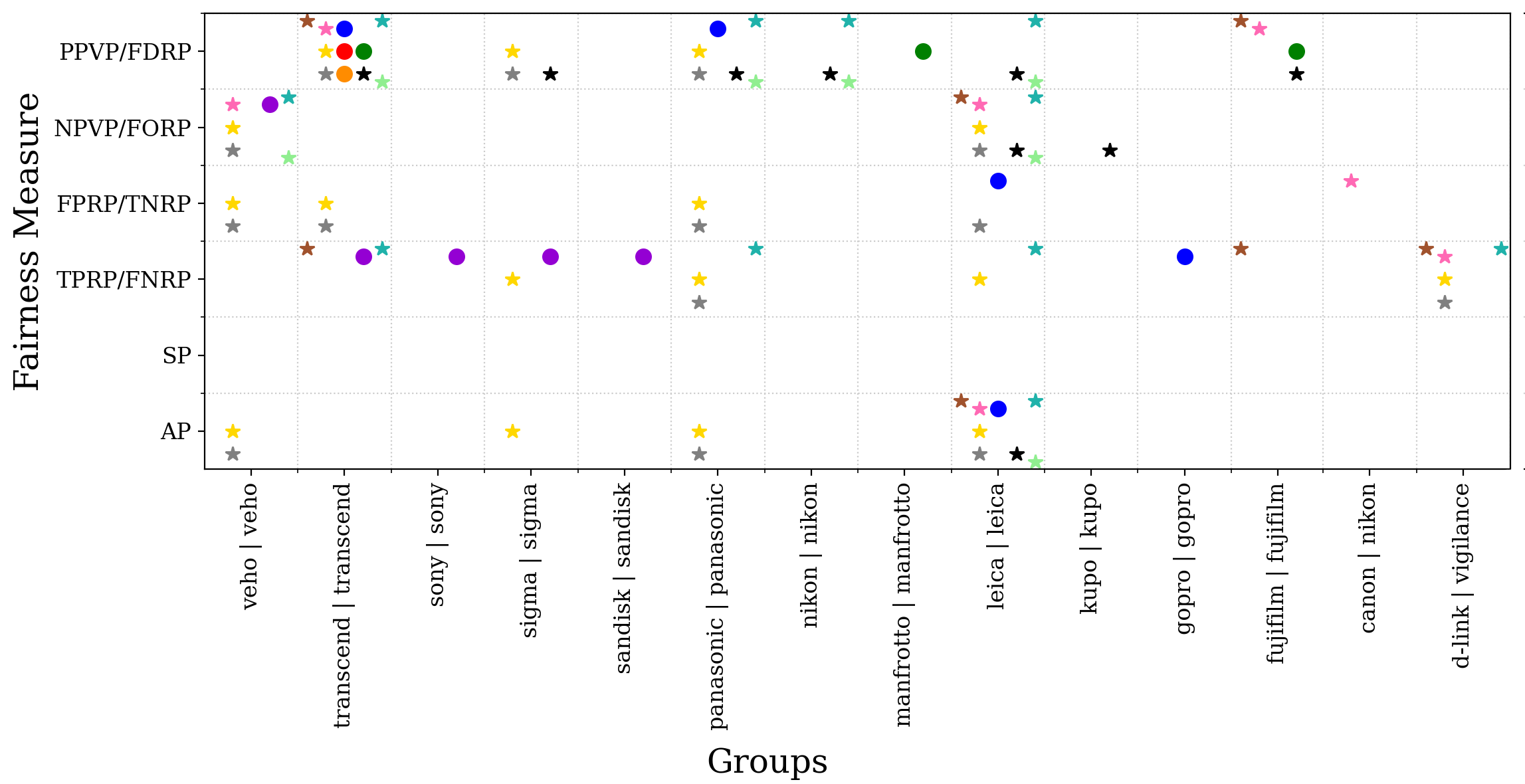}
    \vspace{-8mm}\caption{{\sc Cameras}: Pairwise Fairness.}
\label{fig:cameras-pairwise}
\end{minipage}
\end{figure*}

\subsection{{\sc iTunes-Amazon} Dataset: Pairwise Fairness}
 
In Figure~\ref{fig:itunes-amazon-pairwise}, extensive unfair behavior of neural matchers happens in the pairwise matching of {\tt\small Dance \& Electronic} with {\tt\small Music}  and {\tt\small Dance}.  

\begin{figure*}[!tb]
\centering 
    \includegraphics[width=\textwidth]{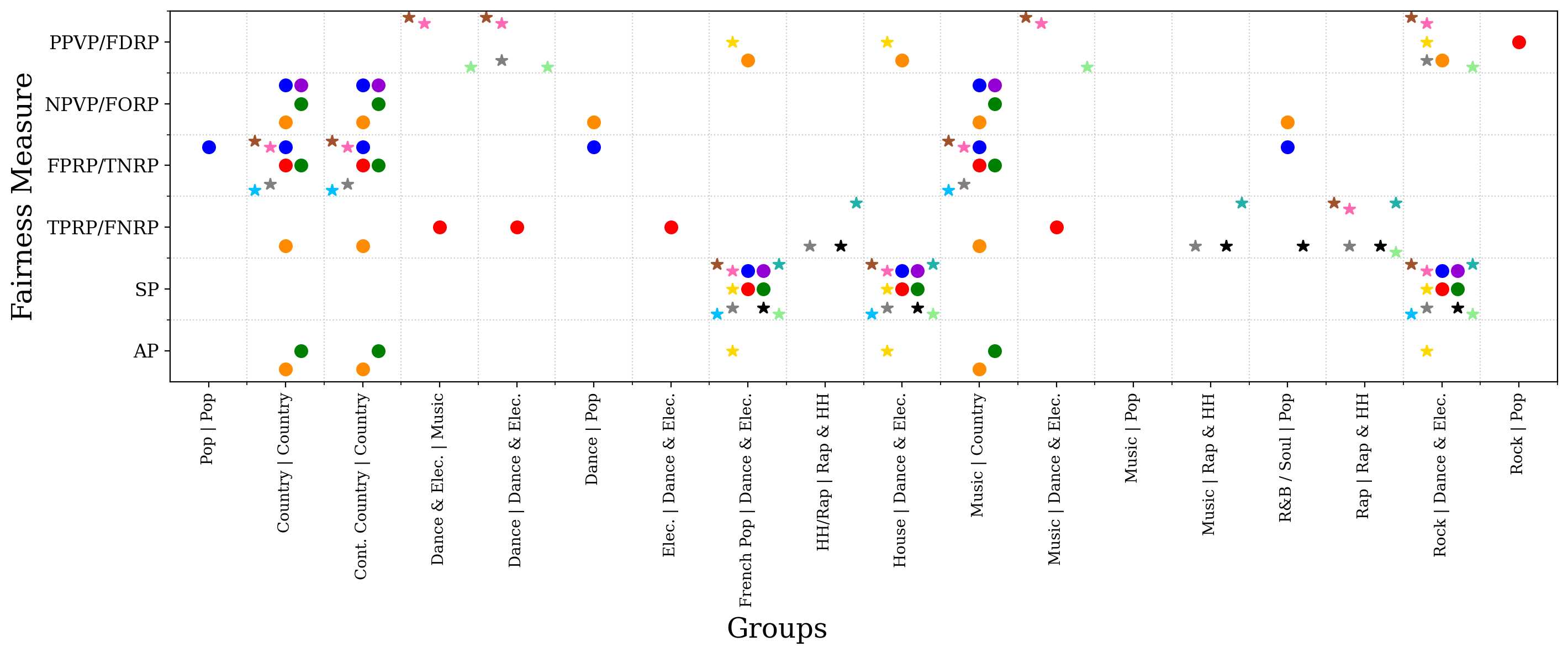}
    \vspace{-8mm}\caption{{\sc iTunes-Amazon}: Pairwise Fairness. (HH: Hip-Hop, Elec.: Electronic, Cont.: Contemporary)}
\label{fig:itunes-amazon-pairwise}
\end{figure*}

\subsection{{\sc Shoes} Dataset: Single and Pairwise Fairness}
The unfair behavior of matchers on {\sc Shoes} dataset is reported in Figures \ref{fig:shoes-single} and \ref{fig:shoes-pairwise}.

\begin{figure*}[!tb]
\begin{minipage}[t]{0.65\linewidth}
\centering 
    \includegraphics[width=\textwidth]{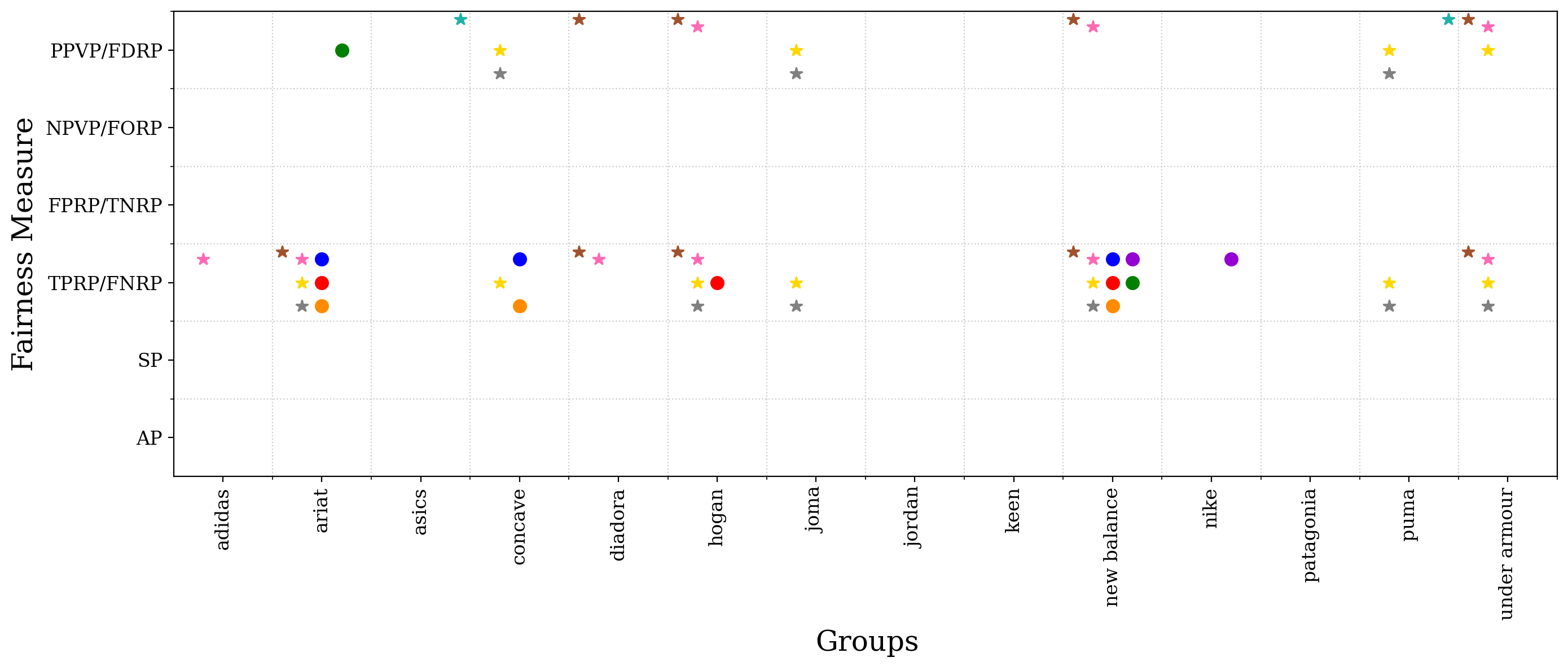}
    \vspace{-8mm}\caption{{\sc Shoes}: Single Fairness}
\label{fig:shoes-single}
\end{minipage} 
\hfill
\begin{minipage}[t]{0.34\linewidth}
\centering 
    \includegraphics[width=.9\textwidth]{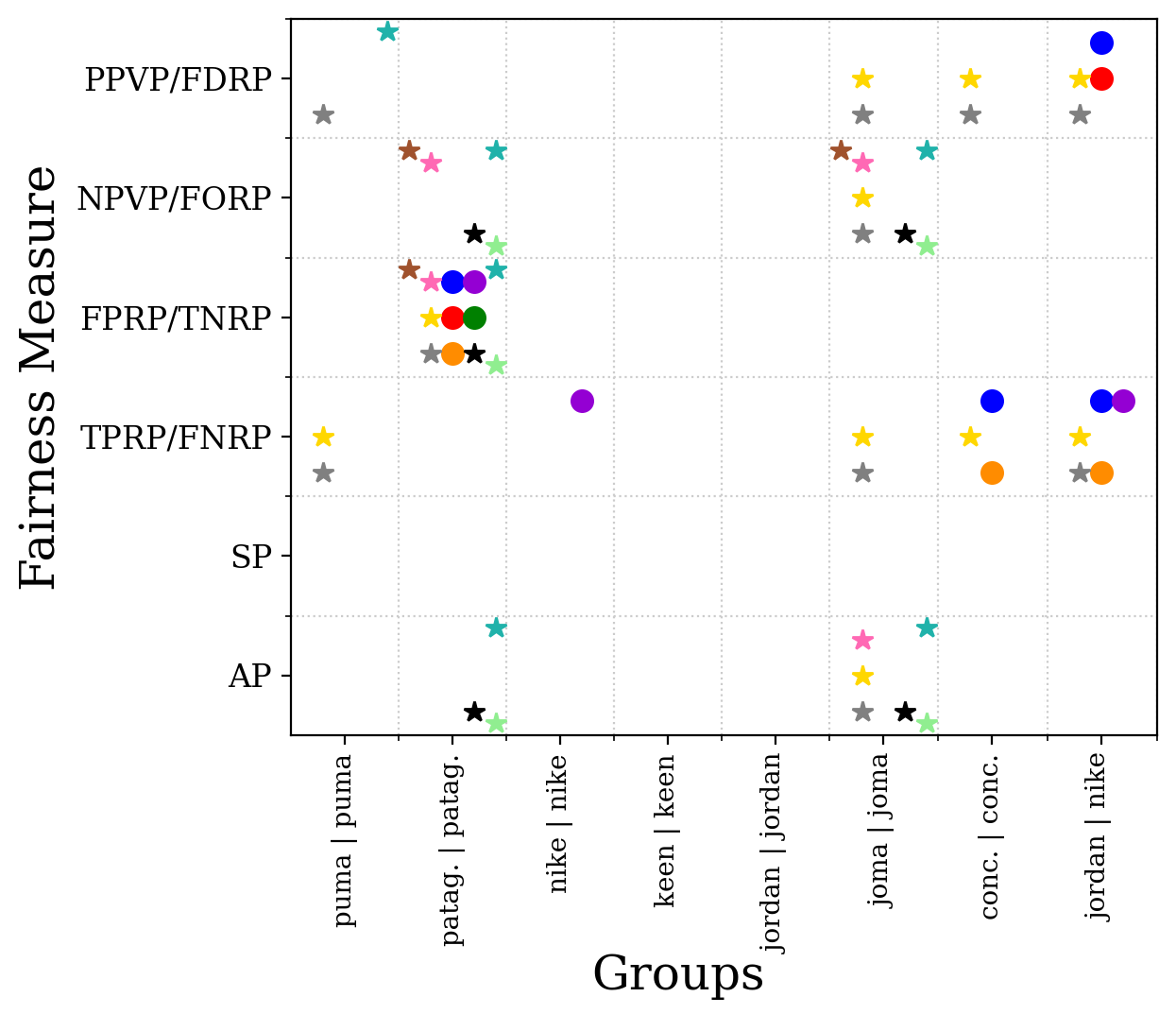}
    \vspace{-4mm}\caption{{\sc Shoes}: Pairwise Fairness}
\label{fig:shoes-pairwise}
\end{minipage} 
\vspace{-3mm}
\end{figure*}

\subsection{Matching Threshold vs. Fairness and Accuracy}
In Figures \ref{fig:heatmap_tprp_dblp_acm}, \ref{fig:heatmap_tprp_dblp_scholar}, \ref{fig:heatmap_tprp_cameras} we show the behavior of matchers at different matching thresholds for {\sc DBLP-ACM}, {\sc DBLP-Scholar} and {\sc Cameras} datasets w.r.t. accuracy (TPR) and fairness (TPRP). Figures \ref{fig:heatmap_ppvp_itunes_amazon}, \ref{fig:heatmap_ppvp_dblp_acm}, \ref{fig:heatmap_ppvp_dblp_scholar}, \ref{fig:heatmap_ppvp_cameras}, show the same thing for {\sc iTunes-Amazon} {\sc DBLP-ACM}, {\sc DBLP-Scholar} and {\sc Cameras} datasets w.r.t. accuracy (PPV) and fairness (PPVP).
\begin{figure*}[ht] 
  \begin{minipage}[b]{0.48\linewidth}
    \centering
    \includegraphics[scale=0.36]{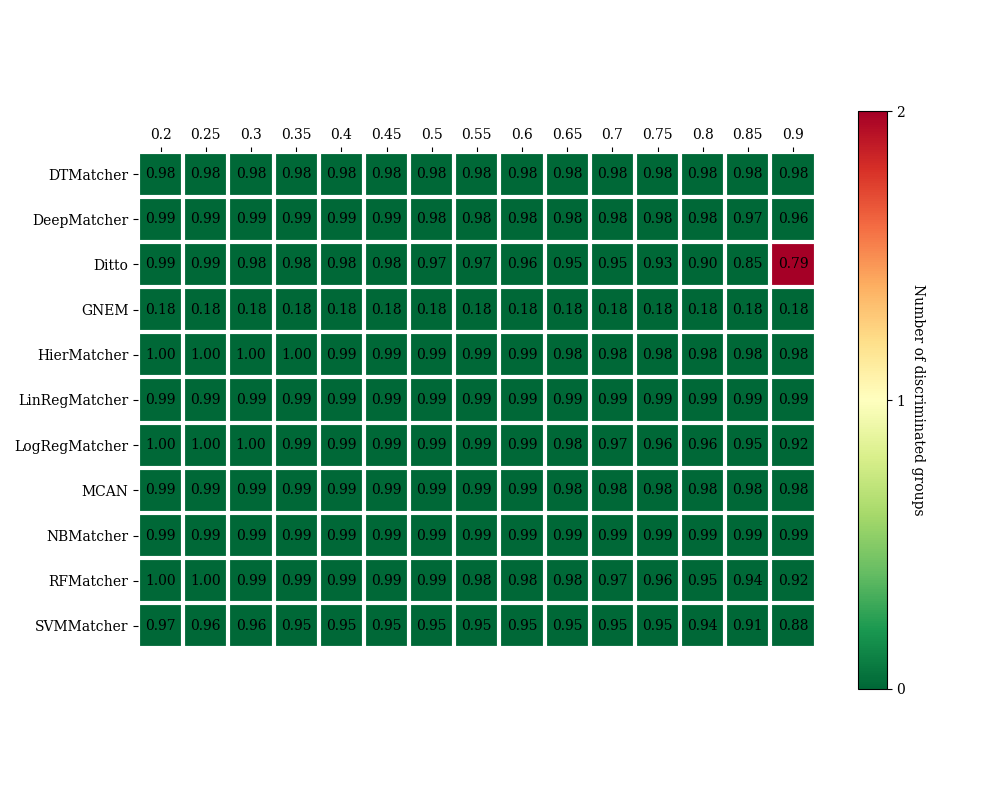} 
    \vspace{-15mm}
    \caption{\rev{The effect of matching threshold on fairness (TPRP) and accuracy (TPR) of the models on {\sc DBLP-ACM} dataset.}} 
    \label{fig:heatmap_tprp_dblp_acm}
  \end{minipage}
  \begin{minipage}[b]{0.48\linewidth}
    \centering
    \includegraphics[scale=0.36]{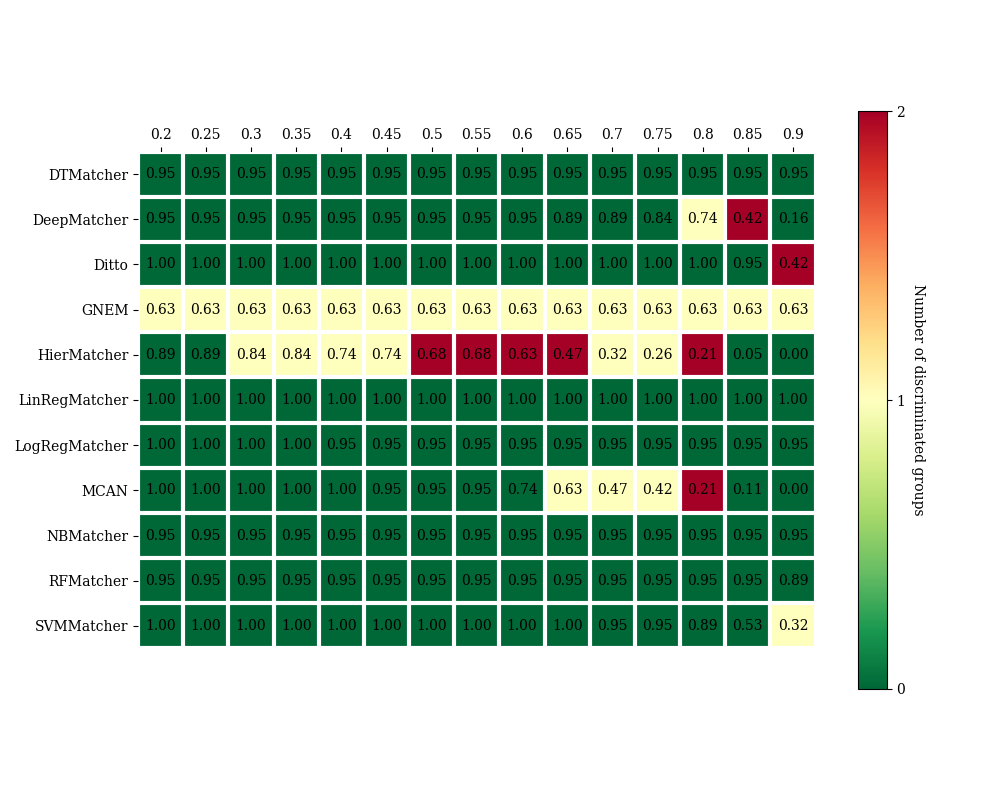} 
    \vspace{-15mm}
    \caption{\rev{The effect of matching threshold on fairness (TPRP) and accuracy (TPR) of the models on {\sc DBLP-Scholar} dataset.}} 
    \label{fig:heatmap_tprp_dblp_scholar}
  \end{minipage}
\end{figure*} 

\begin{figure*}[ht] 
  \begin{minipage}[b]{0.48\linewidth}
    \centering
    \includegraphics[scale=0.36]{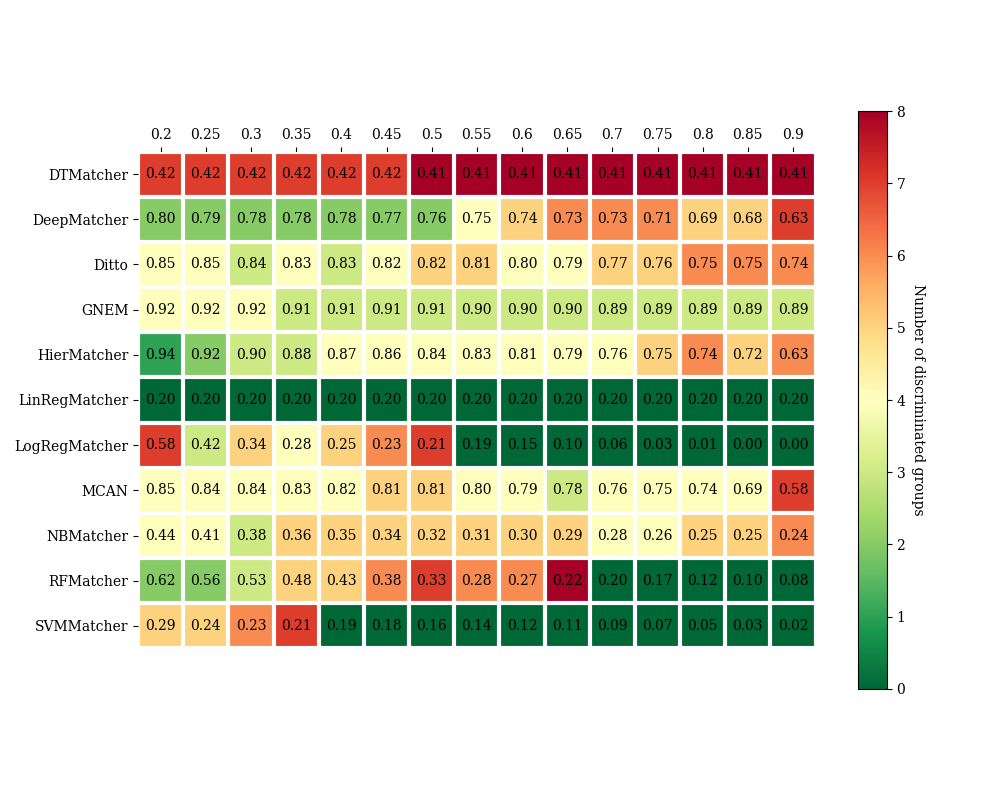} 
    \vspace{-15mm}
    \caption{\rev{The effect of matching threshold on fairness (TPRP) and accuracy (TPR) of the models on {\sc Cameras} dataset.}} 
    \label{fig:heatmap_tprp_cameras}
  \end{minipage}
  \begin{minipage}[b]{0.48\linewidth}
    \centering
    \includegraphics[scale=0.36]{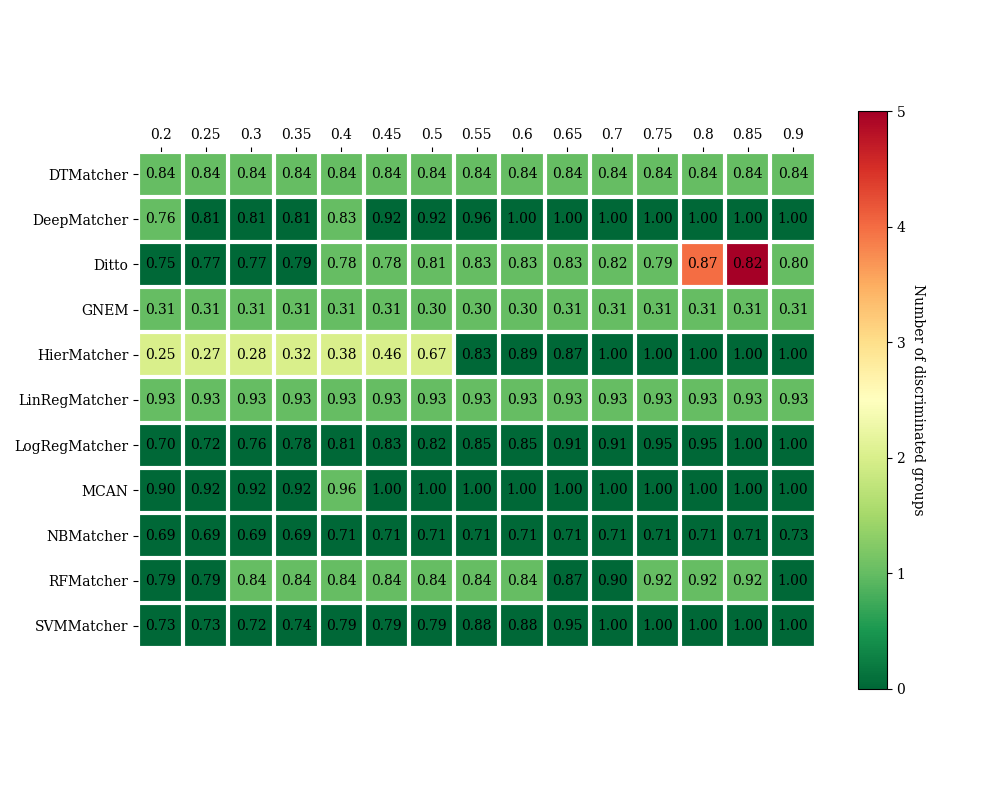} 
    \vspace{-15mm}
    \caption{\rev{The effect of matching threshold on fairness (PPVP) and accuracy (PPV) of the models on {\sc iTunes-Amazon} dataset.}} 
    \label{fig:heatmap_ppvp_itunes_amazon}
  \end{minipage}
\end{figure*} 

\begin{figure*}[ht] 
  \begin{minipage}[b]{0.48\linewidth}
    \centering
    \includegraphics[scale=0.36]{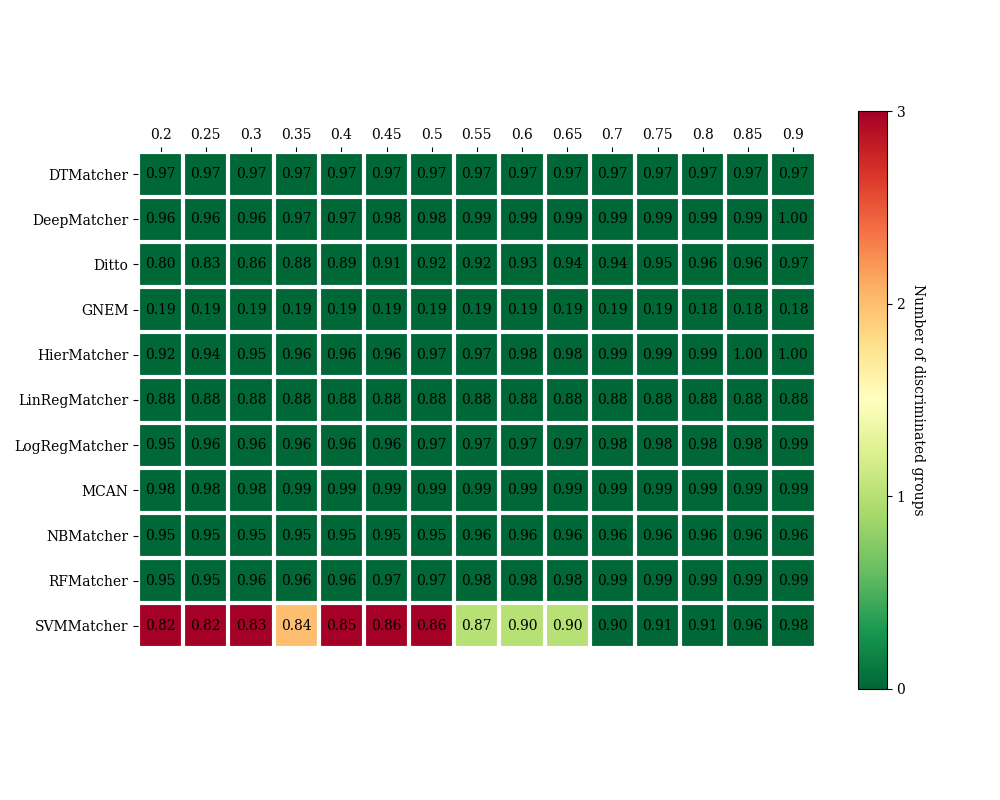} 
    \vspace{-15mm}
    \caption{\rev{The effect of matching threshold on fairness (PPVP) and accuracy (PPV) of the models on {\sc DBLP-ACM} dataset.}} 
    \label{fig:heatmap_ppvp_dblp_acm}
  \end{minipage}
  \begin{minipage}[b]{0.48\linewidth}
    \centering
    \includegraphics[scale=0.36]{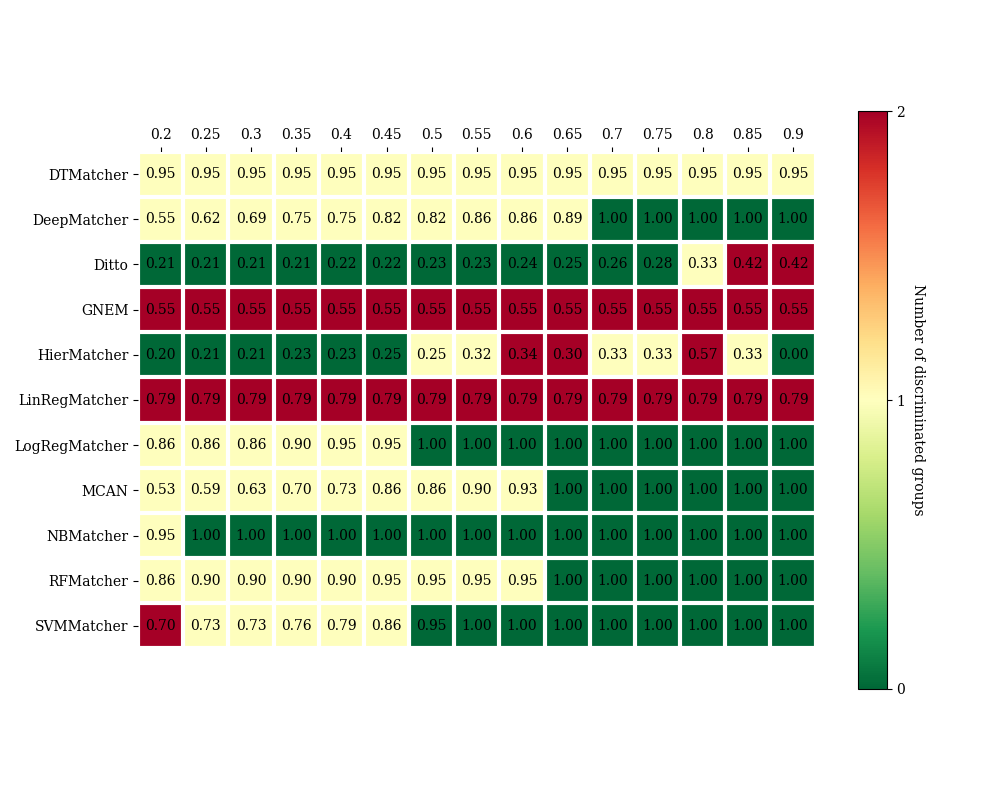} 
    \vspace{-15mm}
    \caption{\rev{The effect of matching threshold on fairness (PPVP) and accuracy (PPV) of the models on {\sc DBLP-Scholar} dataset.}} 
    \label{fig:heatmap_ppvp_dblp_scholar}
  \end{minipage}
\end{figure*} 

\begin{figure*}[ht] 
  \begin{minipage}[b]{0.48\linewidth}
    \centering
    \includegraphics[scale=0.36]{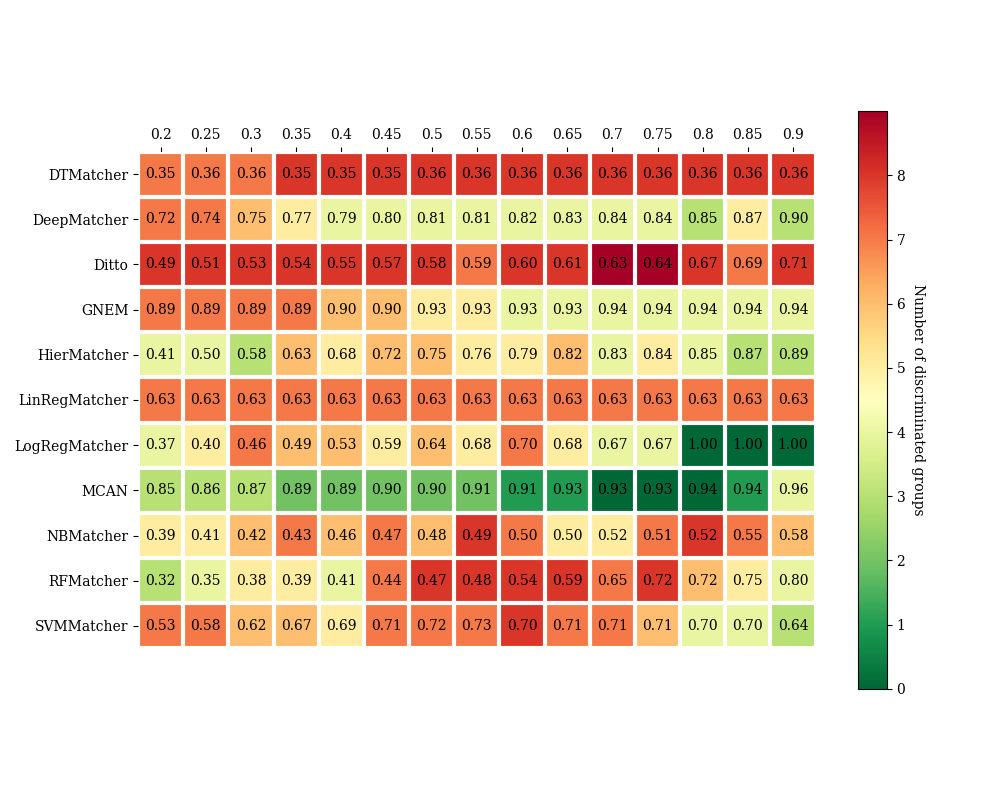} 
    \vspace{-15mm}
    \caption{\rev{The effect of matching threshold on fairness (PPVP) and accuracy (PPV) of the models on {\sc Cameras} dataset.}} 
    \label{fig:heatmap_ppvp_cameras}
  \end{minipage}
\end{figure*} 

}

\end{document}